\documentclass[12pt]{iopart}

\usepackage{graphicx}  
\usepackage{iopams}  
\usepackage[FDR]{harvard}
\usepackage[pdftex,bookmarksopen=false,pdfauthor=Irstea,colorlinks=false,pagebackref,plainpages=false]{hyperref}
\usepackage[usenames,dvipsnames]{color}
\usepackage{multirow}
\usepackage{eurosym}

\newcommand{\x}{\mathbf{x}}
\newcommand{\y}{\mathbf{y}}
\newcommand{\X}{\mathbf{X}}
\newcommand{\Y}{\mathbf{Y}}
\newcommand{\R}{\mathcal{R}}
\newcommand{\W}{\mathbf{W}}
\newcommand{\Q}{\mathcal{Q}}
\newcommand{\K}{\mathcal{K}}
\newcommand{\I}{\mathbf{I}}
\newcommand{\E}{\mathbf{E}}
\newcommand{\w}{\mathbf{w}}
\newcommand{\vv}{\mathbf{v}}

\newcommand{\G}{\mathcal{G}}

\newcounter{algocpt}
\newcommand{\myalgomyname}[3]
{
\noindent
\begin{center}
\noindent
\begin{tabular}{l}
\hline 
 \multicolumn{1}{c}{{\bf Algo #2: }#3} \\
\hline 
\\
#1 
\hline
\end{tabular}
\end{center}
}

\begin{document}

\title[]{A particle filter to reconstruct a free-surface flow from a depth camera}

\author{Benoit Comb\`{e}s$^1$, Dominique Heitz$^1$\footnote{Corresponding author: dominique.heitz@irstea.fr}, Anthony Guibert$^1$ and Etienne M\'{e}min$^2$}

\address{$^1$IRSTEA, UR TERE, 17 avenue de Cucill\'e, F-35044 Rennes cedex, France}
\address{$^2$INRIA, Fluminance group, Campus universitaire de Beaulieu, F-35042 Rennes Cedex, France}

\ead{dominique.heitz@irstea.fr,etienne.memin@inria.fr}

\begin{abstract}
We investigate the combined use of a Kinect depth sensor and of a stochastic data assimilation method to recover free-surface flows. More specifically, we use a Weighted ensemble Kalman filter method to reconstruct the complete state of free-surface flows from a sequence of depth images only. This particle filter accounts for model and observations errors. This data assimilation scheme is enhanced with the use of two observations instead of one classically. We evaluate the developed approach on two numerical test cases: a collapse of a water column as a toy-example and a flow in an suddenly expanding flume as a more realistic flow. The robustness of the method to depth data errors and also to initial and inflow conditions is considered. We illustrate the interest of using two observations instead of one observation into the correction step, especially for unknown inflow boundary conditions. Then, the performance of the Kinect sensor to capture temporal sequences of depth observations is investigated. Finally, the efficiency of the algorithm is qualified for a wave in a real rectangular flat bottom tank. It is shown that for basic initial conditions, the particle filter rapidly and remarkably reconstructs velocity and height of the free surface flow based on noisy measurements of the elevation alone.
\end{abstract}

\vspace{2pc}
\noindent{\it Keywords}: Data-assimilation, Free-surface flow, Particle filter, Ensemble Kalman filter, Depth sensor, Shallow-water equations

\maketitle

\section{Introduction}
The circulation of water in open channel (or free-surface flow) is a key variable in many problems involved in hydrology. Most often, it is not possible to observe the complete free-surface flow (i.e. time and space dependent velocity and elevation) but only temporal sequences of sparse measurements coming from e.g. drifters, altimeters or satellites. Although insufficient in complex situations, these observations can be used as the starting ingredients to obtain a more complete characterization of the flow under study. A  common way to proceed consists in coupling the observations with the conservation laws of the flow using a \emph{data-assimilation} (DA) method. In geophysics, DA has a long history of developments and applications. In fluid dynamics, DA was first used to provide low order representations of the flow \cite{dadamo_etal_2007,cuzol_etal_2007}. Then, for the first time, using variational approach and particle filter, turbulent flow image sequences were coupled  with direct numerical simulations \cite{papadakis_memin_2008} and with vortex particle simulations \cite{cuzol_memin_2009}, respectively. The reader can refer to \citeasnoun{heitz_etal_2010} for a synopsis of these attempts. More recently, DA has been carried out either with sequential approaches \cite{colburn_etal_2011,suzuki_2012} or with variational approaches \cite{gronskis_etal_2013,foures_etal_2014,mons_etal_2014} or even through the interesting combination of both schemes \cite{yang_etal_2014}.

\subsection{Data-assimilation using noise-free dynamic model}
 In the context of free-surface flows, many methods propose to tackle the data-assimilation problem by considering that we know the dynamic model describing the evolution of the flow under study up to some parameters such as the bed roughness, the inflow velocity, etc. The goal of  the DA method is then to estimate these unknown parameters of the model, fitting as best as possible the observations. In addition, to provide an estimate of the free-surface flow, such an approach is able to handle more specific applications such as: the estimation of equivalent topography (bed geometry and roughness) to fit the parameters of a given dynamic model to a particular river then allowing to perform suitable numerical simulations \cite{roux06a}; the forecasting of a given situation by estimating the current flow and the parameters of the model. In the literature, it has been many times shown that such methods successfully perform on synthetic and sometimes real specific application cases \citeaffixed{hostache10a,honnorat10a}{e.g.}. As for now, most of these works focused on the estimation of the initial condition of the flow \cite{tinka09a,titaud10a,lai09a}, the time dependent flow at the open boundaries \cite{strub09b,belanger05a,castaings06a,lai09a,honnorat09a}, the bed roughness coefficients \cite{ding04a} or the equivalent topography, i.e. the geometry plus optionally the roughness coefficient \cite{roux06a,castaings06a,honnorat09a}, mainly from set of sparse trajectories, sparse elevations or a combination of them. The associated dynamic models can consist of simple 1D/2D shallow water optionally coupled together \cite{monnier07a} or of more sophisticated models such as the regular ocean modeling system (ROMS) model \cite{powell08a} or the Mike11-Mike22 flooding model \cite{madsen03a,hartnack05a}. The conceptual bottleneck of DA approaches using noise-free dynamic model is that the behavior of the fluid under study is assumed to be completely explained by tuning the parameters of the model (this is the so-called {\it strong-constraint} paradigm of the 4D-Var formulation). However, in practical situations the complete list of parameters involved can be huge: (time-dependent) wind stress, (time-dependent) topology, bed roughness, (time-dependent) boundary conditions, ... and for any situation involving parameters not dealt by the data assimilation method, the estimation is likely to fail. This in particular explains that such methods have not been exploited in the purpose of fully characterising the flow under study, with the notable exception of \citeasnoun{strub09b}.

\subsection{Data-assimilation using stochastic dynamic model}
 An alternative, and to our opinion a more appropriate, track consists in considering that we have only an estimate of the true dynamic model describing the evolution of the flow under study (in the context 4D-Var technique, this is the so-called {\it weak-constraint} paradigm). To model this view, one considers that the actual state of the system at each time $t$ results from the propagation of the state at time $t-1$ through the estimated (known) dynamic plus a stochastic term.
By using such a modeling, one is tempted to consider the problem of fully estimating the flow understudy (instead of a set of parameters of the dynamic model). Some authors have followed this track and most of them uses ensemble-based methods \cite{salman06a,verlaan96a,madsen03a,hartnack05a,wolf00a} to solve the problem. 

Among them, some authors considered the assimilation of a small set of lagrangian observations, given for example by drifters \cite{salman06a,strub09a}. We think that the application of such observations, limited to particular applications as the acquisition of larges set trajectories, can be in practice a complicated task. 
The other authors proposed to use water level measurements \cite{verlaan96a,madsen03a,hartnack05a,wolf00a}. In these works, the authors consider the assimilation of a very few measurements and are interested on the reliability of the observed components of the flow (velocity or elevation) at some critical points and do not consider the opportunity to characterize the full flow understudy. Two main reasons can explain this situation.
In the above mentioned articles and in many hydrological applications, observations consist of sparse local measure of velocity, flow or elevation from which it seems difficult to estimate a reliable complete characterization of the flow all over the space. Furthermore, the use of dense observations necessarily add a significant computational cost and memory usage into the DA procedure. Thanks to the advent of depth-sensors and of Monte Carlo methods for high-dimensional problems, one can reconsider this question.

In this work, we propose a complete system to characterize free-surface shallow flows. This system is composed of a very affordable commercial depth camera and of the particle filter data-assimilation method proposed by \citeasnoun{Papadakis10a}, dealing with stochastic dynamic model and involved in this study to account for the complexity of real free-surface flows. In section \ref{sec:method}, we present the stochastic data assimilation method reconstructing the spatiotemporal geometry and motion fields of free-surface shallow flows from sequences of depth images, only. We propose an enhancement of the particle filter for unknown inflow boundaries by using two observations instead of one. In section \ref {ssec:kinect}, we discuss the ability of the Kinect depth sensor having the strong advantages to be cheap and of easy use. Then, in section \ref{sec:numerical_valid} and in section \ref{sec:experimental_valid}, we evaluate the potential of our system by showing its ability to estimate free-surface shallow flows spatiotemporal geometry and velocity fields from synthetic and real data, respectively. Finally, we conclude and give some perspectives in section \ref{sec:concl}.

\section{The particle filter to assimilate depth images into a shallow flow model}
\label{sec:method}

The problem of estimating the consecutive states $\x_{1 \ldots t}$ of a physical system from a sequence of partial observations $\y_{1 \ldots t}$ is ubiquitous in geosciences. 
A common approach consists in stating it as a filtering problem {\it i.e.} characterizing the probability distributions associated to the state space of the physical system given the past observations and a stochastic dynamic model. More specifically, one states an observation model and a dynamic model respectively defining the densities $p_{\Y_t}(. |\x_{t})$ and $p_{\X_t}(. | \x_{0}, \ldots \x_{t-1})$  that are then combined via Bayes' rule to define the posterior probability $p_{\X_0, \ldots \X_t}(. | \y_0, \ldots \y_t)$. This last distribution is called {\it the filtering distribution} and generally cannot be given in a computable closed-form solution but can be estimated using Monte Carlo methods.

\subsection{Stating the filtering distribution}
\label{sec:stating}

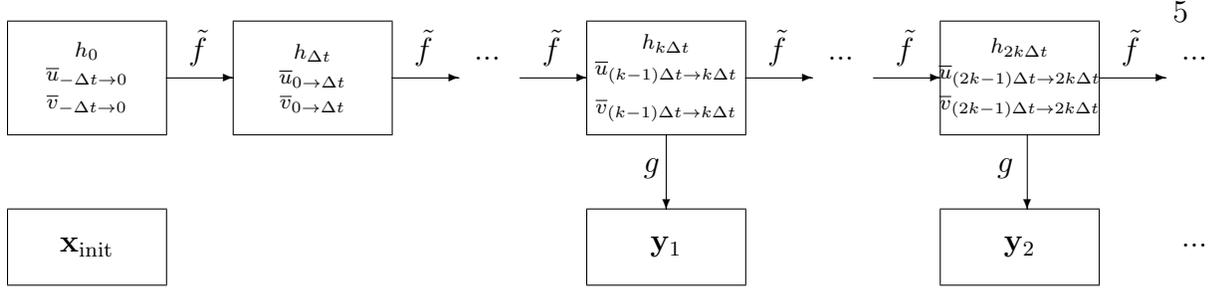
\begin{figure}[t!]
\setlength{\unitlength}{1cm}
\begin{picture}(5,3)
   \put(0,2){\framebox(2.1,1.5)[c]{\scriptsize \shortstack{$h_0$ \\ $\overline
{u}_{-\Delta t \rightarrow 0}$ \\ 
$\overline{v}_{-\Delta t \rightarrow 0}$}}}
   \put(2.1,2.75){\vector(1,0){0.9}} 
    \put(2.4,3){$\tilde{f}$} 

    \put(0,0){\framebox(2.1,1)[c]{$\x_{\rm init}$}}
    
   \put(3,2){\framebox(2.1,1.5)[c]{\scriptsize \shortstack{$h_{\Delta t}$ \\ $\overline
{u}_{0 \rightarrow \Delta t}$ \\ 
$\overline{v}_{0 \rightarrow \Delta t}$}}}
    \put(5.1,2.75){\vector(1,0){0.9}}
    \put(5.4,3){$\tilde{f}$}

   \put(6.2,3){...}
   
   \put(6.8,2.75){\vector(1,0){0.9}} 
   \put(7.1,3){$\tilde{f}$} 
   \put(7.7,2){\framebox(2.1,1.5)[c]{\scriptsize \shortstack{$h_{k\Delta t}$ \\ $\overline
{u}_{(k-1)\Delta t \rightarrow k\Delta t}$\\
\\ $\overline
{v}_{(k-1)\Delta t \rightarrow k\Delta t}$}}}
   \put(9.8,2.75){\vector(1,0){0.9}}
   \put(10.1,3){$\tilde{f}$} 
   
   \put(8.75,2){\vector(0,-1){1}}
    \put(8.45,1.5){$g$} 
    
    \put(7.7,0){\framebox(2.1,1)[c]{$\y_1$}}

     \put(10.9,3){...}

    \put(11.5,2.75){\vector(1,0){0.9}} 
  \put(11.8,3){$\tilde{f}$} 
 \put(12.4,2){\framebox(2.1,1.5)[c]{\scriptsize \shortstack{$h_{2k\Delta t}$ \\ $\overline
{u}_{(2k-1)\Delta t \rightarrow 2k\Delta t}$\\ $\overline
{v}_{(2k-1)\Delta t \rightarrow 2k\Delta t}$}}}
  \put(14.5,2.75){\vector(1,0){0.9}}
   \put(14.8,3){$\tilde{f}$} 
   
   \put(13.55,2){\vector(0,-1){1}}
 \put(13.15,1.5){$g$} 

    \put(12.4,0){\framebox(2.1,1)[c]{$\y_2$}}

	\put(15.6,3){...}

	\put(15.6,0.5){...}
\end{picture}
\caption{Schematic view of the modeling. Each state of the system consists of the elevation and the velocity components for the given time $k \Delta t$. The evolution of the system is handled by a function $\tilde{f}$ which is unknown (we simply know a simplified version $f$). An observation $\y_t$ is generated from the states each $k$ time step of the dynamic. The unknown quantities to estimate ({\it i.e.} the states $\x_t$) consist in the first row of this scheme while the known variables consist in the second row.}
\label{fig:model}
\end{figure}

The overall model is graphically represented in figure \ref{fig:model} and presented in details in the three next subsections.

\subsubsection{The dynamic model}\label{ssec:dynamic}

\begin{figure}[htbp]
\begin{center}
{\resizebox{10cm}{!}{\input{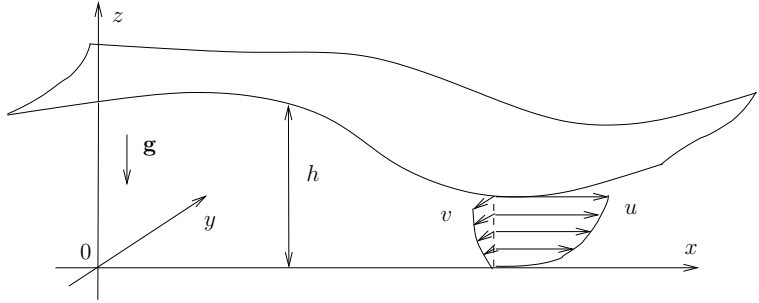}}}
\caption{Free surface flow configuration.}
\end{center}
\label{fig:surface}
\end{figure}

The shallow-water model (often referred to as Saint-Venant model) is practically suited to handle many hydrological situations such as tides, storm surges, river and coastal flows, lake flows, tsunamis and more generally is theoretically supported by the linear wave theory (or airy wave theory) when the ratio water depth under wave length is smaller than $5.10^{-2}$~\cite{vreugdenhil95a}.
 Let briefly summarise it. We consider a liquid in a gravitational field  in open channel (or free surface) contained in a rectangular cuboid of length $L$.  
 The free surface level is supposed to be a time and space differentiable function $h(x,y,t)$. The bed topology is flat as sketched in figure \ref{fig:surface}. In addition, we assume that the fluid is only subject to gravity, viscous stress and that the liquid is Newtonian, incompressible and of constant viscosity $\nu$. 

The shallow water approximation consists in considering that $\epsilon=h_{max}/L \ll 1$ (calling $h_{max}$ the maximal water level) and that the dimensionless parameters follow the asymptotical dominance as $\epsilon \rightarrow 0$: $\epsilon/Re=O(\epsilon)$, $1/(\epsilon Re)=O(1)$ and $Fr^2=O(1)$, with $Re=u_{max}h_{max}/\nu$ the Reynolds number and $Fr=u_{max}/\sqrt{gh_{max}}$ the Froude number. Then, performing a dimensional analysis of the Navier-Stokes equations for an incompressible flow according to these dominances, integrating the velocity components $u$ and $v$ along the water depth $[0,h]$, considering that the pressure is hydrostatic and empirically replacing the two bottom stress component  $ \nu \,\partial u / \partial z (x,y,0)$  and $ \nu \, \partial v /\partial z (x,y,0)$ by $(gn^2/h^{1/3})\overline
{u} |(\overline
{u},\overline
{v})| $ and $(gn^2/h^{1/3}) \overline
{v} |(\overline
{u},\overline
{v})|$  (from Groves and Groen formula and Manning formula, $n$ being the Manning bed roughness factor, $\overline
{u}$ and $\overline{v}$ being described in a few lines), the shallow water equations read
\numparts
\begin{eqnarray}
\hspace{-0.5cm} \frac{\partial h}{\partial t} +  \frac{\partial h\overline
{u}}{\partial x} + \frac{\partial h\overline
{v}}{\partial y} =0  \label{eq:shallow_watera} \\
\hspace{-0.5cm} \frac{\partial (h\overline
{u})}{\partial t} + \frac{\partial}{\partial x}\left(h \overline
{u}^2 + \frac{1}{2}g h^2 \right) + \frac{\partial (h \overline
{u} \overline
{v})}{\partial y}  + \frac{gn^2}{h^{1/3}} \overline
{u} |(\overline
{u},\overline
{v})| = 0 \label{eq:shallow_waterb} \\
\hspace{-0.5cm} \frac{\partial (h \overline
{v})}{\partial t} + \frac{\partial (h \overline
{u}\overline
{v})}{\partial x} + \frac{\partial}{\partial y}\left( h\overline
{v}^2 + \frac{1}{2}g h^2\right)  +\frac{gn^2}{h^{1/3}} \overline
{v} |(\overline
{u},\overline
{v})| = 0, \label{eq:shallow_waterc}
\end{eqnarray}
\endnumparts
where $\overline
{u}=\int_{0}^{h(x,y,t)} u(x,y,z) dz$ and $\overline
{v}=\int_{0}^{h(x,y,t)} v(x,y,z) dz$ are the depth integrated horizontal velocities.

On that basis, we consider that the dynamic of the free surface flow under study is completely described at time $t$ by the state variable $\x_t=[\overline
{u}(x,y,t),\overline
{v}(x,y,t),h(x,y,t)]$, satisfying the discrete time integration of the shallow water system dynamics $f$:
\begin{equation}
\x_{t}=f(\x_{t-\Delta t}). \label{eq:dynamicExact}
\end{equation}
 As previously mentioned, this numerical model is based on physical assumptions and on numerical truncation errors.  In other terms,  it is attempt to produce only an approximation of the actual dynamic of the flow. We model this aspect by considering the evolution of the state  as a stochastic process (in the following we use capital letters to denote random vectors and caligraphic letters to denote matrices):
\begin{equation}
\X_{t}=f(\X_{t-\Delta t })  + \W^f_{t}, 
\label{eq:dynamicStoch}
\end{equation}
with $\X_0 \sim \mathcal{N}(\x_{\rm init},\R_0) $ and $\W^f_t  \sim \mathcal{N}({\bf 0},\R_t)$ (that denotes that the random variable $\W^f_t$ has a normal distribution with mean ${\bf 0}$ and  covariance matrix $\R_t$). 
In practice, we model the spatial components of $\W^f_t$ as a stationnary random field whose covariance coefficients evolve as a function of $\exp(-r^2/r_h ^2)$ where $r$ is the distance between the sites of the field and $r_h$ is a bandwith parameter \cite{Evensen94a}. The correlation between the different times and between the three components $h$, $\overline{u}$ and $\overline{v}$ of $\W^f_t$ is chosen null. We call (\ref{eq:dynamicStoch}) the {\it dynamic model} and we note $p_{\X_t}(.|x_{t-\Delta t})$ the subsequent conditional probability density function that we call {\it prior}.
We consider that the discrete-time variables $\X_0, \X_{\Delta t}, \X_{2\Delta t},\ldots \X_T$  are connected in a first order Markov chain  (i.e. $p( \x_t | \x_{t-\Delta t} ,\x_{t-2 \Delta t},  \ldots \x_0)= p(\x_t |\x_{t-\Delta t})$) with an initial distribution $p_{\X_0}(.)$. Each of these variables $\X_t$ is latent (or unobserved) and fully characterizes the state of the system under study at time $t$. We call $n_x$ the size of each variable $\X_t$.

\subsubsection{The observation model}
\label{sec:obsModel}

We consider that we acquire an observation $\y_t$ of the geometry of the free surface $z=h(x,y,t)$ using a perspective range camera (often refer to as  {\it 2.5D camera}). We call $g$ the function such that $\y_t=g(\x_t)=g(h(.,.,t))$.
There is no simple formula giving the value of the observation $\y_t$ for a given elevation function $h$. A first solution consists in using a ray casting procedure to simulate the operator $g$.  However, such a solution can be very time consuming considering the number of observations that we will have to perform to solve the filtering problem. 
To simulate the range camera operator, the observation $\y_t$ can be back-projected into the state grid $\x_t$ using an appropriate matrix $\G$ converting camera frame coordinates to object frame coordinates. 
This way $g(\x_t)$ simply consists in subsampling the elevation component of $\x_t$. We tested both methods and we do not noticed any significant difference. For that reason, we used the back-projection approach in order to speed-up the computations. 

In the following, the values $\y_t$ are considered to be rearranged in a vector form (to facilitate the algebraic formulations). As the acquisition process is subject to noise, we model the observation of the state $\x_t$ as a stochastic process given by
\begin{equation}
\Y_t = g(\X_t) + \W^g_t, 
\label{eq:obsModel}
\end{equation}
 where $\W^g_t \sim \mathcal{N}({\bf 0},\Q_t)$. In practice, we consider that there are no spatial correlation, thus $\Q_t$ is a diagonal matrix. We call (\ref{eq:obsModel})  the {\it observation model}.  We consider the set of discrete-time variables $\Y_1 ,\ldots, \Y_t, \ldots , \Y_T $ conditionally independent provided that $\X_0 \ldots \X_t, \ldots , \X_T$ are known. We note $p_{\Y_t }(.|x_t)$ the subsequent conditional probability density function. We call $n_y$ the size of each variable $\Y_t$.

\subsubsection{The filtering model}
\label{ssec:overall_model}
The overall model formally reads
\begin{equation}
\label{eq:overall_model}
\cases{\X_0 & $\sim\mathcal{N}(\x_{\rm init},\R_0)$\\
 \X_t | \X_{t-\Delta t}=\x_{t-\Delta t} & $\sim\mathcal{N}(f(\x_{t-\Delta t}),\R_t)$\\
 \Y_t | \X_t=\x_t & $\sim\mathcal{N}(g(\x_t) , \Q_t).$}
\end{equation}

Then, given an initial guess of the system at time $t=0$ ($\x_{\rm init}$,$\R_0$) and the set of observations $\y_{1,\ldots,T}$, we want to estimate the filtering laws $p_{\X_{0 \ldots t}}(.| \y_{1 \ldots t})$ (for $t \in [0,T]$) or one of their estimator such as the root mean square estimator
\begin{equation}
\hat{\rm \E}(\X_{0 \ldots t}) = \int p(\x_{0 \ldots t} | \y_{1 \ldots t}) \x_{0 \ldots t} \mbox{ } d\x_{0 \ldots t}\,.
\label{eq:RMSE}
\end{equation}

In the present study this estimator provides the elevation $\hat{h}$ and the depth integrated horizontal velocities $\hat{\bar{u}}$ and $\hat{\bar{v}}$.

\subsection{Solving the filtering problem : the Weighted Ensemble Kalman Filter}
\label{sec:solving}

From a theoretical point of view, the evolution of the conditional density $p_{\X_{0 \ldots t}}(.| \y_{1 \ldots t})$ is described by the Fokker-Planck (or forward Kolmogorov) equation \cite{Jazwinski70a}. However, its direct numerical evaluation is intractable in real situations. In this context, the representation of the conditional pdf as an ensemble or a set of particles and the integration through time of each of the ensemble elements leads to particularly tractable algorithms. The sequential importance sampling (SIS) filters \cite{doucet00a,gordon93a} and the ensemble Kalman filters (EnKF) \cite{evensen03a}, including its numerous variants, are probably the most used methods. In this work, we followed the recent works of \citeasnoun{Papadakis10a} who proposed the Weighted Ensemble Kalman Filter (WEnKF), a sequential importance sampling filter relying on an ensemble Kalman filter. It combines the best of particle filter and EnKF. The algorithm is based on the Ensemble Kalman Filter updates of samples in order to define a proposal distribution for the particle filter that depends on the history of measurement. In the present study we propose to enhanced the WEnKF scheme with the use of two observations $\y_t$ and $\y_{t+1}$ instead of one. This provides a better coupling of elevation and velocity component
estimations when considering unknown inflow boundary conditions.

In the following, we simplify the notations with $\Delta t=1$ so that e.g. $X_{t+1}=X_{t+\Delta t}$.
 In \sref{sec:SIS}, we recall the foundation of the sequential importance sampling filter. In \sref{sec:proposal}, we discuss the use of an ensemble Kalman to build a specific sequential importance sampling filter. In \sref{sec:proposalMultiObs}, we propose to modify the stochastic process underlying the ensemble Kalman filter procedure so that it uses two observations $\y_t$ and $\y_{t+1}$ instead of one. In \sref{sec:gain}, we discuss a method to build efficiently the Kalman gain, which is the main ingredient of the ensemble Kalman filter. 

\subsubsection{Bayesian filtering of a non-linear process using sequential importance sampling}
\label{sec:SIS}

The sequential importance sampling method stands on the point that we can compute any estimator of $p_{\X_{0 \ldots t}} (.|\y_{1 \ldots t}) $
using sequences of samples $(\x^i_t)$ iteratively drawn from a proposal density function $q$ (optionally parametrized in $\y_{1 \ldots T}$). We denote this general proposal as $q_{\X^Q_t}(.|\x_{0,\ldots,t-1};\y_{1 \ldots T})$ (calling $\X^Q_t$ the associated random vector). In particular, we can write the root mean square estimator (RMSE) associated to the filtering distribution as
\begin{equation}
\hat{\rm \E}(\X_{0 \ldots t}) = \lim\limits_{N \to \infty} \frac{\frac{1}{N} \sum_i w_t(\x_{0 \ldots t}^i) \x_{0 \ldots t}^i }{ \frac{1}{N} \sum_i w_t(\x_{0 \ldots t}^i)} ,
 \label{eq:sis:evalest}
\end{equation} 
where each sample $\x_{k}^i$ ($k \in [1,\ldots, t]$, $i \in [1,\ldots, N]$) is generated from the distribution of density $q_{\X^Q_k}(.|\x^i_{0,\ldots,k-1};\y_{1 \ldots T})$ and where the weights $w_t$ are built recursively as
\begin{equation}
 w_t(\x_{0 \ldots t}^i)=w_{t-1}(\x_{0 \ldots t-1}^i) \frac{p(\y_t |\x_{t}^i) p(\x_t^i | \x_{t-1}^i)}{q(\x_t^i |\x_{0..t-1}^i ; \y_{1 \ldots T})},  \label{eq:sis:weightUpdate}
\end{equation}
with for $t=0$, $w_0(\x_0^i)=1/N$ and each sample $\x_0^i$ is drawn according to the distribution of density $p_{X_0}(.)$.

The combination of the equations~\eref{eq:sis:evalest} and \eref{eq:sis:weightUpdate} is the keystone of any SIS filters and leads to the following algorithm:
\myalgomyname{
init:\\
\hspace{0.3cm} for $i=1:N$\\
\hspace{0.3cm}\hspace{0.3cm} $w_1^i = \frac{1}{N}$ \\
\hspace{0.3cm}\hspace{0.3cm} draw $\x_0^i$ with density $p_{\X_0}$\\
for $t=1:T$\\
\hspace{0.3cm} Sampling: for $i=1:N$\\
\hspace{0.3cm}\hspace{0.3cm} draw $\x_t^i$ with density $q_{\X^Q_t}(.|\x^i_{0,\ldots,t-1};\y_{1 \ldots T})$ \\
\hspace{0.3cm} Weighting: for $i=1:N$ \\
\hspace{0.3cm} \hspace{0.3cm} $w_t^i = w_{t-1}^i (p(\y_t | \x_t^i) p(\x_t^i | \x_{t-1}^i))/q(\x_t^i | \x_{0:t-1}^i; \y_{1 \ldots T})$ \\
\hspace{0.3cm} Normalising: for $i=1:N$\\
\hspace{0.3cm} \hspace{0.3cm} $\hat{w}_t^i =  w_{t}^i/\sum_j w_{t}^j$ \\
\hspace{0.3cm} $\hat{E}[\X_{0 \ldots t}] = \sum_i \hat{w}_t^i \x_{0 \ldots t}^i$ .\\
}{SIS}{basic sequential importance sampling algorithm}

\subsubsection{The proposal distribution $q$} 
\label{sec:proposal}

In practice, any proposal $q_{\X^Q_{t}}(.| \x_{t-1,\ldots,0} ; \y_{1 \ldots T})$  can be used and the different instanciations of the SIS filter mainly differ by this choice. There are two main characteristics to consider when choosing the proposal $q_{\X^Q_{t}}(.| \x_{t-1,\ldots,0} ; \y_{1 \ldots T})$. Firstly, it must be possible to sample it a moderate time (sampling $q$ is needed for the SIS algorithm). Secondly, it has to be as close as possible to the target pdf $p_{\X_t}(.|\y_{1 \ldots t})$ to provide a reliable estimation of the target estimator (for a given number of particle $N$).
A classical choice for $q$ consists of $q_{\X^Q_{t}}(.| \x_{t-1,\ldots,0} ; \y_{1 \ldots T})=p_{\X_{t}}(.| \x_{t-1})$, resulting in the so-called {\it bootstrap} algorithm \cite{gordon93a}. Another classical choice consists of $q_{\X^Q_{t}}(.| \x_{t-1,\ldots,0} ; \y_{1 \ldots T})=p_{\X_{t}}(.| \x_{t-1} ,\y_t)$, resulting in the so-called {\it optimal SIS} algorithm \cite{doucet00a}. While theoretically supported, these instances of the SIS algorithm are often considered as practically less efficient than the ensemble Kalman filter. However it seems apropriate to mention that unlike the SIS algorithm, the ensemble Kalman filter does not converge, as the number of ensemble members tends to infinity, to the target estimator. In this work, we combine the benefits of the two algorithms by using an ensemble Kalman filter to build the proposal distributions $q$ of a SIS filter.

Interestingly, \citeasnoun{legland09a} showed that the ensemble Kalman filter can be seen as a Monte Carlo simulation of the following stochastic process:
\begin{equation*}
\X^{Q}_t=\X^{Q^f}_t+ \K_t (\y_t + \W_{t}^g - g(\X^{Q^f}_t) ),
 \label{eq:process}
\end{equation*}
with $\X^{Q^f}_t=f(\X^Q_{t-1})+\W_{t}^f$ ($ ^f$ standing for forecast) and where $\K_t$ the Kalman gain associated to samples $\X^{Q^f}_t$ reads
\begin{equation*}
\K_t={\rm cov}(\X^{Q^f}_t,g({\X}^{Q^f}_t))(\R_t+{\rm cov}(g(\X^{Q^f}_t),g(\X^{Q^f}_t)))^{-1}.
\end{equation*}

In practice, the Kalman gain $\K_t$ is estimated from the set of samples used in the simulation process. As a consequence the simulated random variables $\X^{Q}_t$ are not independent and identically distributed. However, it has been shown by \citeasnoun{legland09a} that when $N$ tends to infinity, $\K_t$ tends to its deterministic limits and the simulated random variables tend to be independent and identically distributed according to the normal distribution:
 \begin{eqnarray}
 \X^Q_{t}|\X^Q_{t-1}=\x_{t-1};\y_{1,\ldots,T} \sim  \\
  \mathcal{N}(f(\x_{t-1})+\K_t (\y_t-g(f(\x_{t-1}))),(\I+\K_t \G) \R_t(\I+\K_t \G)^T +\K _t \Q_t \K_t^T ), \nonumber
 \end{eqnarray}
 $\G$ being the matrix associated to the observation operator $g$ (see Section \ref{sec:obsModel}). This finally permits to define properly a proposal distribution $q_{\X^Q_{t}}(.| \x_{t-1,\ldots,0} ; \y_{1 \ldots T})$ and a procedure (the ensemble Kalman filter) to sample it (in the limit of large samples). Then, one can easily restate Algo SIS as:
\myalgomyname{
for $i=1:N$ \\
 \hspace{0.3cm}  draw $\x_0^i$ with density $p_{\X_0}$ \\
  \hspace{0.3cm} $w_0^i=\frac{1}{N}$\\
 for $t=1 \ldots T$\\
 \hspace{0.3cm} {\bf 1) Draw} $\x_t^i$: \\
\hspace{0.3cm} {\bf 1.1) Forecast step:}\\
\hspace{0.3cm} for $i=1:N$ \\
\hspace{0.3cm}  \hspace{0.3cm}  $\x^{i^f}_{t} = f(\x_{t-1}^i) +w_t^i $  with $w_t^i$ generated from $\mathcal{N}(0,\Q_t)$ \\
\hspace{0.3cm}  {\bf 1.2) Analysis step:}\\
\hspace{0.3cm} \hspace{0.3cm} for $i=1:N$ \\
\hspace{0.3cm} \hspace{0.3cm} \hspace{0.3cm} $\y^i_t=\y_t + v_t^i$ with $v_t^i$ generated from $\mathcal{N}(0,\R_t)$ \\
\hspace{0.3cm} \hspace{0.3cm} build $\K_t =  {\rm cov}((\x^{f^i}_t),(g(\x^{f^i}_t)))  \times (  {\rm cov}((g(\x^{f^i}_t)),(g(\x^{f^i}_t))) + \R_t)^{-1}$ \\
\hspace{0.3cm} \hspace{0.3cm} for $i=1:N$ \\
\hspace{0.3cm} \hspace{0.3cm} \hspace{0.3cm} $\x^i_{t}=\x^{i^f}_{t} + \K_t (\y^i_t-g(\x^{i^f}_{t}))$ \\
 \hspace{0.3cm}  \hspace{0.3cm} \hspace{0.3cm} set $\x_{0 \ldots t}^i=[\x_{0 \ldots t-1}^i,\x_t^i]$ \\
 \hspace{0.3cm} {\bf 2) Weight samples:}\\
\hspace{0.3cm} for $i=1:N$ \\
\hspace{0.3cm}  \hspace{0.3cm} $w_t^i=w_{t-1}^i (p(\y_t | \x_{t}^i) p(\x_t^i | \x_{t-1}^i))/q(\x_t^i | \x_{t-1}^i; \y_{t})$ \\
 \hspace{0.3cm} {\bf 3) Compute estimators:}\\
 \hspace{0.3cm} $\hat{E}[\x_{1 \ldots t}] = \sum_i (w_t^i/\sum_j w_t^j) \x_{1 \ldots t}^i$\\
}{WEnKF}{Sequential importance sampling with Kalman proposal}

A resulting challenge consists in performing the step 2) of this algorithm : evaluating the three densities $p_{\Y_t}(.|\x_t)$,  $p_{\X_t}(.|\y_t)$ and $q_{\X^Q_t}(. | \x_{t-1}; \y_{t})$. Whereas $p_{\Y_t}(.|\x_t)$, can be evaluated very efficiently, the two other terms are more difficult to address. Following previous works \cite{Hoteit08a,Papadakis10a}, we chose to drop these two terms. More justifications about this choice can be found in \cite{Papadakis10a}. Further work is needed to evaluate the theoretical impact of this choice.

\subsubsection{Using two observations for a better coupling of the state variables} 
\label{sec:proposalMultiObs}

To improve the previously stated algorithm, we slightly modify the proposal distribution $q$ to incorporate two observations $\y_t$ and $\y_{t+1}$ instead of one.
By doing this, we do not simply consider the relationship linking the observation of the current state $g(\x_t)$ and the state $\x_t$ itself to compute the Kalman gain but also introduce the next observation $\y_{t+1}$. 
Our motivation is to provide a better coupling between the estimation of the elevation and the velocity components of the state variables by analyzing through the observation $\y_t$ and $\y_{t+1}$ their combined effects over the time interval $[t, t+1]$ instead of a simply considering successively each observation independently from each others. In practice, this yields to the following stochastic process:
\begin{equation}
\X^{Q}_t=\X^{Q^f}_t+
\K_{t,t+1}
 ([\y_t, \y_{t+1}] + [\W_t^g,\W_{t+1}^g] - [g(\X^{Q^f}_t) , g(\X^{Q^{ff}}_{t+1})] ),
 \label{eq:processTwoObs}
\end{equation}
where $[\vv_1,\vv_2]$ denotes the concatenation of vector $\vv_1$ and $\vv_2$ and with,
\begin{eqnarray*}
\X^{Q^f}_t=f(\X^Q_{t-1})+\W_{t}^f, \\
\X^{Q^{ff}}_{t+1}=f(\X^{Q^f}_t)+\W_{t}^{ff},
\end{eqnarray*}
with $ \W_{t}^{ff}\sim \mathcal{N}({\bf 0},\R_t)$ and where $\K_{t,t+1}$ is the Kalman gain expressing the linear relationship between the observations $\y_t$ and $\y_{t+1}$ and the state variable $\X_t^{Q^f}$. It writes:
\begin{eqnarray*}
\K_{t,t+1}= & {\rm cov}(\X^{Q^f}_t,[g(\X^{Q^f}_t),g(\X^{Q^{ff}}_{t+1})]) \times \\ &\left ({\rm cov}\left ([g(\X^{Q^f}_t),g(\X^{Q^{ff}}_{t+1})],[g(\X^{Q^f}_t),g(\X^{Q^{ff}}_{t+1})]\right)+\R_{2n_y}\right )^{-1},
 \label{eq:gainTwoObs}
\end{eqnarray*}
where $\R_{2n_y}$ is the $2n_y \times 2n_y$ block diagonal matrix composed of two copies of the matrix $\R$. We expect this process to give more suited samples to approximate the filtering distribution $p_{\X_{0 \ldots T}}(.|\y_{1,\ldots T})$ and we use it as a generator for the proposal distribution $q$. The question of whether this process properly defines a density $q$ is not addressed in this work and is a future research direction. As observed in the validation section, such a choice improved quantitatively the estimations.

\subsubsection{Computing the Kalman gain $\K_t$}
\label{sec:gain}
 To our opinion, the main challenge in the EnKF procedure consists in computing efficiently reliable Kalman gain $\K_{t,t+1}$ (or $\K_t$ in the original version). Indeed, this matrix is crucial as it determines how to convert the errors from the observation grid into the state grid and ultimately how to compute the correction term associated to $\x^i_t$ regarding the observation $\y_t$ and the past uncertainty. We assume the covariance matrix $\R_t$ associated to the observation noise as known and hence focus our interest on how to estimate the covariance matrices ${\rm cov}\left (\X^{Q^f}_t,[g(\X^{Q^f}_t),g(\X^{Q^{ff}}_{t+1})]\right )$ and  ${\rm cov}\left ([g(\X^{Q^f}_t),g(\X^{Q^{ff}}_{t+1})],[g(\X^{Q^f}_t),g(\X^{Q^{ff}}_{t+1})]\right )$ (or ${\rm cov}\left (\X^{Q^f}_t,g({\X}^{Q^f}_t)\right )$ and  ${\rm cov}\left (g({\X}^{Q^f}_t),g({\X}^{Q^f}_t)\right)$ in the basic version).

In the original EnKF implementation, these covariance matrices are estimated as the empirical covariance matrices of the samples\footnote{This choice leads to the algebrical reformulation exploited by the reduced rank, and the EnTKF formulation.}.
However, the empirical covariance is only a rough estimate of the true covariance when $N<<n_x$ (this is a case of the so-called {\it large $p$-small $n$} inference problems). Indeed, the number of degree of freedom ($n_x(n_x+1)/2$) is too high to express correctly the covariance matrices using only $N$ samples. However, $\X_t$ (and $g(\X_t)$) represents a space-structured random vectors and we can infer a structure on their covariance matrix. Adding a structure to or regularizing the covariance matrices will decrease the number of degree of freedom of the statistic. Then, the estimated regularised covariance matrices are likely to better characterize the sample statistic (for a given sample size) and thus to result in a more appropriate Kalman gain and analysis. In this context, a pragmatic solution consists in smoothing the covariance matrix and imposing a complete independence between spatially distant points of the state grid \cite{houtekamer01a}.  Such a localization increases virtually the size of the ensemble. In this work, we choose to use a strategy similar to the one proposed by \citeasnoun{houtekamer01a}, that consists in performing the product between the empirical covariance coefficients and a slowly positive definite radial basis function decreasing with point-to-point distance and canceling at a fixed cut-off distance $h_{\rm correl}$. In practice, we use a \citeasnoun{buhmann03a} function.  Then all matrices involved in the analysis step (step 1.2 in Algo SIS-KF) become sparse and the analysis step becomes a sparse linear algebra problem that can be solved efficiently using appropriate optimization algorithms \cite{golub96a}.

\section{Numerical validation}\label{sec:numerical_valid}
In this section, we investigate the performance of the WEnKF method to estimate the dynamic of a free-surface flow from synthetic sequences of depth data.

The robustness of the proposed method to simulated depth data quality is evaluated for two flow configurations: a collapse of a water column as a toy-example and a flow in an suddenly expanding flume as a more realistic flow. Furthermore, we illustrate the interest of using two observations instead of one observation into the correction step (see section \ref{sec:proposalMultiObs}). We note oneObs and twoObs the two WEnKF algorithms using in the correction step one observation and two observations, respectively.

The numerical dynamical model used to simulate the two flow configurations involves a finite volume implementation of the shallow-water system of equations (\ref{eq:shallow_watera},\ref{eq:shallow_waterb},\ref{eq:shallow_waterc}) as proposed by \citeasnoun{bradford_etal_2002}. Time integration is performed with a second-order Runge-Kutta scheme.  
A no-slip boundary condition is applied on the walls.

The true states $\tilde{\x}_{1 \ldots t}$ are calculated with this numerical model running from an initial condition $\tilde{\x}_{0}$. Based on that true state we build the synthetic depth data sequences taking every fourty time step $\Delta t$ of the numerical model integration. Then, to simulate the depth sensor image quality, we add a white noise of standard deviation $\sigma_{\rm obs}$ and introduce a percentage of outliers $p_{\rm out}$ in the depth images. That provides the observations $\y_{1\ldots t}$. Note that for the sake of simplicity the observations are created with the same spatial resolution as the numerical simulation grid.

The initial state $\x_{\rm init}$ is obtained by deteriorating the initial true state $\tilde{\x}_0$. For that purpose we add a random noise with a covariance having a large range (as for the noise on the dynamical model) such that the initial state perturbation parameter
\begin{equation}
\mathcal{E}_{\rm init}=\frac{||\tilde{\x_0}-\x_{\rm init}||}{||\tilde{\x_0}||},
\end{equation}
 is equal to a given quantity (typically 0.1), where $||.||=\sqrt{<. ,.>}$ is the $L_2\mbox{-norm}$ acting on the whole flow domain. The following states $\x_{1\ldots t}$ are estimated by assimilating the observations $\y_{1\ldots t}$ given the perturbated initial state $\x_{\rm init}$.
 
To assess the accuracy of the free-surface flow reconstruction we compare the estimated flows $\hat \x_{1\ldots t}$ to the true ones $\tilde{\x}_{1\ldots t}$ such as the assimilation errors read
\begin{equation}
\mathcal{E}_{\hat{\bar{u}},\hat{\bar{v}}}=\frac{||\hat \w-\tilde{\w}||}{u_0} ~~\mbox{and}~~\mathcal{E}_{\hat h}=\frac{||\hat h-\tilde{h}||}{h_0},
\label{eq:error_estim}
\end{equation}
where  $\hat \x =(\hat \w,\hat h)=(\hat{\bar{u}},\hat{\bar{v}},\hat h)$ and $\tilde \x =(\tilde \w,\tilde h)=(\tilde{\bar{u}},\tilde{\bar{v}},\tilde h)$ are the estimated (via assimilation) and true state components, respectively, and where $u_0$ and $h_0$ are a characteristic velocity and a characteristic length scale, respectively. In addition, we define the free-run errors
\begin{equation}
\mathcal{E}_{\bar{u},V}=\frac{||\w-\tilde{\w}||}{u_0} ~~\mbox{and}~~\mathcal{E}_{h}=\frac{||h-\tilde{h}||}{h_0},
\end{equation}
where  $\x =(\w,h)=(\bar{u},\bar{v},h)$ are the free-run state components calculated with the shallow-water numerical model from the initial condition $\x_{\rm init}$ up to time $t$. Then, to give an insight of the gain obtained using the observations driving the model instead of running the model alone, we compute the ratios between assimilation and free-run errors for the velocity $\mathcal{E}_{\hat{\bar{u}},\hat{\bar{v}}}/\mathcal{E}_{\bar{u},\bar{v}}$ and for the elevation $\mathcal{E}_{\hat h}/\mathcal{E}_{h}$.
 
The same set of parameters was used to evaluate our method for both synthetic flow configurations. The number of particles $N$ was set to 100. The initial standard deviations of the covariance matrix $\R_0$ were equal to $0.05\, h_0$ and $0.25\,u_0$ for the elevation and the velocity components, respectively. Then the standard deviations of $\R_t$ were set to $0.04\, h_0$ and $0.06\,u_0$ for the elevation and the velocity components, respectively. The bandwith parameter $r_h$ indicating the spatial decrease of $\R_0$ and $\R_t$ covariance coefficients was set to $2\,h_0$. The covariance matrix $\Q_t$ was set diagonal (no spatial dependance on noise) with constant standard deviation equal to $0.114~h_0$ (this value was estimated for the Kinect depth sensor, see section \ref{ssec:kinect}). Finally, $h_{\rm correl}$, the cut-off distance of the radial basis function used for the localization process described in section \ref{sec:gain}, was set equal to $0.6\,h_0$.

\subsection{Collapse of a water column}
\label{ssec:toy}

In this simple case, a small circular column of water placed at the center of a square container collapses under gravity and generates a wave. The initial elevation $h_0$ of the water column above the initial container water level $h_\infty$ was such that $h_0=0.5\, d_0$, where the water column diameter $d_0=2\mbox{~cm}$ and with $h_\infty=1.5\,d_0$. The width of the container was equal to $L=10\,d_0$. The initial velocity in the container was zero. To non-dimensionnalize the results we define the characteristic time scale $t_0=\sqrt{h_0/g}$ and the characteristic velocity $u_0=\sqrt{gh_0}$. The collapse of the water column was simulated with the shallow-water numerical model in a computational domain $L_x \times L_y = L \times L$ discretized on a square grid of $n_x \times n_y= 200 \times 200$ points and integrated in time with a time step $\Delta t\,u_0/h_0=0.006$. 

Here, we considered the initial state $\x_{\rm init}$ as the ``perfect'' case described above, i.e. with flat and static initial water surfaces (top of the water column and around in the container), whereas the initial true state $\tilde{\x}_0$, considered as the ``real'' case, was more complex, i.e. with a large scale random extra component providing non smooth initial water surfaces with local non-zero velocities. The observation $\y_{1 \ldots t}$ were built from snapshots of the true state elevation component taken every $40 \Delta t\,u_0/h_0$, leading to an observations Strouhal number $St_{\rm obs}=h_0/(40 \Delta t\,u_0) = 4.15$.

\begin{figure}[h!]
\centering
\includegraphics[trim = 40mm 80mm 51mm 74mm, clip,width=0.31\textwidth]{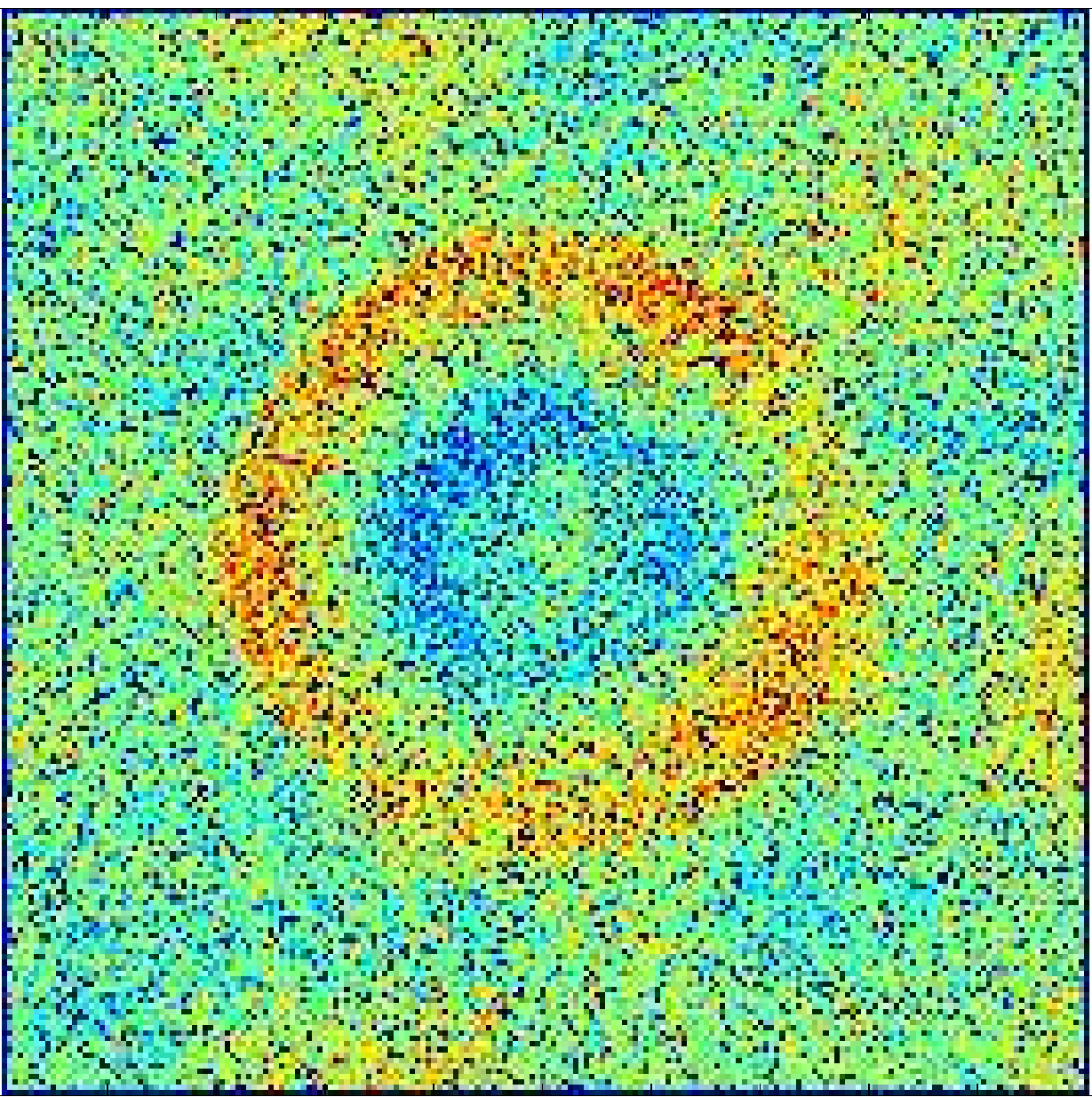}
\includegraphics[trim = 40mm 80mm 51mm 74mm, clip,width=0.31\textwidth]{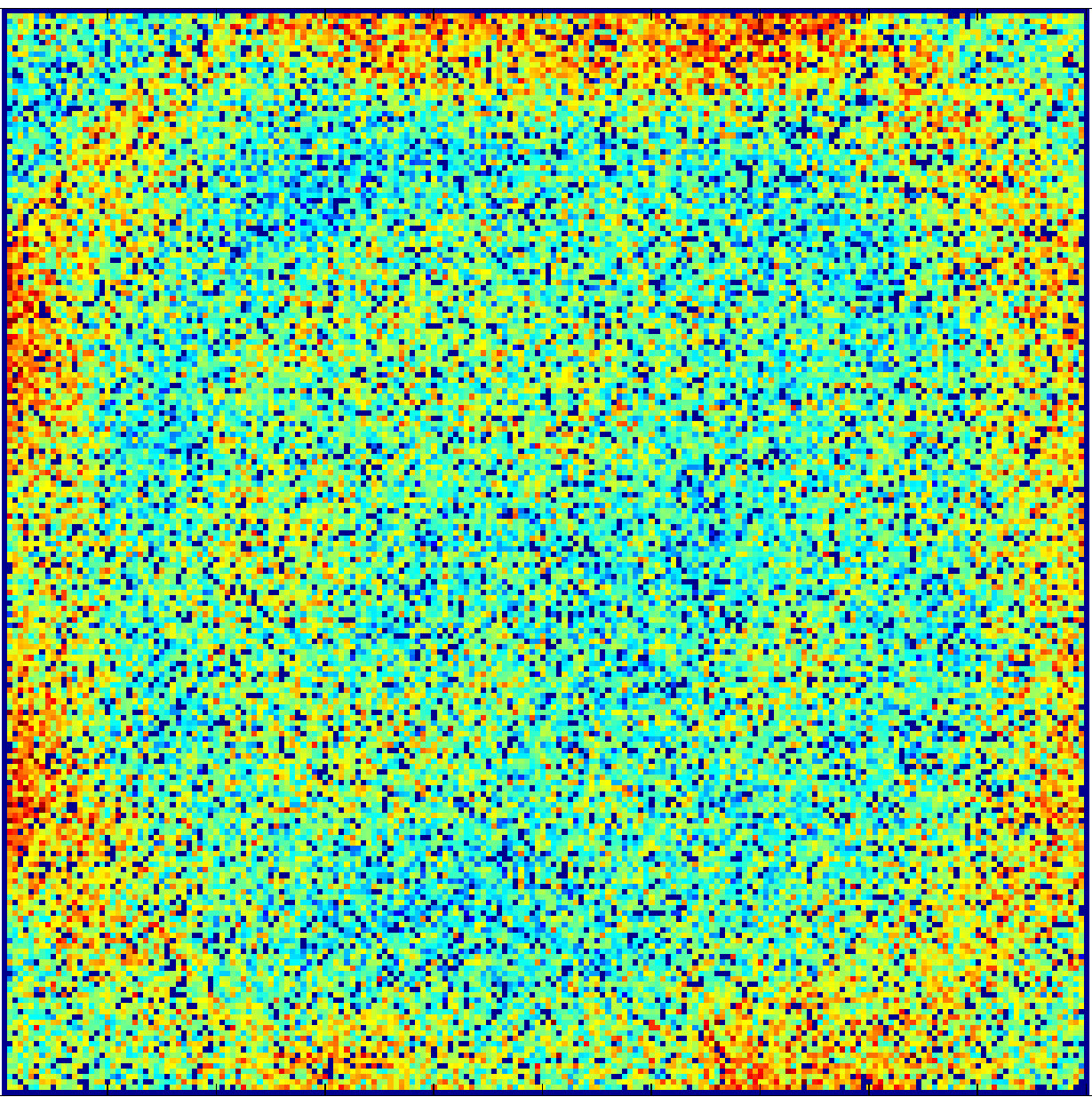}
\includegraphics[trim = 40mm 80mm 51mm 74mm, clip,width=0.31\textwidth]{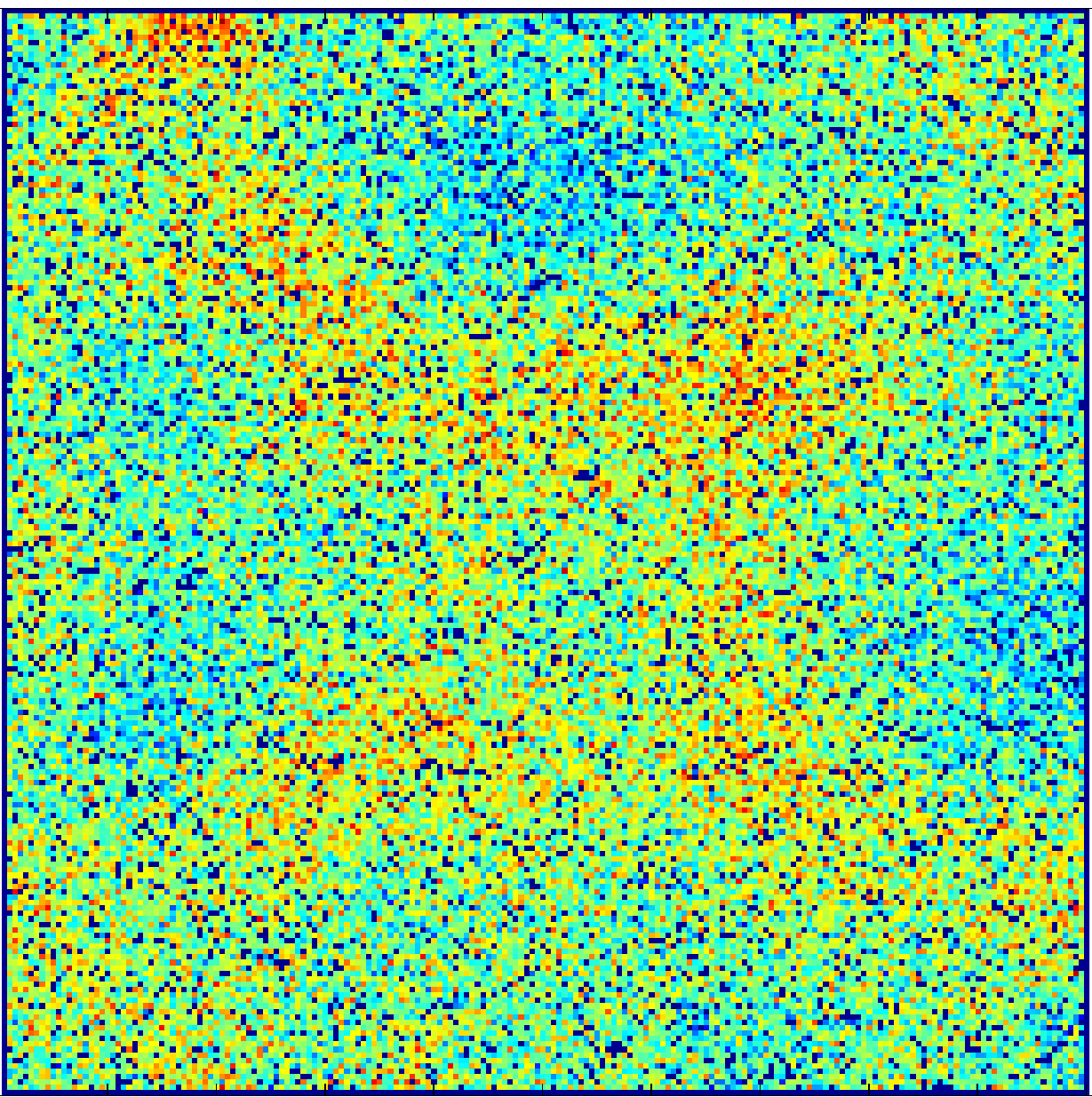}\\
\includegraphics[trim = 49mm 80mm 36mm 90mm, clip,width=0.31\textwidth]{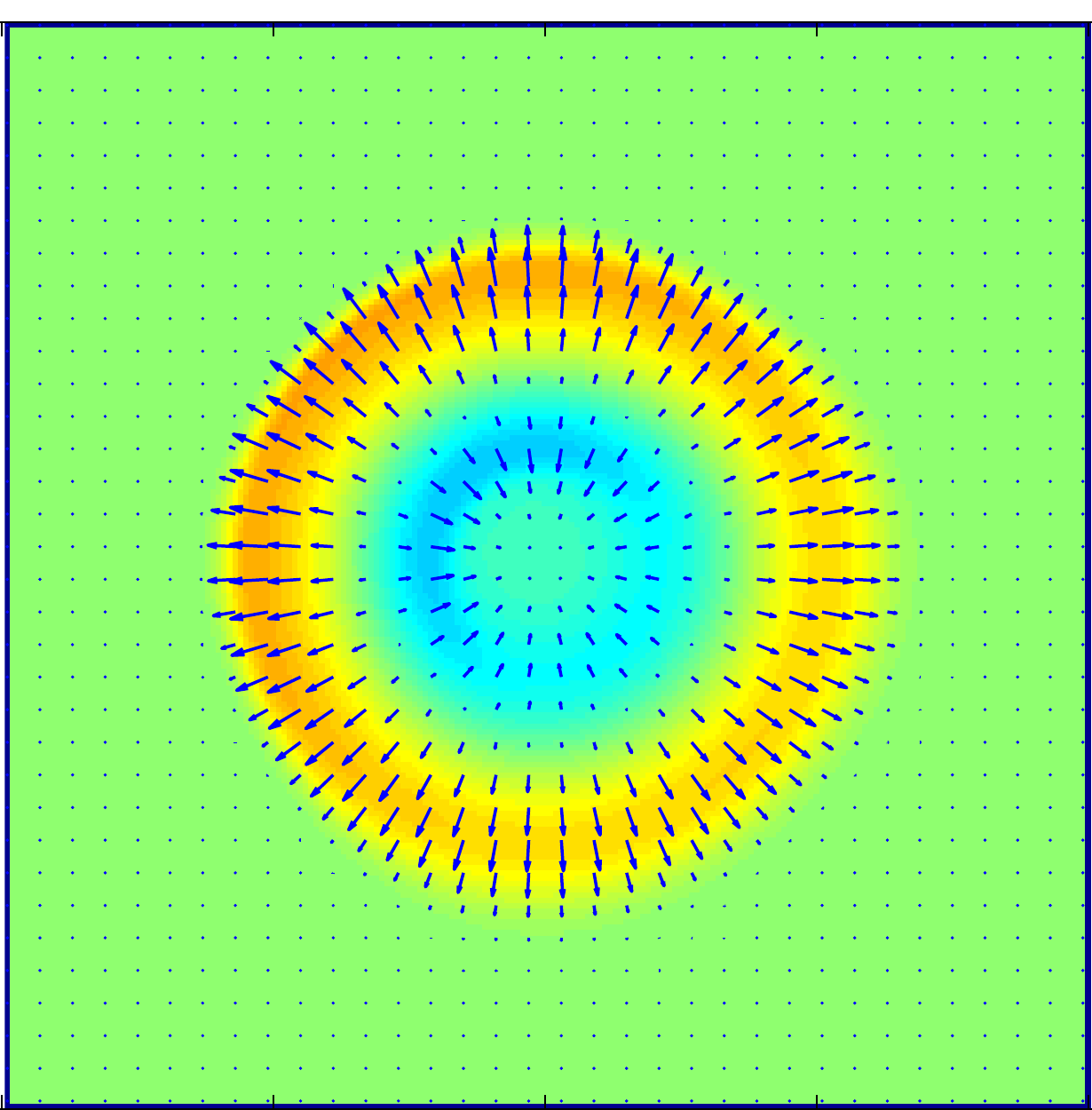}
\includegraphics[trim = 49mm 80mm 36mm 90mm, clip,width=0.31\textwidth]{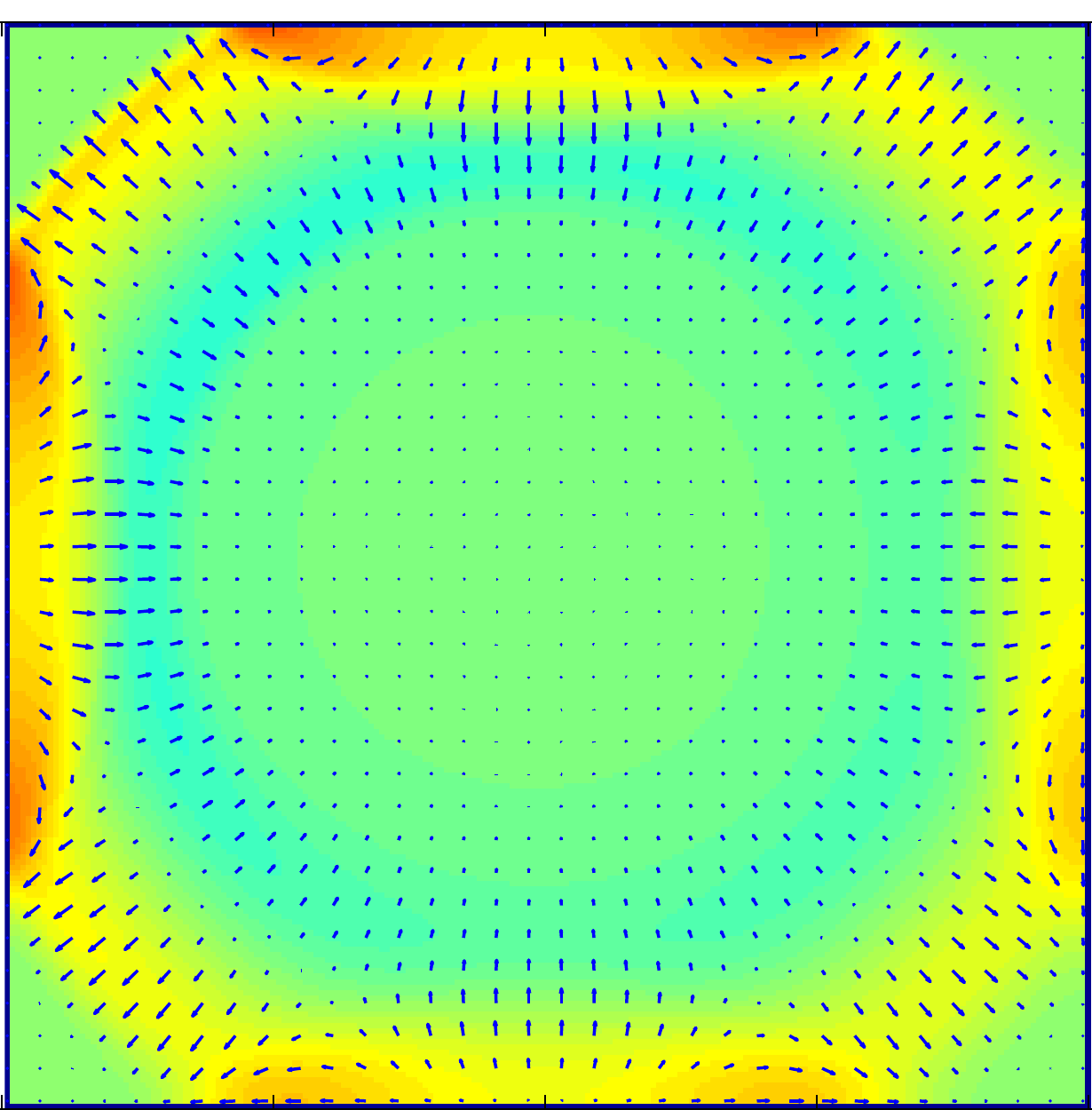}
\includegraphics[trim = 49mm 80mm 36mm 90mm, clip,width=0.31\textwidth]{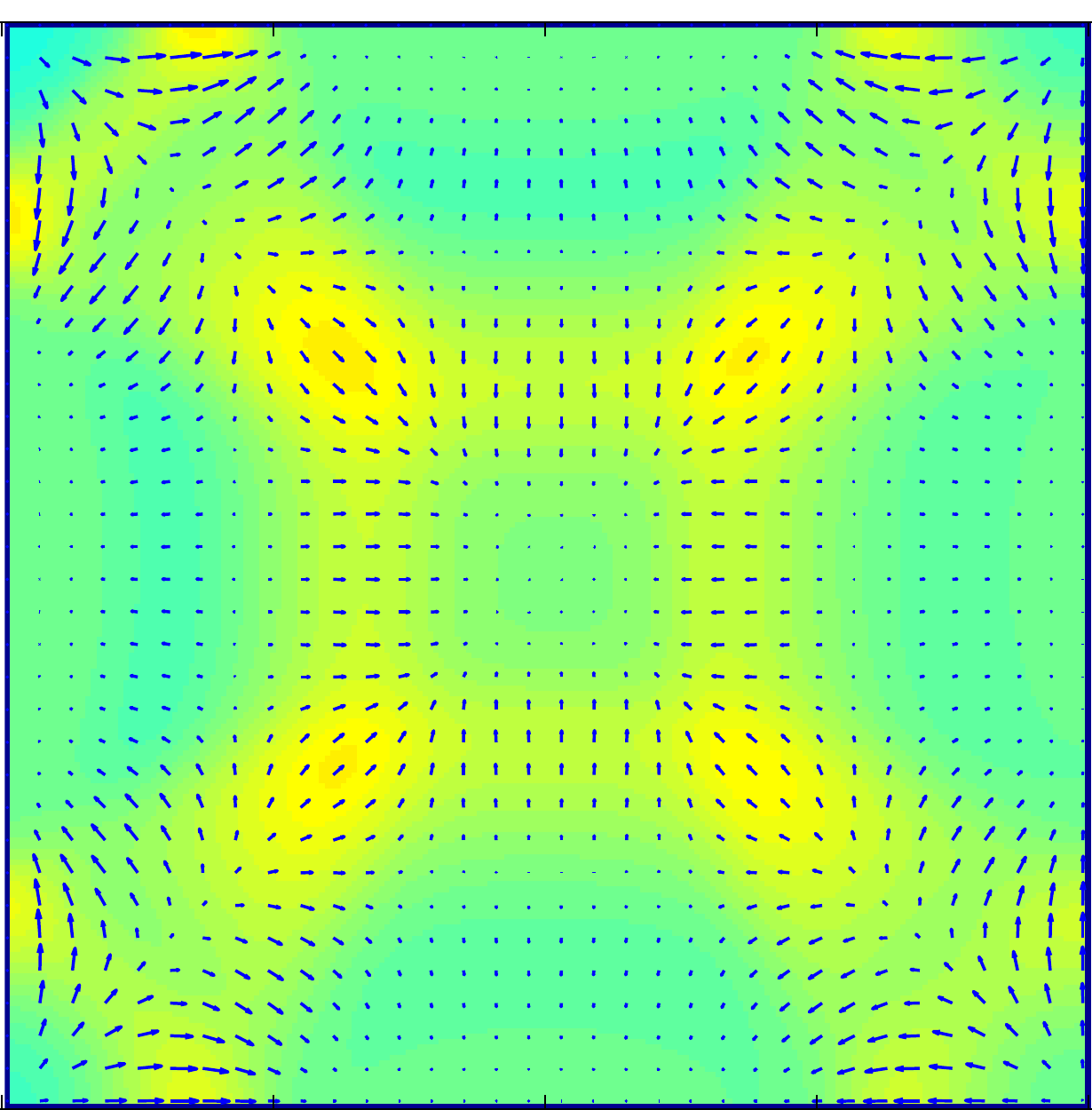}\\
\includegraphics[trim = 68mm 99mm 55mm 110mm, clip,width=0.31\textwidth]{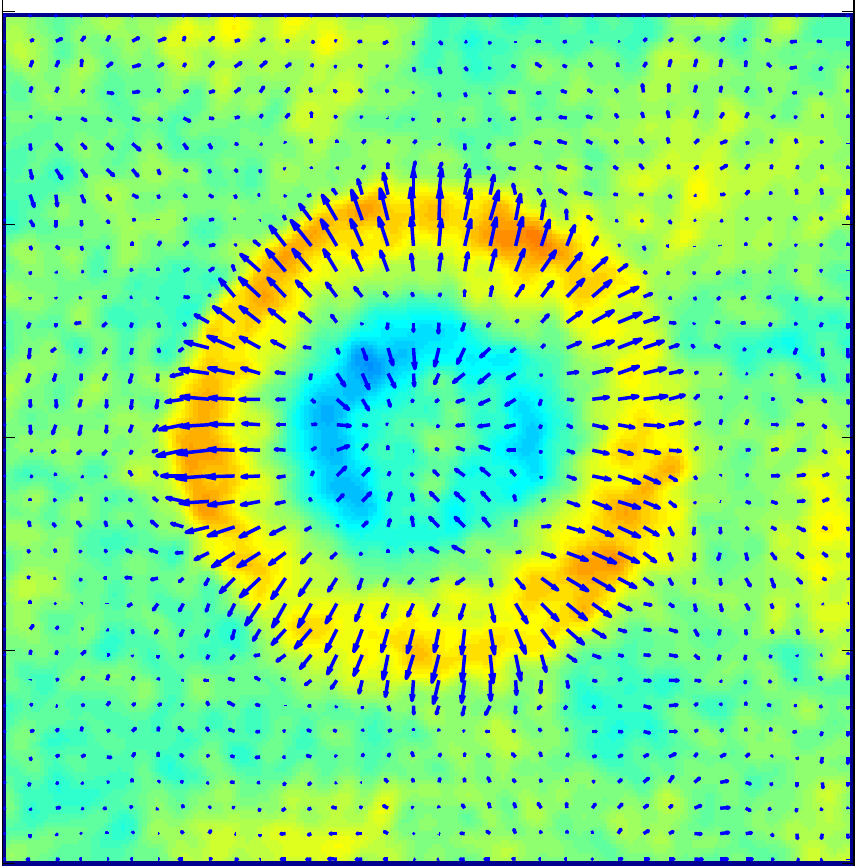}
\includegraphics[trim = 68mm 99mm 55mm 111mm, clip,width=0.31\textwidth]{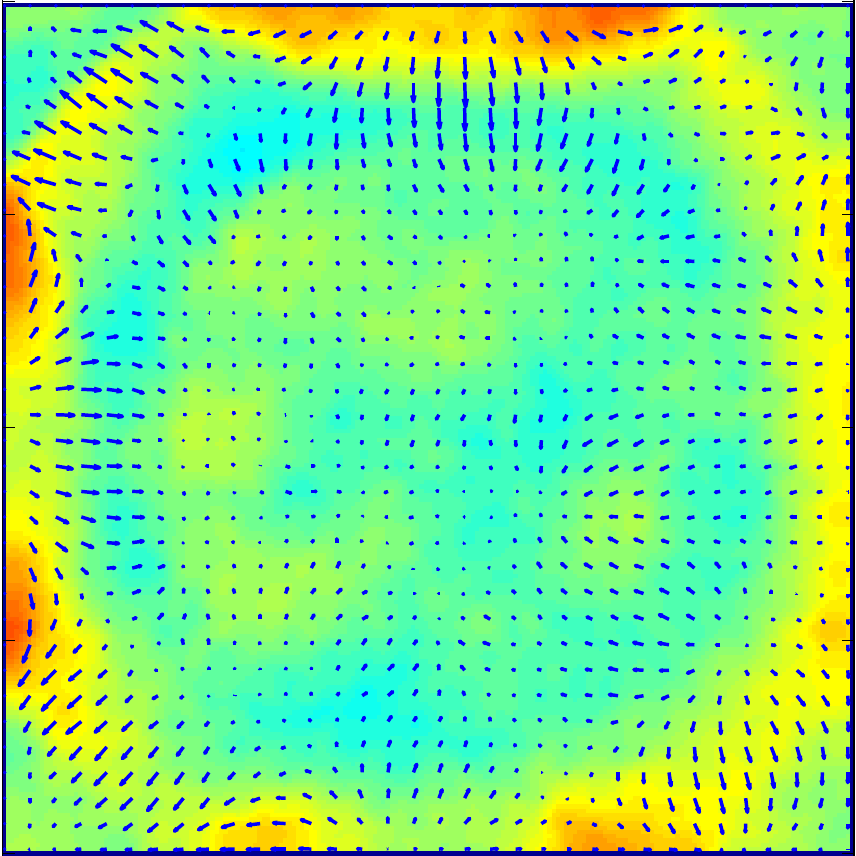}
\includegraphics[trim = 68mm 99mm 55mm 111mm, clip,width=0.31\textwidth]{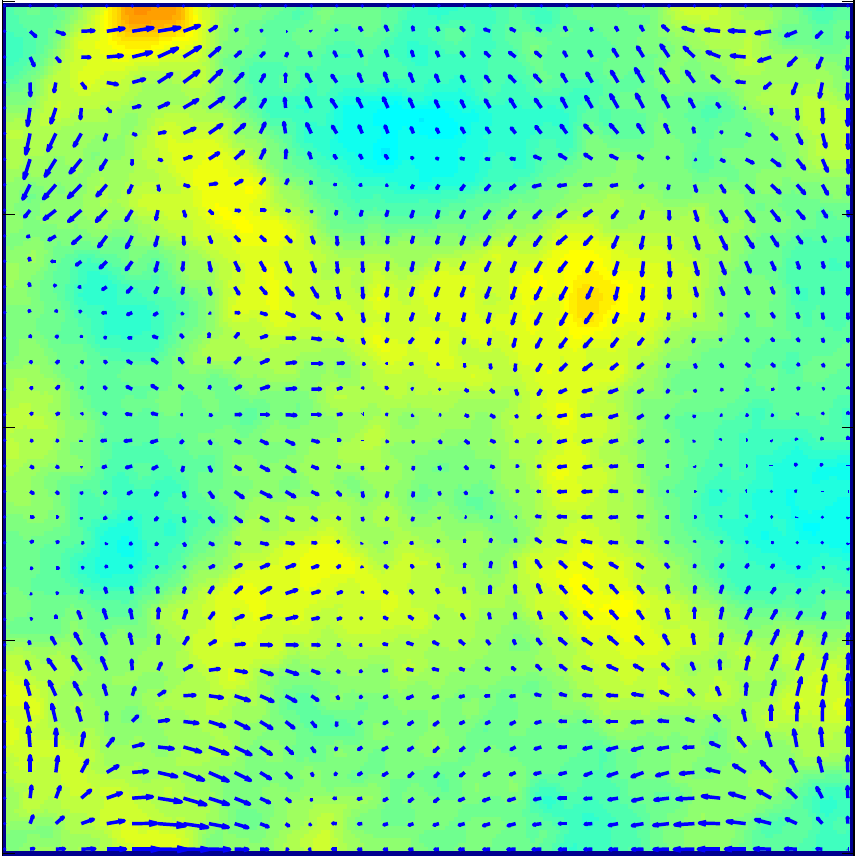}
\includegraphics[trim = 68mm 99mm 55mm 110mm, clip,width=0.31\textwidth]{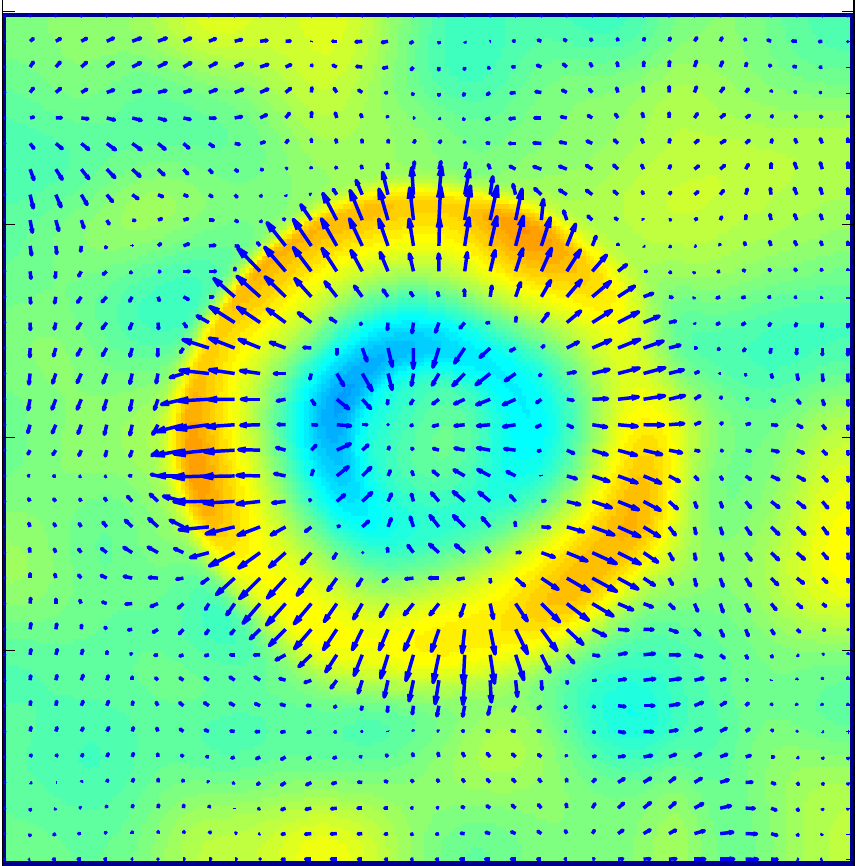}
\includegraphics[trim = 68mm 99mm 55mm 111mm, clip,width=0.31\textwidth]{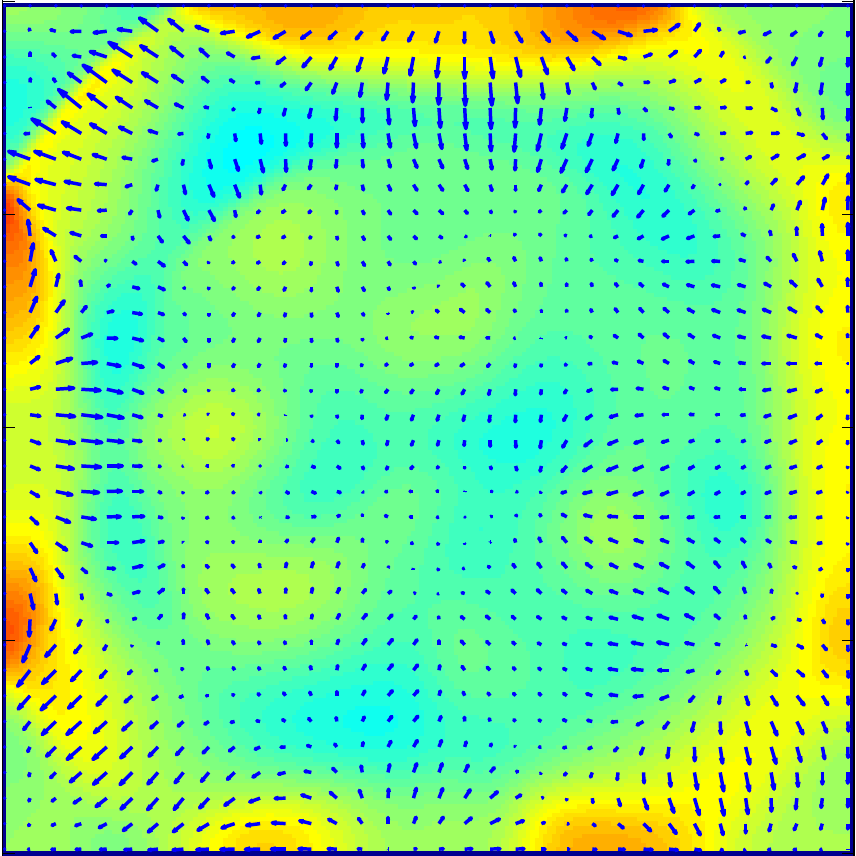}
\includegraphics[trim = 68mm 99mm 55mm 111mm, clip,width=0.31\textwidth]{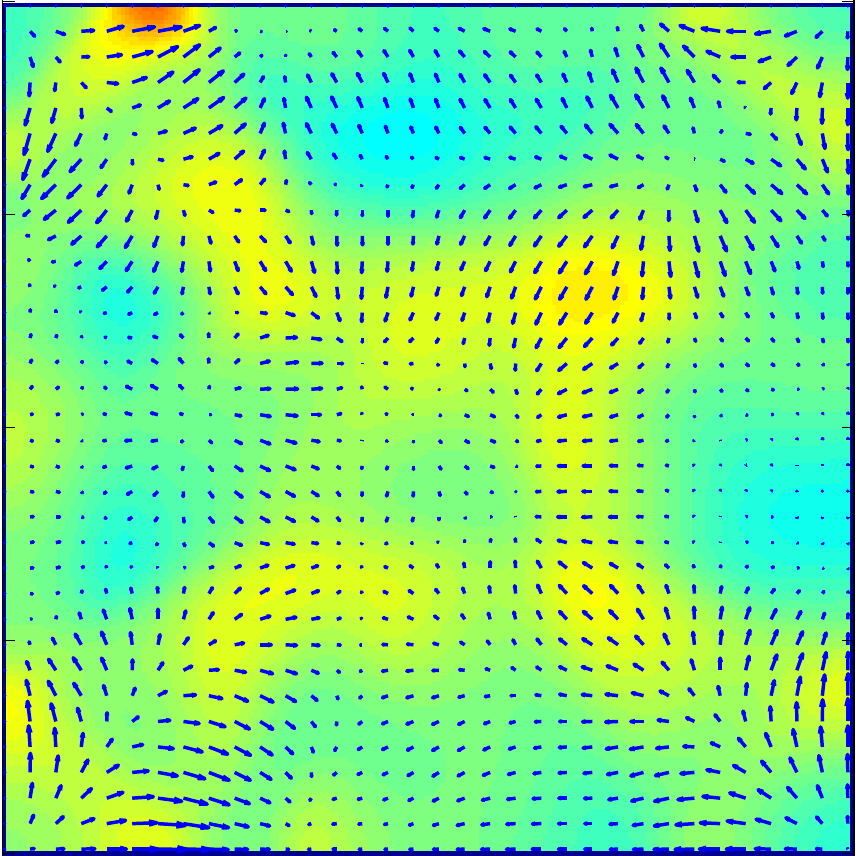}\\
\includegraphics[trim = 35mm 185mm 35mm 100mm, clip,width=0.4\textwidth]{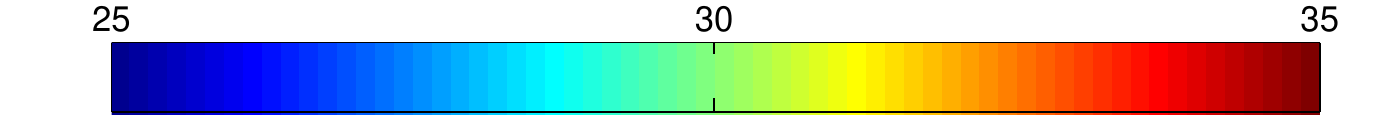}\\
\caption{Comparisons of elevation colormaps in millimeters and of free-surface motion vector fields, for the water column collapse. From top to bottom: observations $\y$, free-run states $\x$, assimilated states $\hat \x$, true states $\tilde \x$. From left to right: $t/t_0=3.17$, $t/t_0=6.34$ and $t/t_0=9.51$. Perturbation parameters on the observations and on the initial state were set to $\sigma_{\rm obs}/h_0=0.16$, $p_{\rm out}=0.1$ and $\mathcal{E}_{\rm init}=0.1$.}
\label{fig:validToy:fig1}
\end{figure}

Figure \ref{fig:validToy:fig1} displays the time evolution of the observations, the free-run simulation, the particle filter estimations and the true at three time steps. The results given by the proposed WEnKF method showed a good agreement with the true for both the elevation and the velocities. As expected for a sequential data assimilation scheme, the agreement improved in time. The free-run states, i.e. the simulation starting from $\x_{\rm init}$, rapidly diverged from the truth whereas the data assimilation scheme, starting from an ensemble of simulations around $\x_{\rm init}$ rapidly recovered the true states over time. It should be noted here that the proposed data assimilation approach reconstructs the full states of the free-surface flow (i.e. elevations and velocities) based on depth observations alone. This is quite remarkable, given the limited input information. Following those encouraging results, obtained for parameters fixed to median values, it was legitimate to analyse the sensitivity of the method to the data quality and to the initialisation. To this end we investigated varying parameter horizons, ranging from $\sigma_{obs}=0.015\,d_0$ to $\sigma_{obs}=0.12\,d_0$, from $p_{\rm out}= 0$ to $p_{\rm out}=0.35$ and from $\mathcal{E}_{\rm init}=0.025$ to $\mathcal{E}_{\rm init}=0.5$.

\begin{figure}[t!]
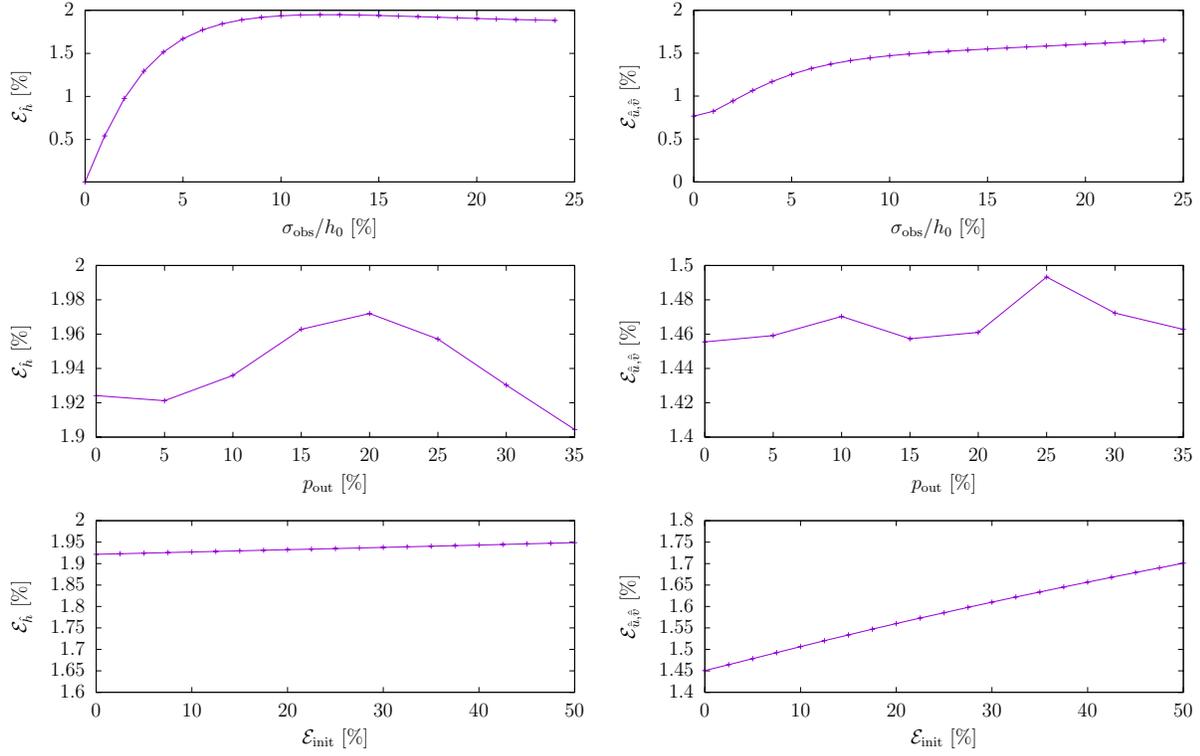

{\resizebox{8cm}{!}{\input{sigmaobs-h.tex}}}
{\resizebox{8cm}{!}{\input{sigmaobs-vit.tex}}}\\
{\resizebox{8cm}{!}{\input{outliers-h.tex}}}
{\resizebox{8cm}{!}{\input{outliers-vit.tex}}}\\
{\resizebox{8cm}{!}{\input{initialcondition-h.tex}}}
{\resizebox{8cm}{!}{\input{initialcondition-vit.tex}}}\\
\caption{Results for the collapse of a water column, for twoObs at $t/t_0=9.51$: Left, elevation component errors $\mathcal{E}_{\hat h}$; Right, velocity components errors $\mathcal{E}_{\hat{\bar{u}},\hat{\bar{v}}}$ as defined in \eref{eq:error_estim}. From top to bottom, errors as a function of: the non-dimensional level of noise $\sigma_{obs}/h_o$ with $p_{out}=0.1$ and $\mathcal{E}_{\rm init}=0.1$; the rate of outliers $p_{\rm out}$ with $\sigma_{obs}/h_0=0.1$ and $\mathcal{E}_{\rm init}=0.1$; the rate of initial perturbation $\mathcal{E}_{\rm init}$ with $p_{out}=0.1$ and $\sigma_{obs}/h_0=0.1$.}
\label{fig:validToy:curve}
\end{figure}
\begin{figure}[h!]
\centering
{\small (a)}\\
\includegraphics[trim = 32mm 80mm 18mm 80mm, clip,width=0.285\textwidth,height=0.31\textwidth]{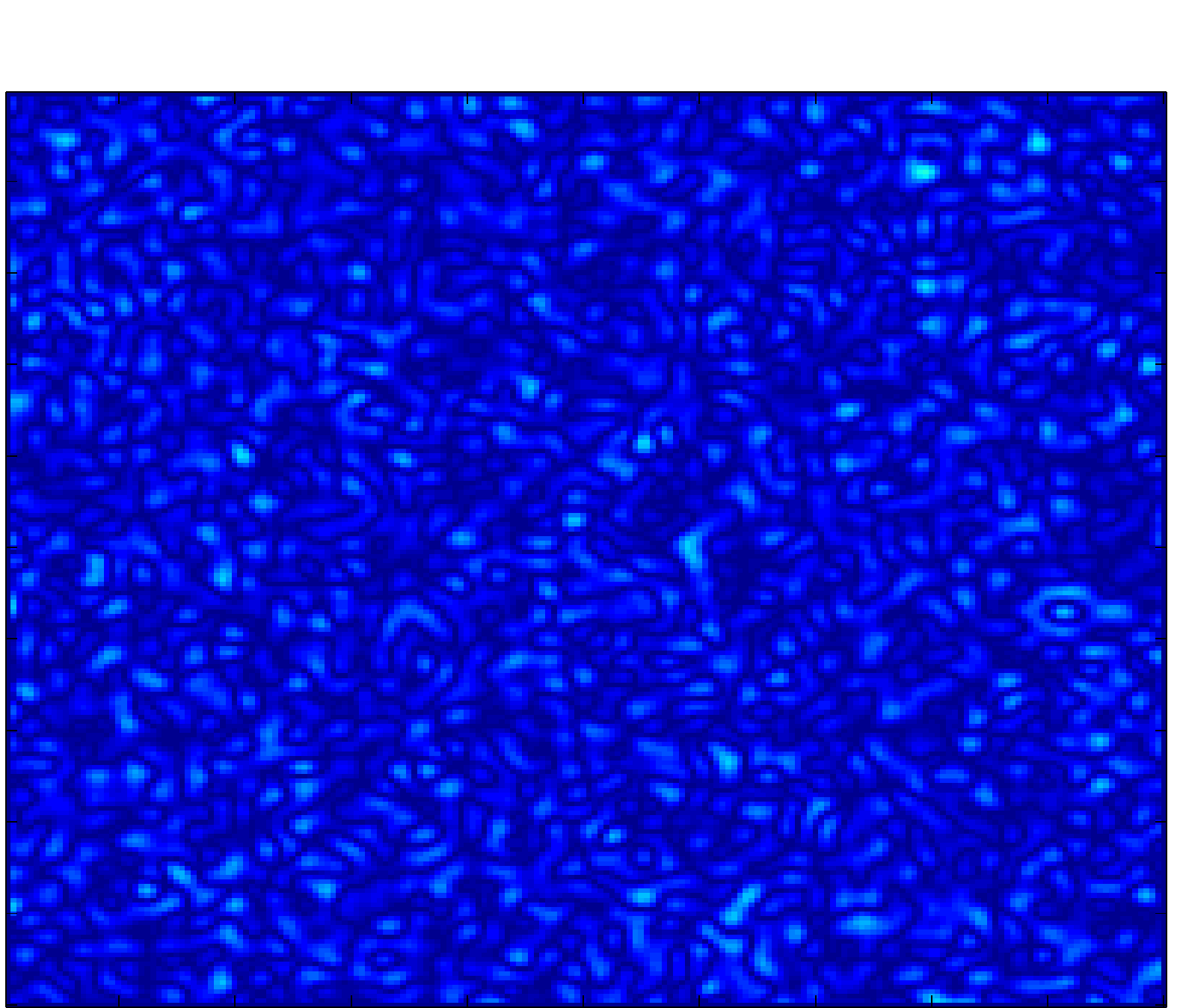}
\includegraphics[trim = 32mm 80mm 18mm 80mm, clip,width=0.285\textwidth,height=0.31\textwidth]{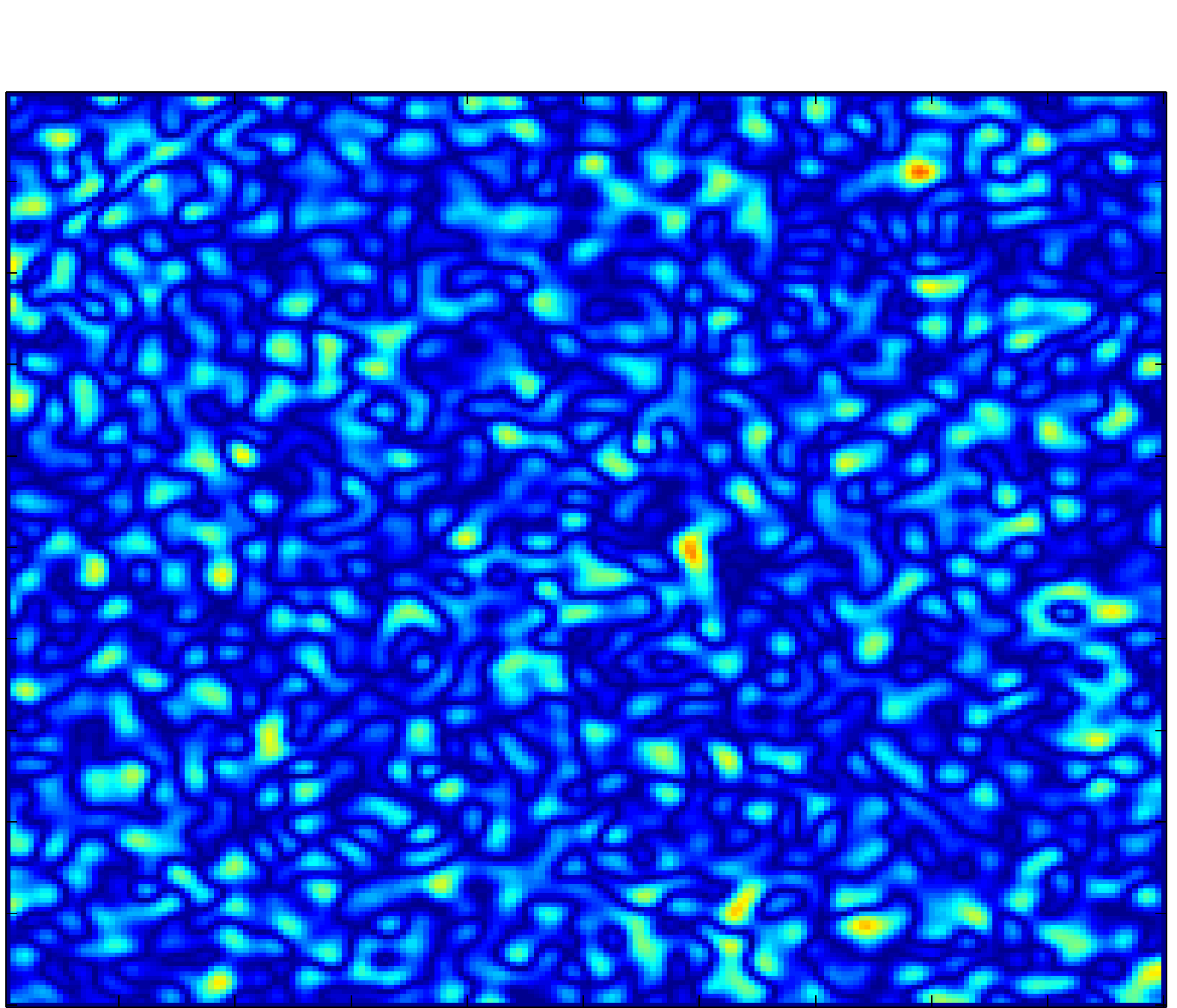}
\includegraphics[trim = 32mm 80mm 18mm 80mm, clip,width=0.285\textwidth,height=0.31\textwidth]{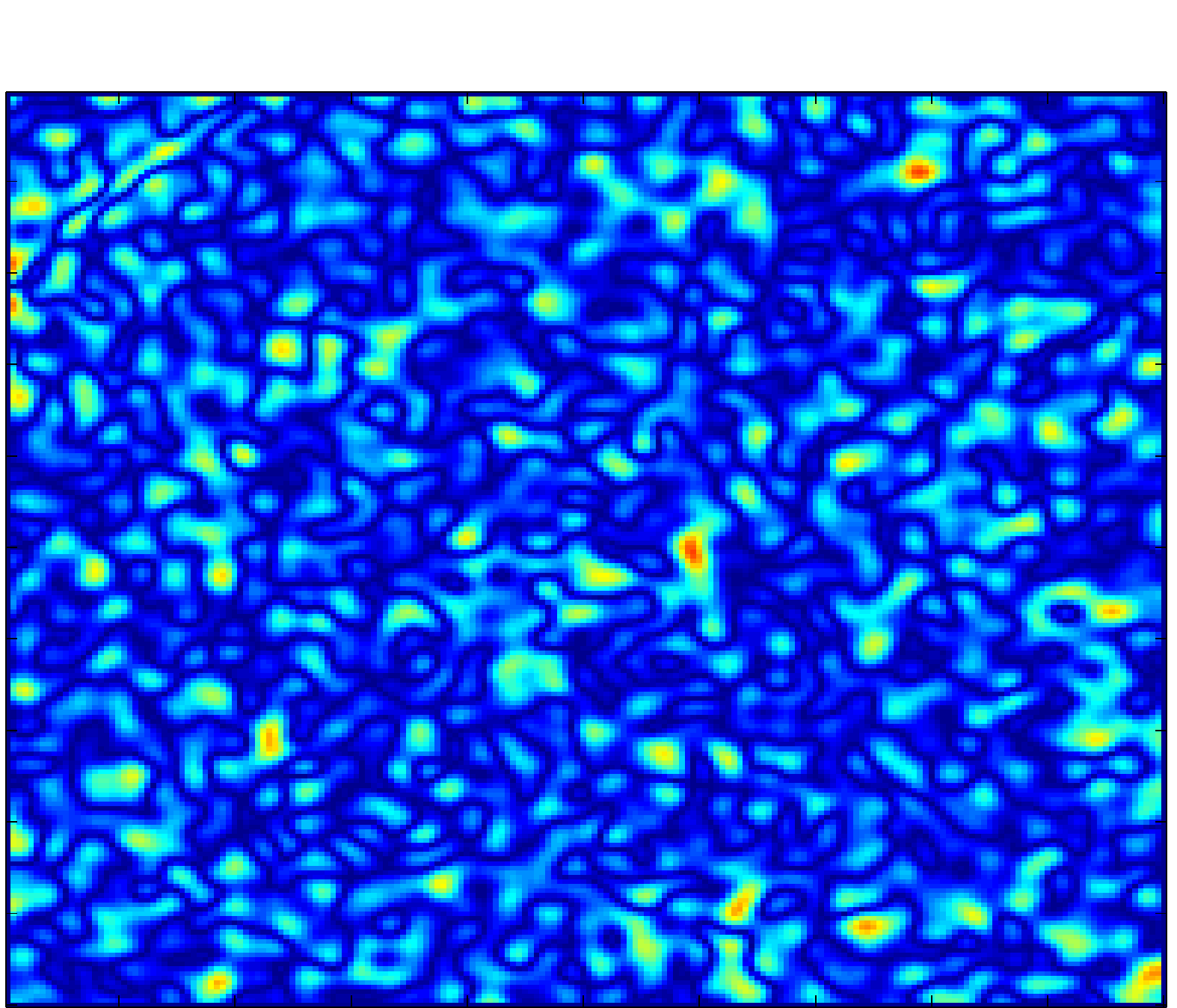}
\includegraphics[trim = 43mm 185mm 35mm 100mm, clip, angle=90,width=0.0265\textwidth]{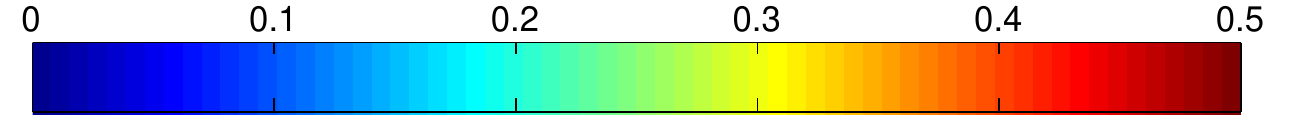}\\
\includegraphics[trim = 32mm 80mm 18mm 80mm, clip,width=0.285\textwidth,height=0.31\textwidth]{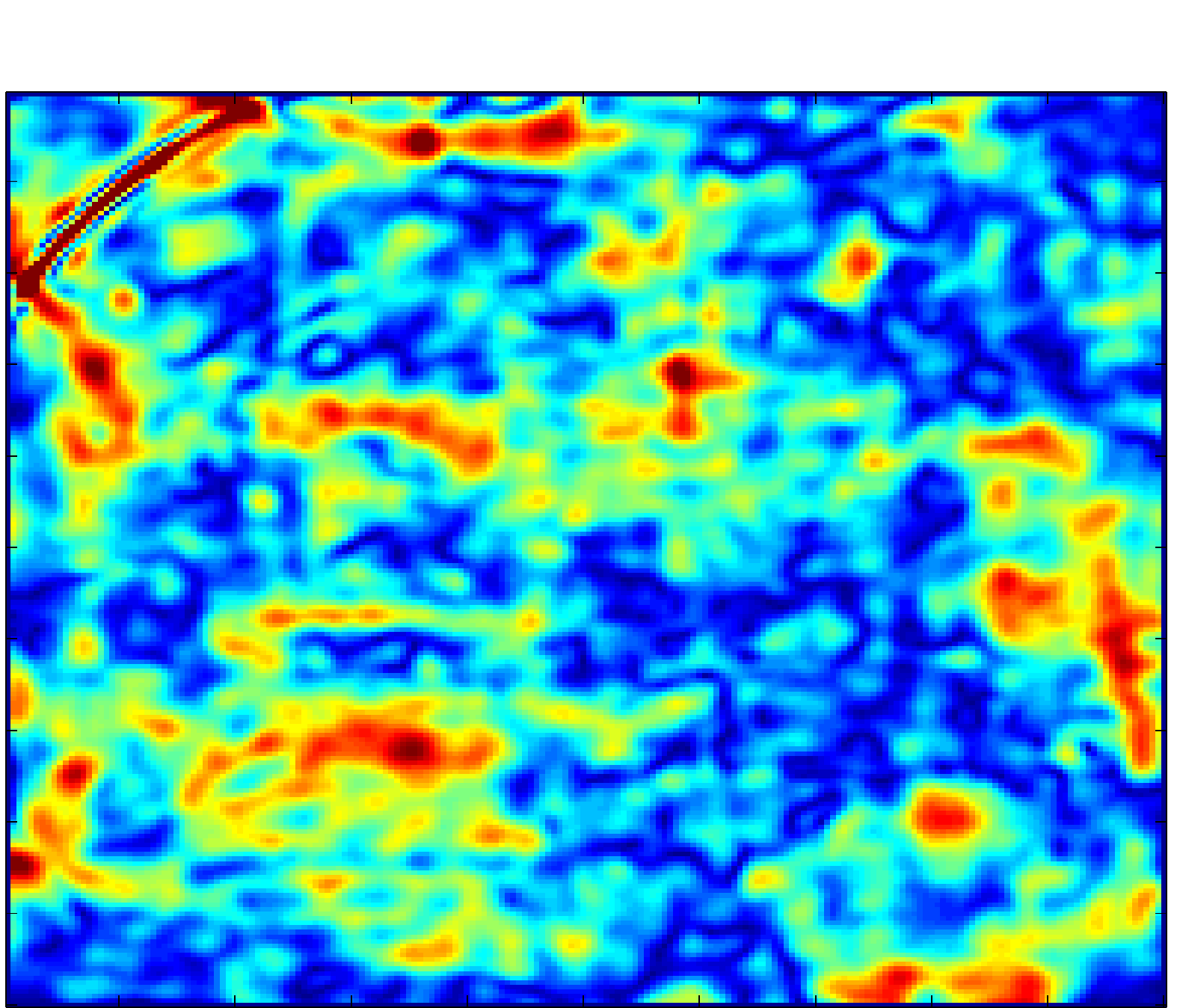}
\includegraphics[trim = 32mm 80mm 18mm 80mm, clip,width=0.285\textwidth,height=0.31\textwidth]{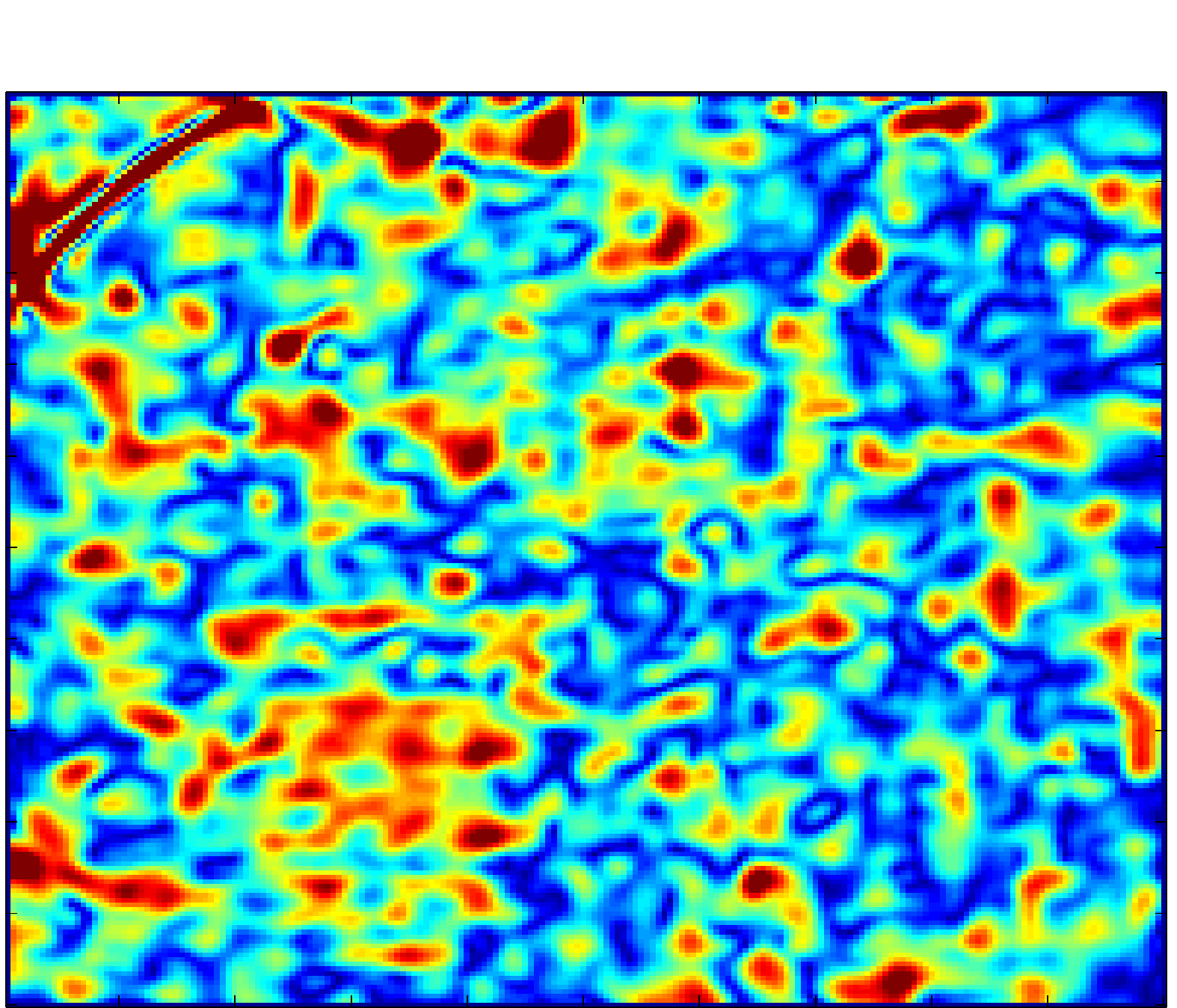}
\includegraphics[trim = 32mm 80mm 18mm 80mm, clip,width=0.285\textwidth,height=0.31\textwidth]{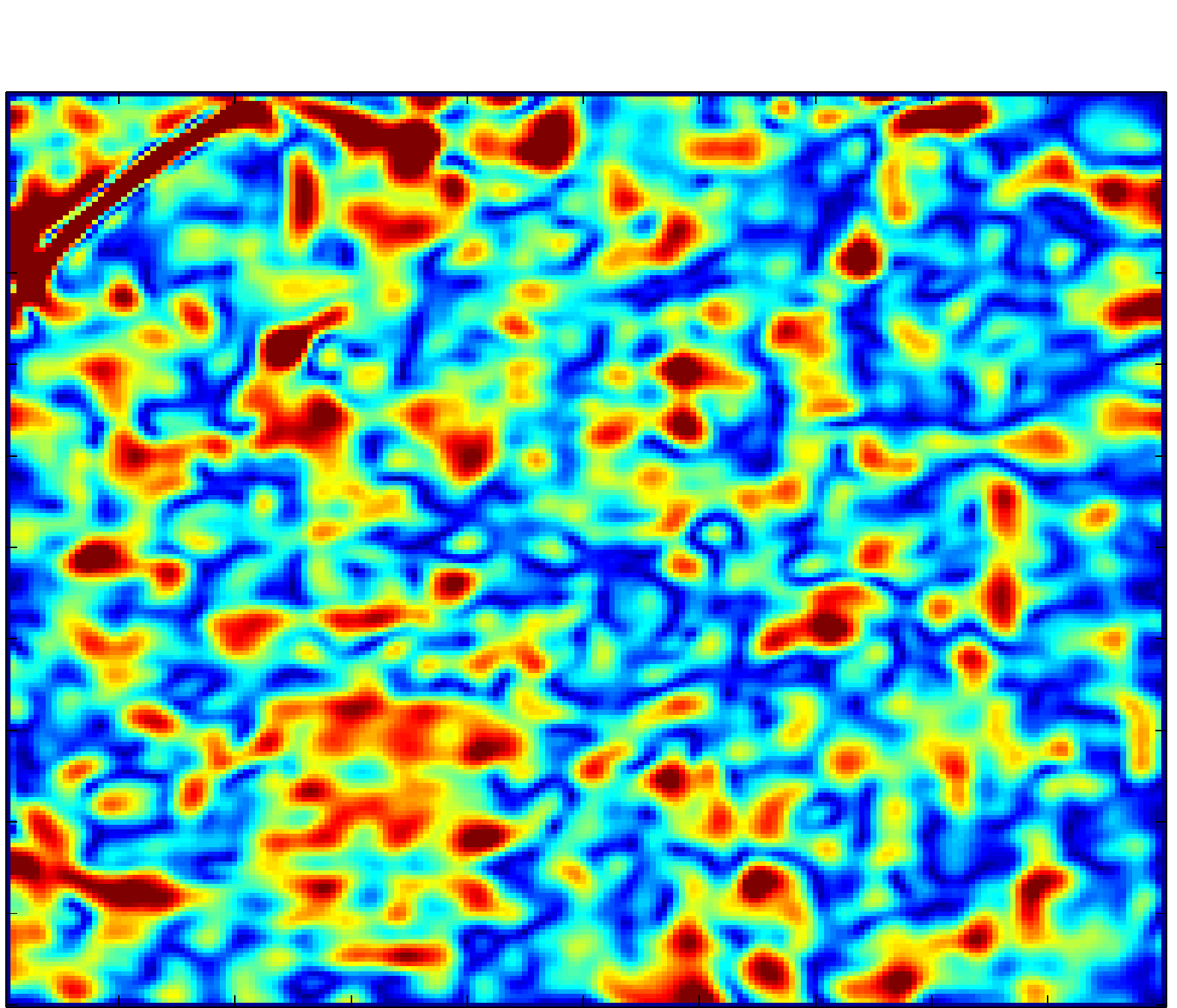}
\includegraphics[trim = 43mm 185mm 35mm 100mm, clip, angle=90,width=0.0265\textwidth]{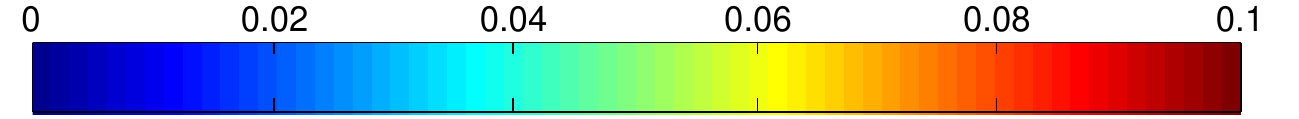}\\
{\small (b)}\\
\includegraphics[trim = 32mm 80mm 18mm 80mm, clip,width=0.285\textwidth,height=0.31\textwidth]{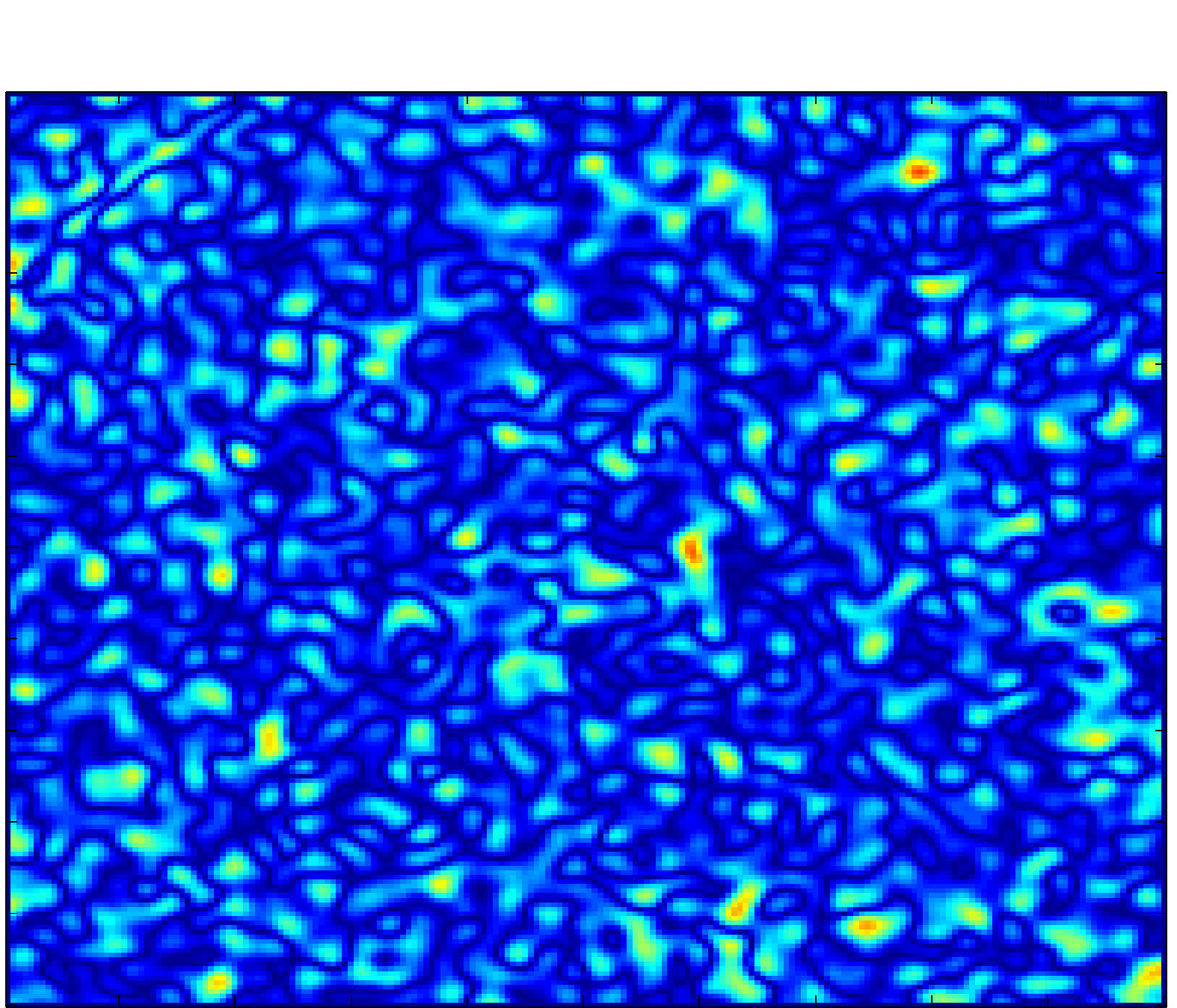}
\includegraphics[trim = 32mm 80mm 18mm 80mm, clip,width=0.285\textwidth,height=0.31\textwidth]{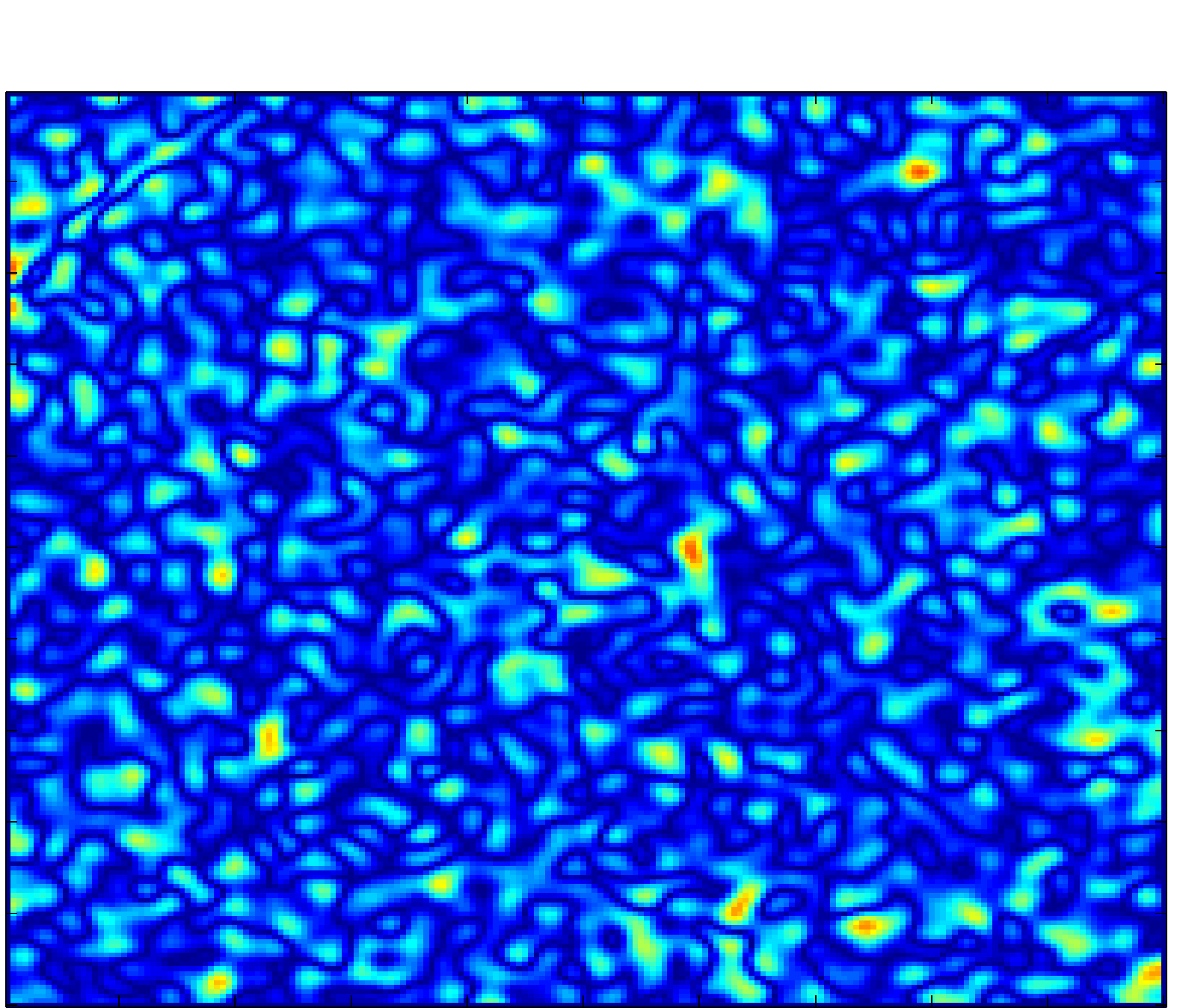}
\includegraphics[trim = 32mm 80mm 18mm 80mm, clip,width=0.285\textwidth,height=0.31\textwidth]{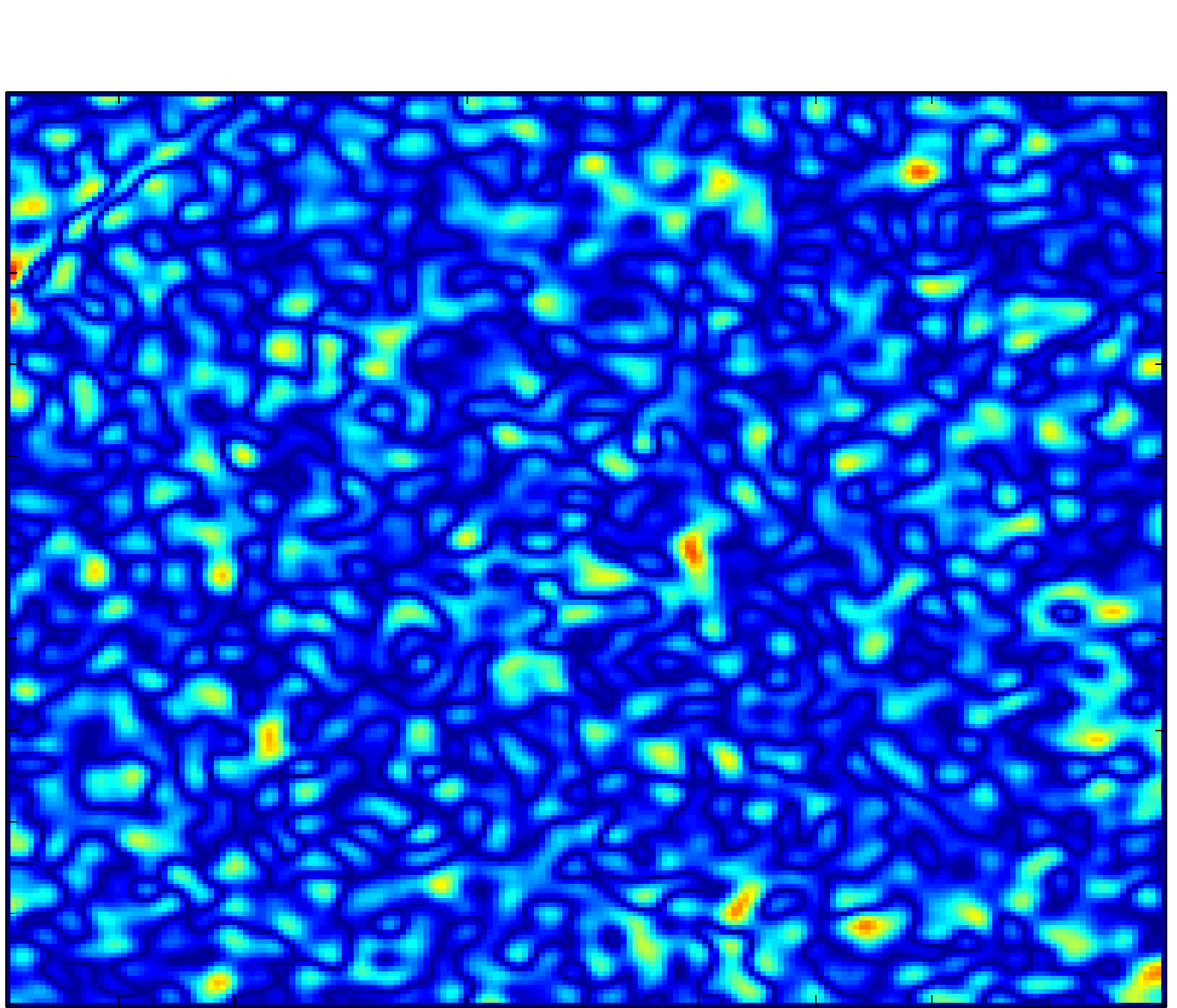}
\includegraphics[trim = 43mm 185mm 35mm 100mm, clip, angle=90,width=0.0265\textwidth]{barToyErrorElevation.pdf}\\
\includegraphics[trim = 32mm 80mm 18mm 80mm, clip,width=0.285\textwidth,height=0.31\textwidth]{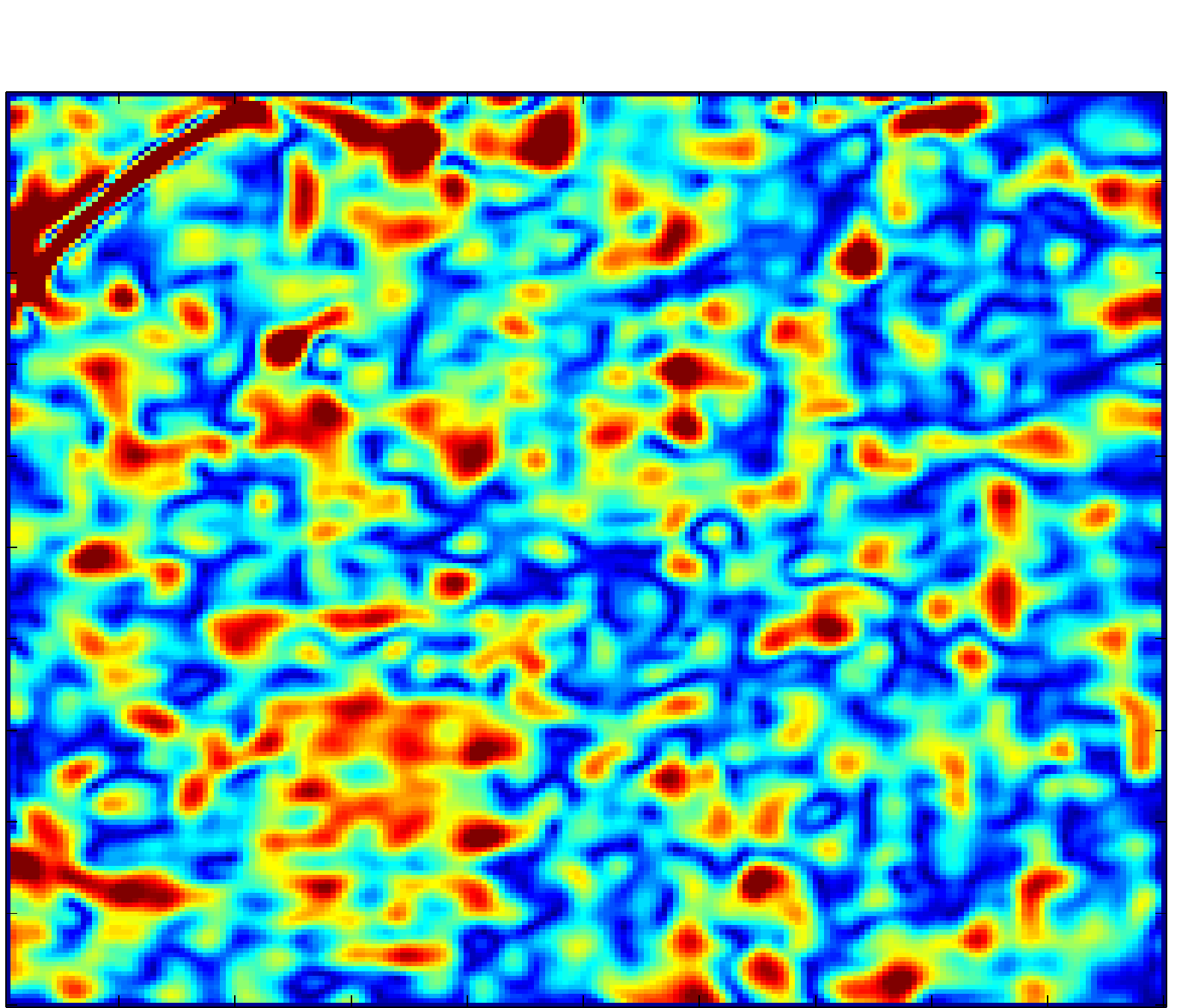}
\includegraphics[trim = 32mm 80mm 18mm 80mm, clip,width=0.285\textwidth,height=0.31\textwidth]{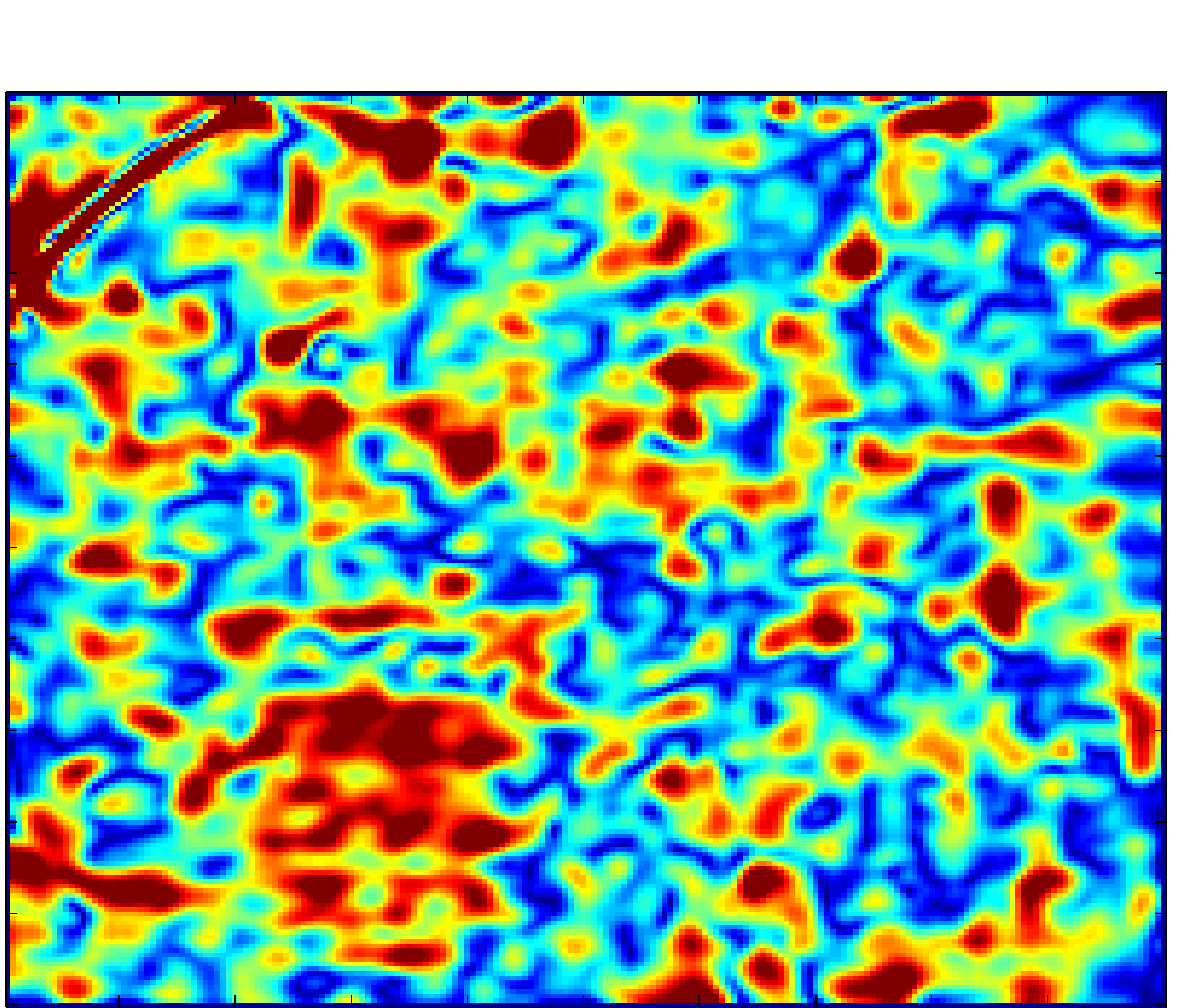}
\includegraphics[trim = 32mm 80mm 18mm 80mm, clip,width=0.285\textwidth,height=0.31\textwidth]{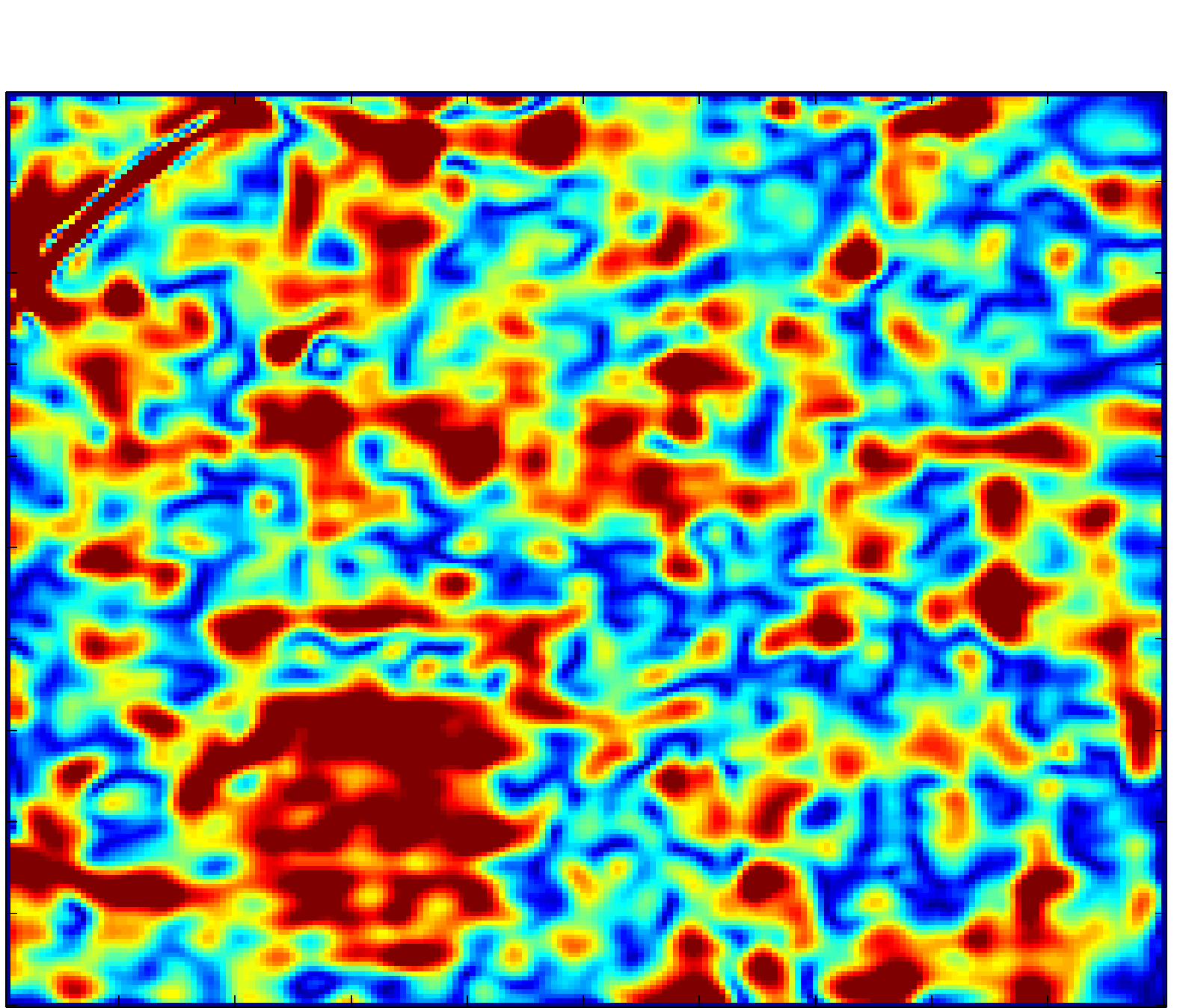}
\includegraphics[trim = 43mm 185mm 35mm 100mm, clip, angle=90,width=0.0265\textwidth]{barToyErrorDisplacement.pdf}
\caption{Error maps for the water column collapse at time $t/t_0=6.34$: (a), with increasing additive noise level $\sigma_{obs}/d_0 \in [0.01, 0.04, 0.08]$; (b), with increasing perturbations of the initial condition $\mathcal{E}_{\rm init}\in [0.1, 0.2, 0.3]$. First rows: elevation errors $||h-\tilde{h}||~\mbox{in~[mm]}$. Second rows:  velocity errors $||\w-\tilde{\w}||~\mbox{in~[cm/s]}$.}\label{fig:validToy:errorMap1}
\end{figure}

Figure~\ref{fig:validToy:curve} represents data assimilation errors as a function of the non-dimensional level of noise $\sigma_{obs}/h_o$, the rate of outliers $p_{\rm out}$ and the rate of initial perturbation $\mathcal{E}_{\rm init}$, for twoObs at $t/t_0=9.51$. Whatever the parameter values, the three estimated state components had the same low error level, below $2\%$. The rate of outliers (given realistic ratio below $35\%$) did not influence the estimations. On the contrary the errors increased with the level of noise in the observations. For $\sigma_{obs}/h_0$ larger than $10\%$ and $5\%$ the errors reached a plateau slightly decreasing for the elevation $h$ and slightly increasing for the two velocity components $\bar{u}$ and $\bar{v}$, respectively. Note that this changing slope might be correlated with the noise standard deviations of the observation and dynamical models (covariance matrices $\R_t$ and $\Q_t$) equal to $11\%$ and $6\%$ of $h_0$, respectively. This asymptotic behavior, also reflected in figure \ref{fig:validToy:errorMap1}a, indicated the good robustness of the WEnKF scheme. The slope difference seen between the elevation and the velocity component errors could be due to the fact that the elevation was the observed component. Another feature of the observed component was clearly exhibited when plotting the errors as a function of the initial perturbation level $\mathcal{E}_{\rm init}$ (see figures \ref{fig:validToy:curve} and \ref{fig:validToy:errorMap1}b): the elevation component estimations were not sensitive to $\mathcal{E}_{\rm init}$ whereas the velocity component estimation errors slightly increased with $\mathcal{E}_{\rm init}$. Note that the noise standard deviation of the observation model

As a conclusion of this first simple flow configuration experiment, the proposed particle filter method exhibited very interesting reconstruction and robustness properties. Furthermore, we indicate that we did not observed any significant differences between the methods oneObs and twoObs. This result can be explained by the fact that in the present configuration, the only source of departure from the free-run flow to the true flow came from the unknown initial state. As a result, one did not need a complex coupling between the model and the observation after the first few time integration because by itself the model was able to explain the flow. The two schemes (oneObs and twoObs) will be compared further more extensively for more realistic flow configurations when adding unknown varying inflow conditions and a stochastic forcing term. In the following we fix data quality and initialisation parameters to $p_{out}=0.1$, $\sigma_{obs}=0.03\,d_0$ and $\mathcal{E}_{\rm init}=0.2$, respectively.

\subsection{Flow in an suddenly expanding flume} 

\begin{figure}[h!]
\centering
{\resizebox{10cm}{!}{\input{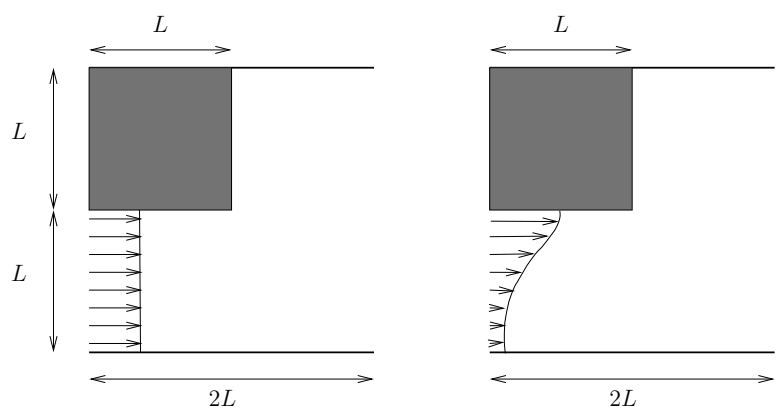}}}
\caption{Schematic view of the suddenly expanding flume for two inflow configurations: left, uniform inlet velocity profile; right, half bell-shape inlet velocity profile.}
\label{fig:scheme}
\end{figure}

In this case study, we considered the flow in a suddenly expanding flume as described in figure \ref{fig:scheme}. The flume consisted of a $L=10~\mbox{cm}$ long, $L$ wide approach flume before a sudden expansion and a $L$ long, $2L$ wide expanding flume. The inflow water surface elevation and velocities were oscillatory in phase with a frequency of $1~\mbox{Hz}$ and amplitudes of $1~\mbox{cm}$ and $0.22~\mbox{m/s}$, respectively. The inflow mean elevation and mean velocity were $H_{\rm in}=1~\mbox{cm}$ and $U_{\rm in}=0.22~\mbox{m/s}$, respectively. The corresponding inflow Froude number $Fr=U_{\rm in}/\sqrt{g\,H_{\rm in}}=0.7$ was lower than unity which indicates that the flow was subcritical, i.e. behaved like a fluvial motion. A characteristic of this flow lies in the sudden expansion where the flow separates and generates a two dimensional vortex. In addition we considered two cases for the oscillatory inflow: one with a uniform inlet velocity profile and another with a half bell-shape (square-cosinus) inlet velocity profile leading to a more complex free surface flow dynamic. Both expanding flume flow configurations were simulated with the shallow-water numerical model in a computational domain $L_x \times L_y = 2L \times 2L$ discretized on a square grid of $n_x \times n_y= 200 \times 200$ points and with a time step $\Delta t\,u_0/L=0.006$, where $u_0=\sqrt{g\,H_{\rm in}}$.
 
The true inflow states $\tilde{\x}_{{\rm in},0 \ldots t}$ were varying periodically in elevation and in velocity. For the assimilation methods, the variations in time of the inflow conditions were considered as unknown and in practice were fixed to their mean values $H_{\rm in}$ and $U_{\rm in}$. We stress the fact that to show the ability of the stochastic model \eref{eq:dynamicStoch} to handle unexpected situations, the inflow components were not considered as an unknown vector parameter of an augmented model system like in \citeasnoun{gronskis_etal_2013}, but just as state variable inlet values. Moreover we considered the case, named \emph{stochastic forcing}, where we add a random large scale extra component to the state vectors when generating true states $\tilde \x_{0 \ldots t}$ before each time integration. Hence, each true states included the \emph{stochastic forcing} transported by the dynamical model during one time step $\Delta t$ and more. This was done to mimic the complexity of real free surface flows which cannot be simulated with the deterministic shallow water model (\ref{eq:shallow_watera},\ref{eq:shallow_waterb},\ref{eq:shallow_waterc}). In contrast with data assimilation methods considering ``perfect'' models, the stochastic data-assimilation scheme \eref{eq:overall_model} carried out in this study better modeled such complex situations. The noisy observations $\y_{1 \ldots t}$ were built from the true state elevation component taken every $40 \Delta t\,u_0/L$, thus leading to an observations Strouhal number $St_{\rm obs}=L/(40 \Delta t\,u_0) = 4.17$. 

\begin{figure}[h!]
\centering
{\small (a)}\\
\includegraphics[trim = 49mm 78.5mm 36mm 92mm, clip,angle=90, width=0.31\textwidth]{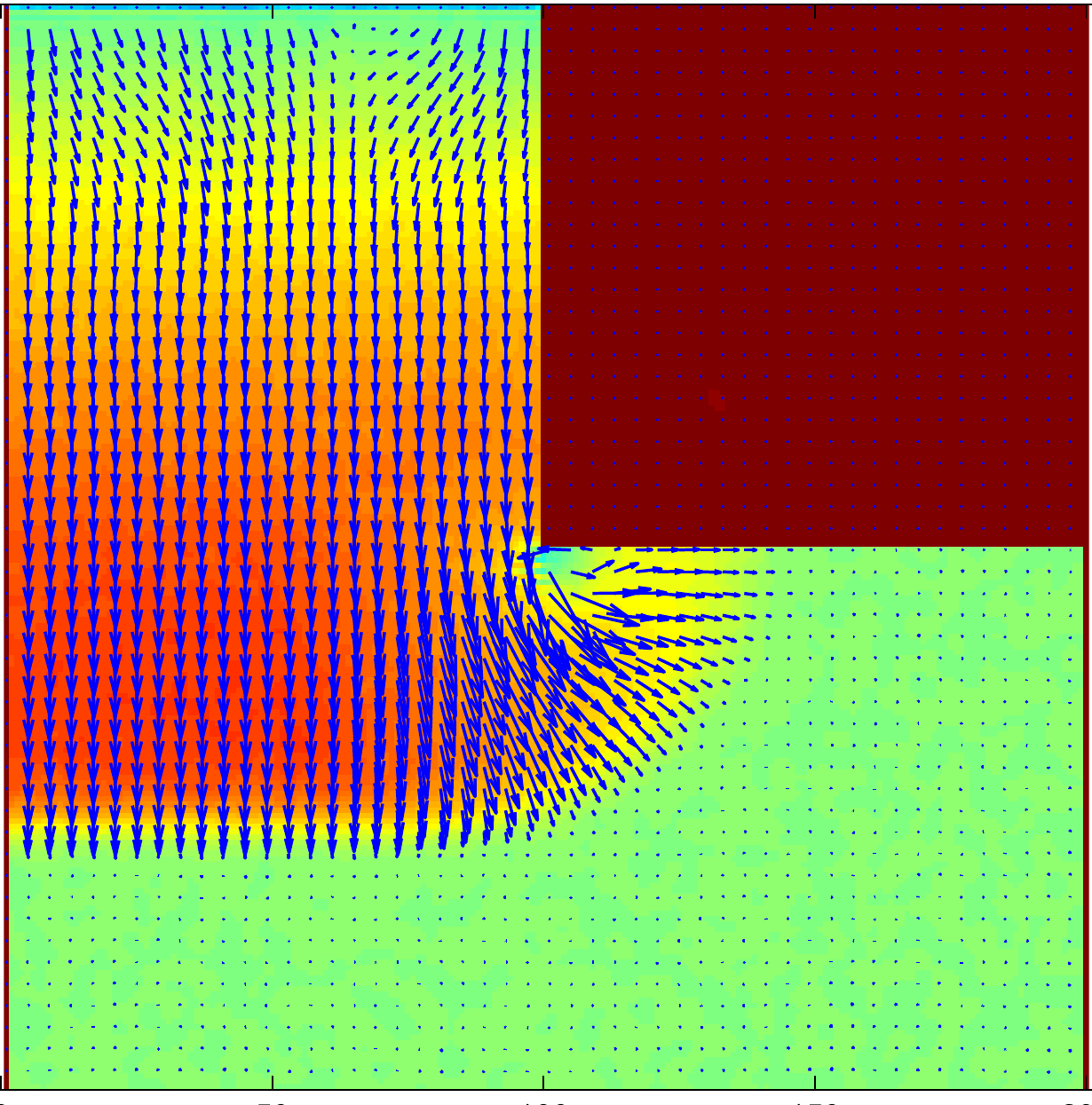}
\includegraphics[trim = 49mm 78.5mm 36mm 92mm, clip,angle=90, width=0.31\textwidth]{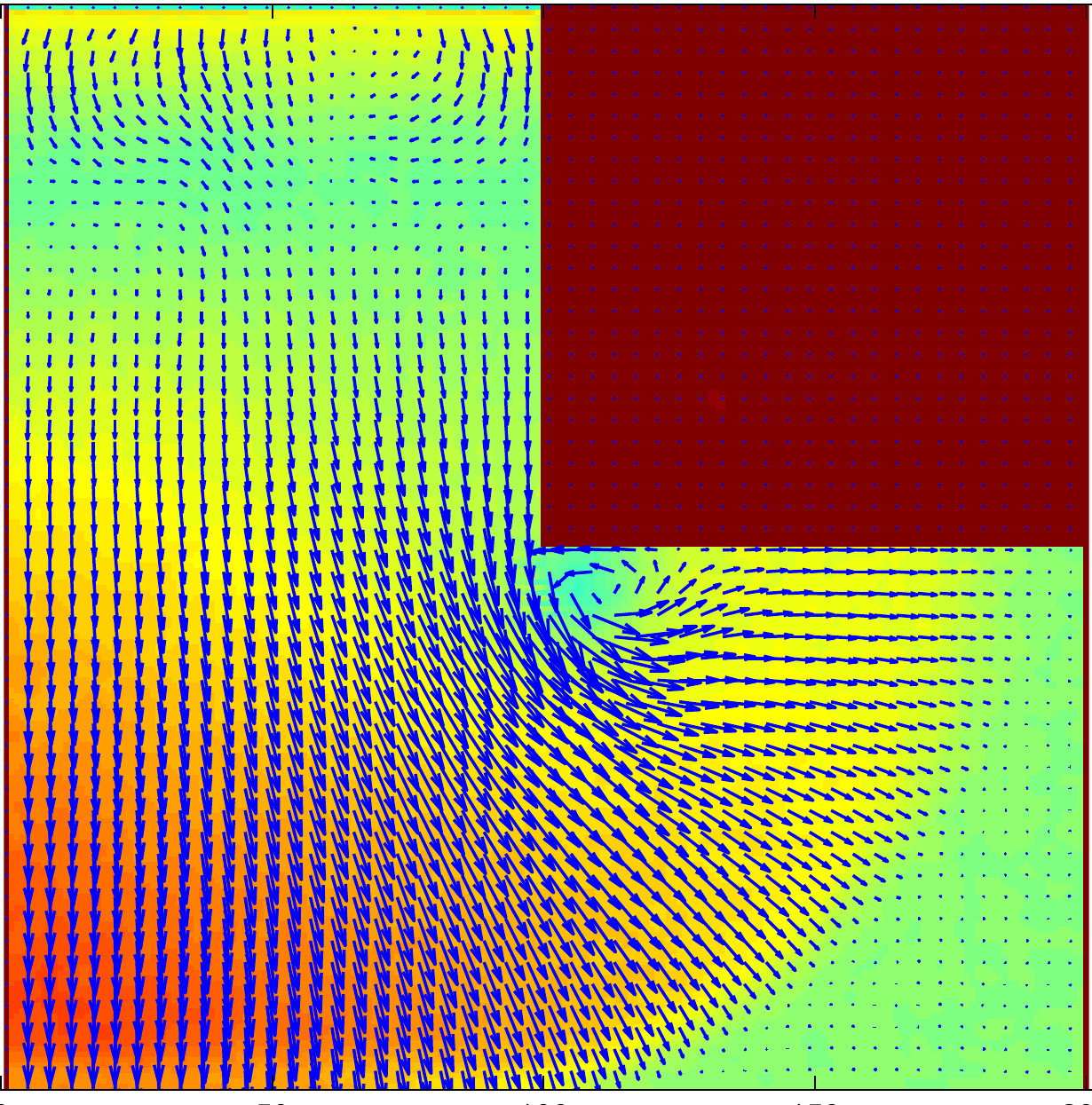}
\includegraphics[trim = 49mm 78.5mm 36mm 92mm, clip,angle=90, width=0.31\textwidth]{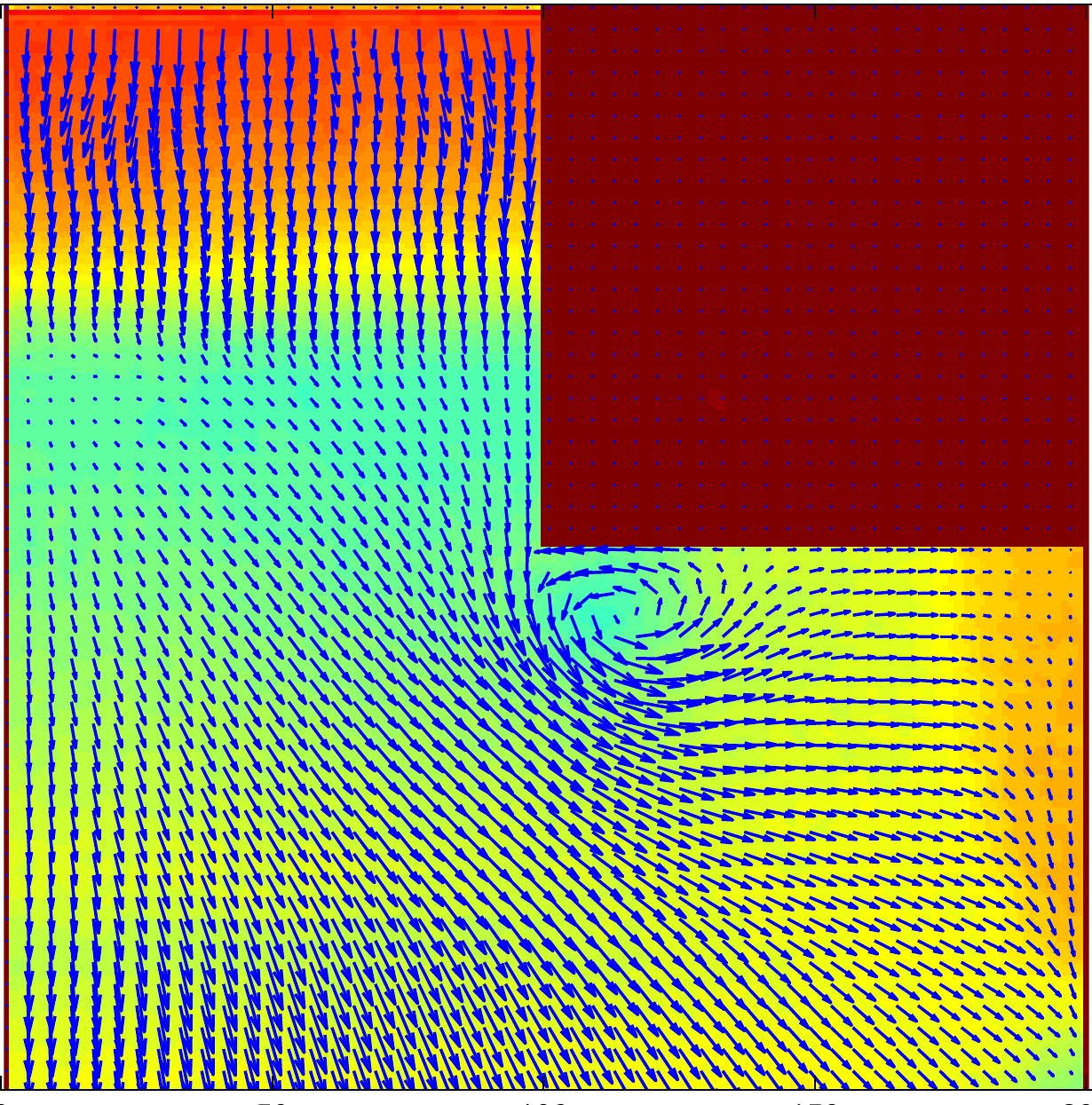}\\
\includegraphics[trim = 49mm 78.5mm 36mm 92mm, clip,angle=90, width=0.31\textwidth]{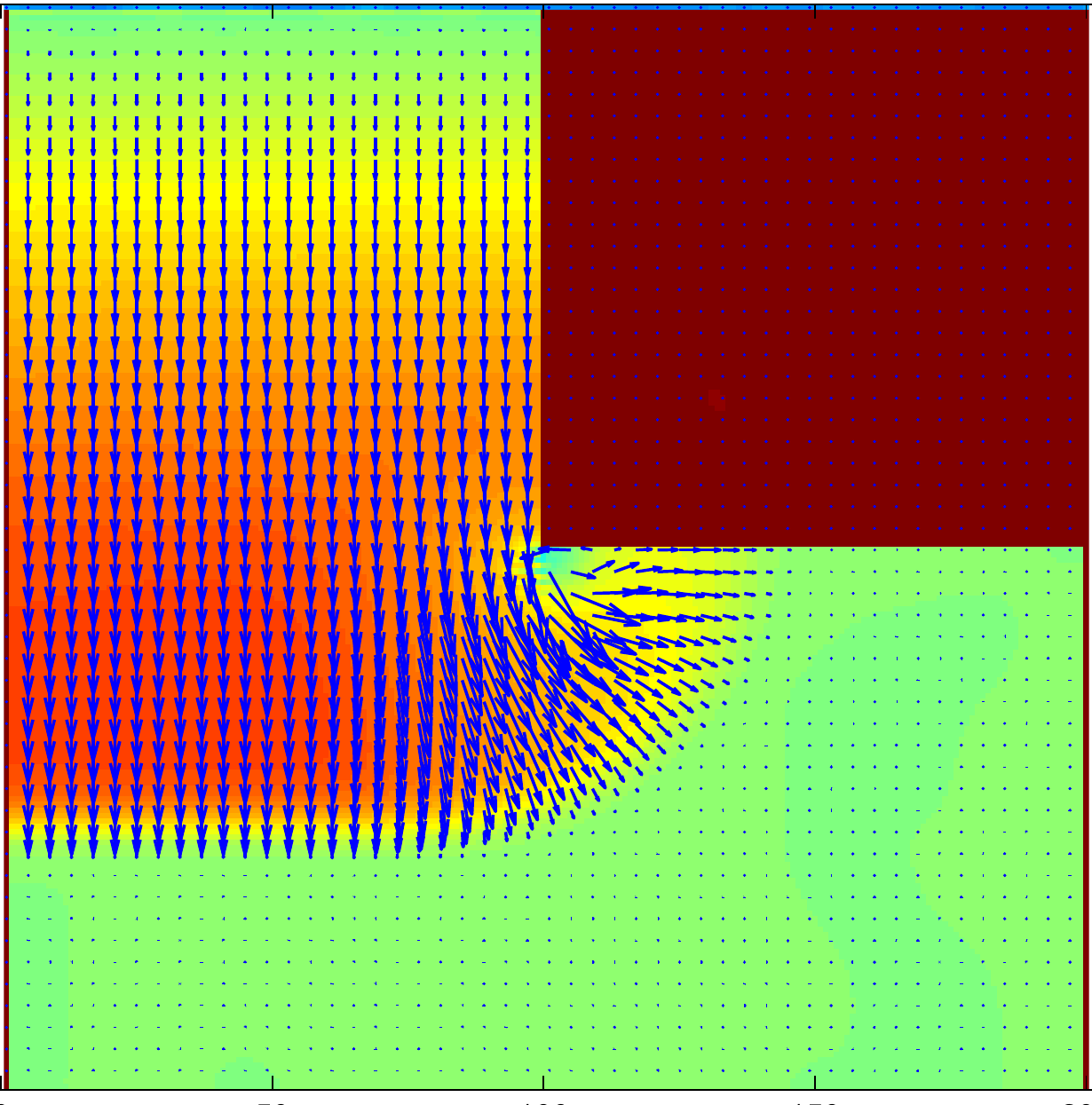}
\includegraphics[trim = 49mm 78.5mm 36mm 92mm, clip,angle=90, width=0.31\textwidth]{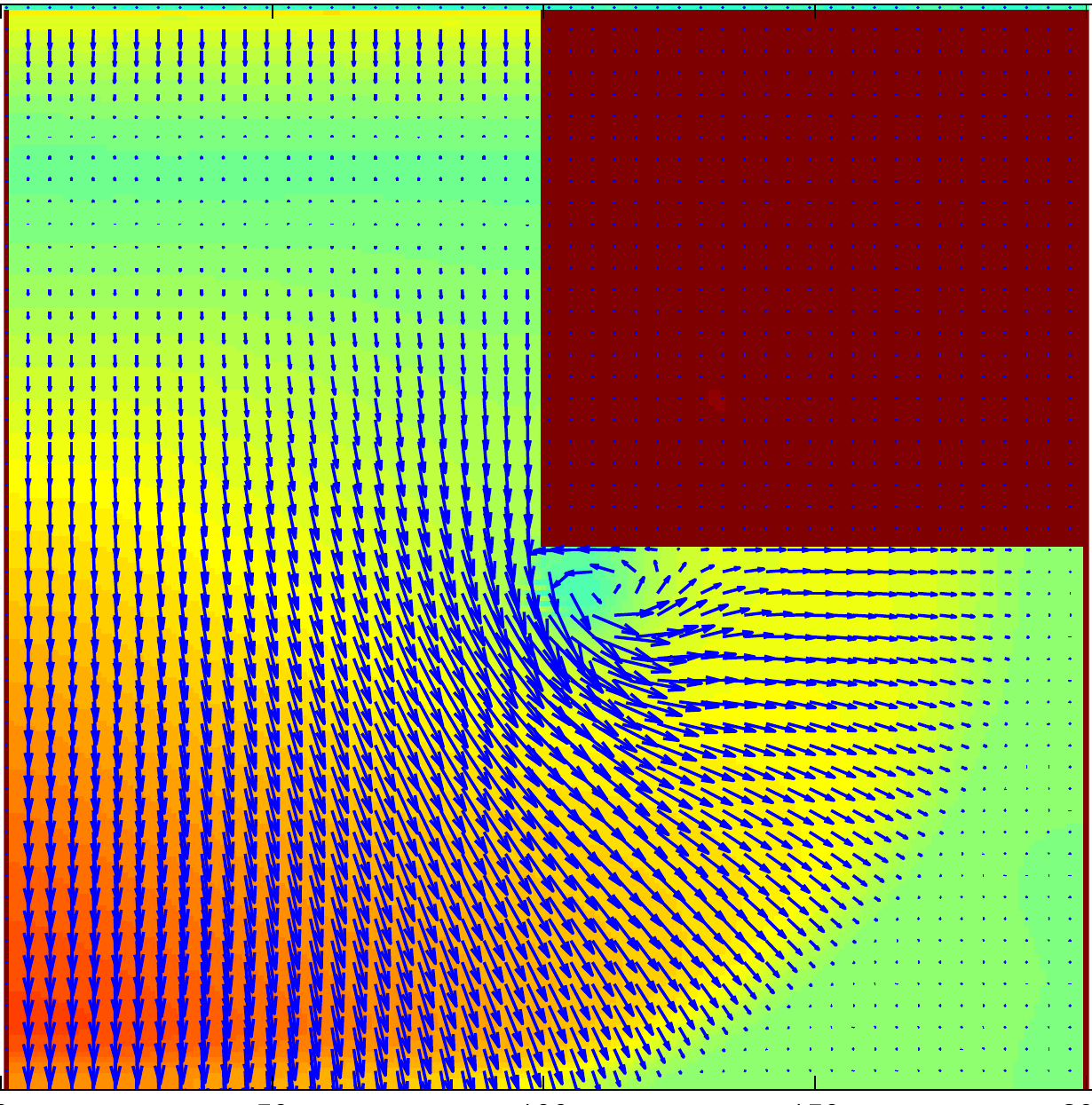}
\includegraphics[trim = 49mm 78.5mm 36mm 92mm, clip,angle=90, width=0.31\textwidth]{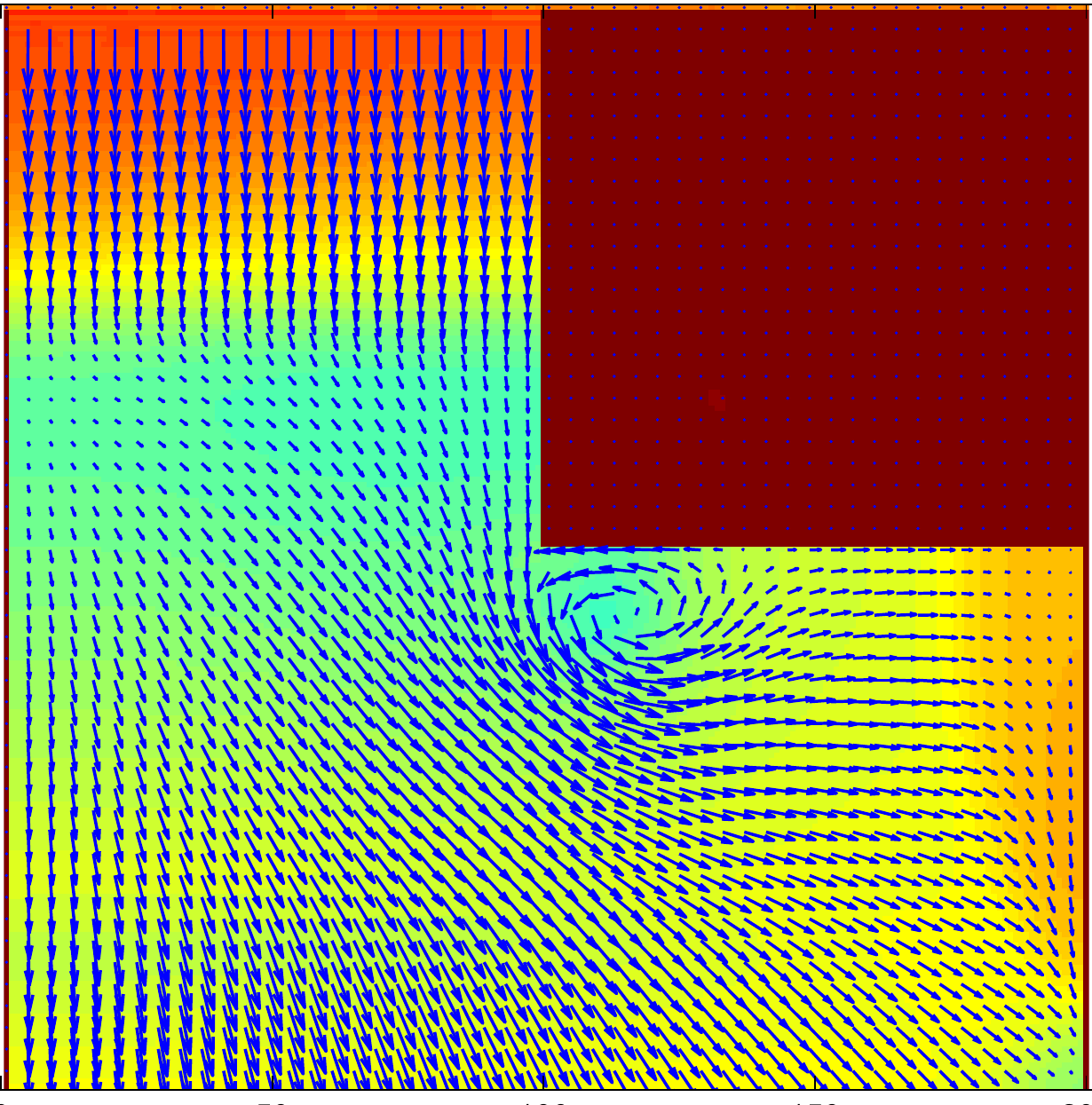}\\
{\small (b)}\\
\includegraphics[trim = 49mm 78.5mm 36mm 92mm, clip,angle=90, width=0.31\textwidth]{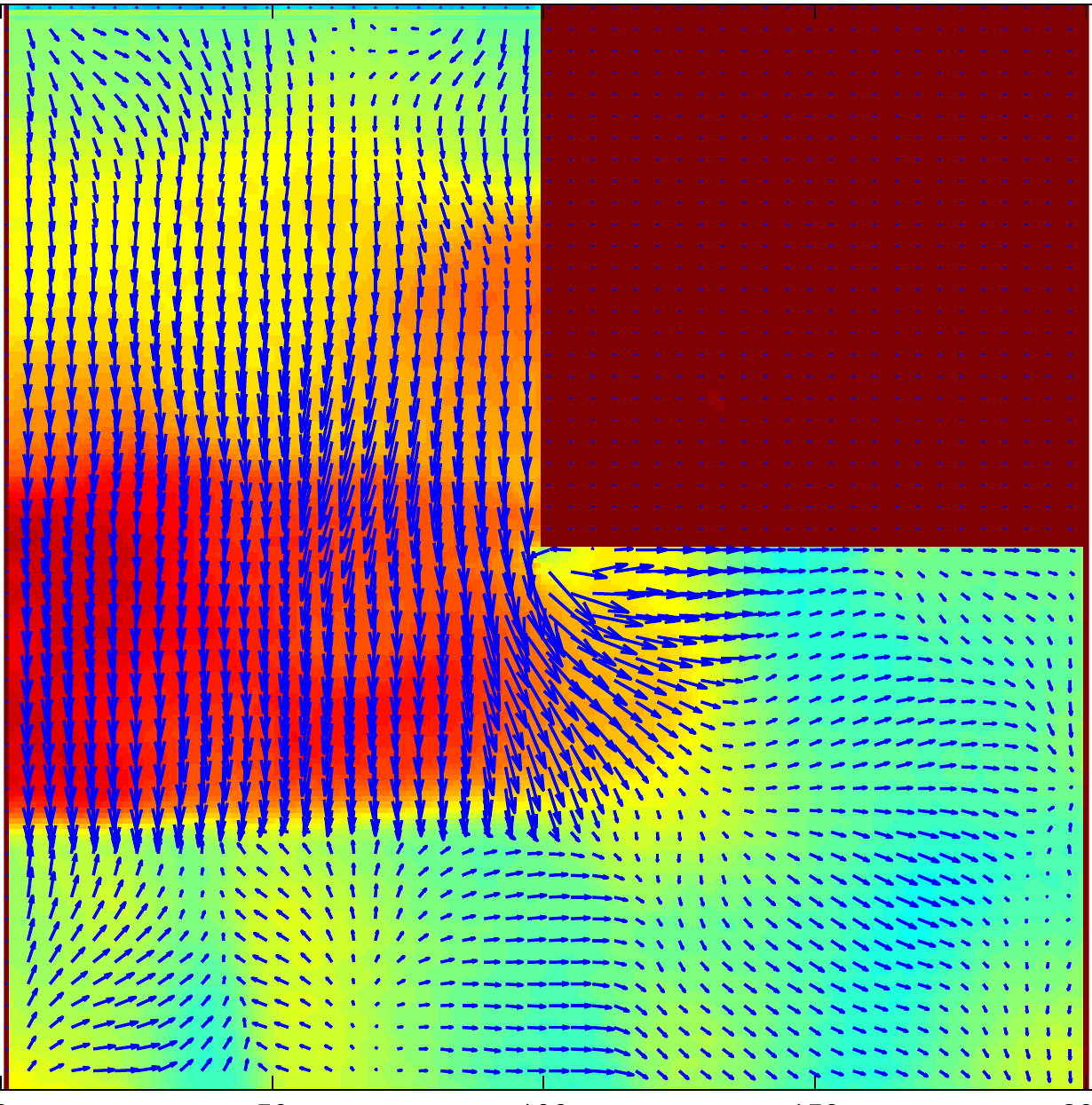}
\includegraphics[trim = 49mm 78.5mm 36mm 92mm, clip,angle=90, width=0.31\textwidth]{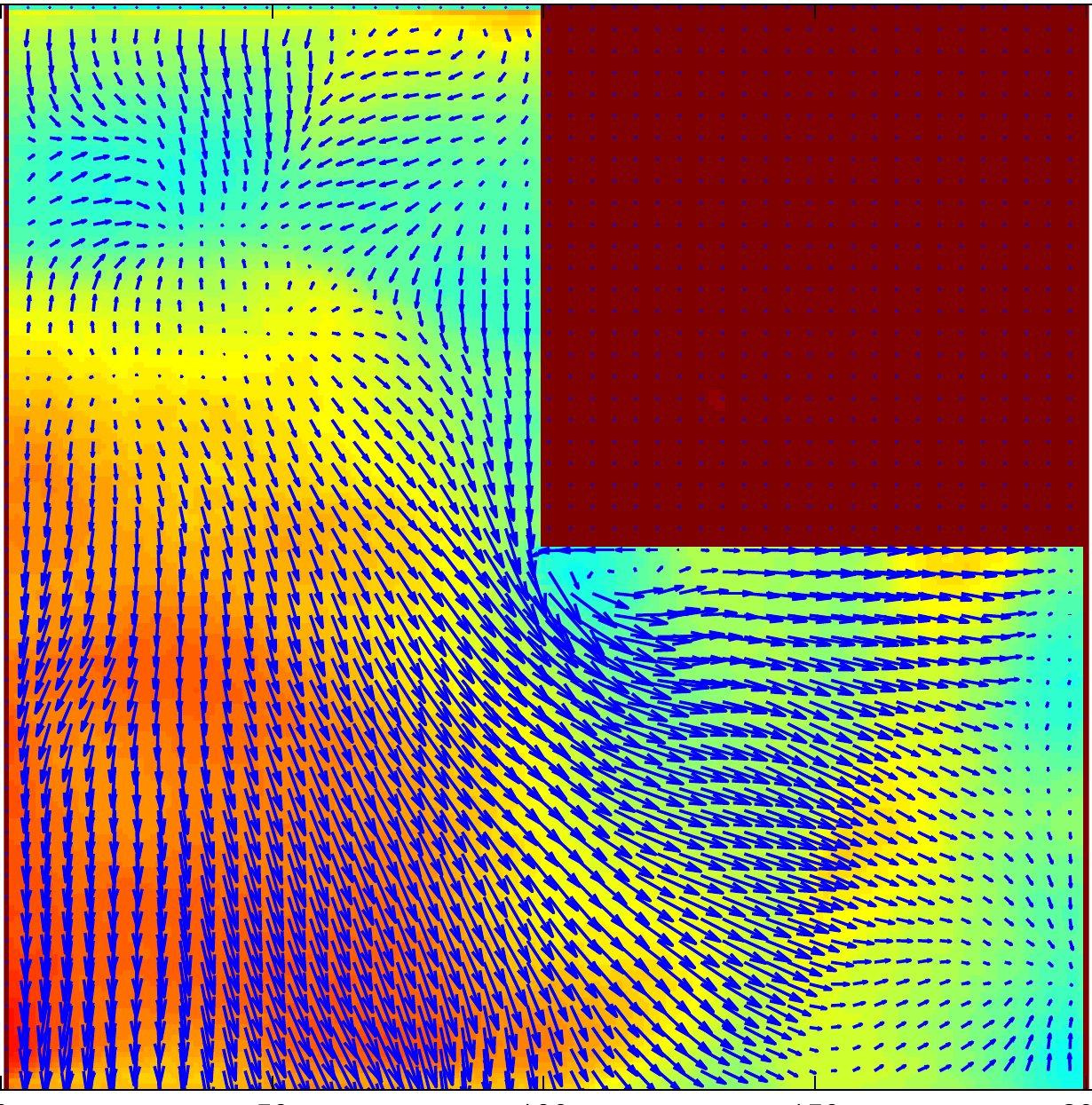}
\includegraphics[trim = 49mm 78.5mm 36mm 92mm, clip,angle=90, width=0.31\textwidth]{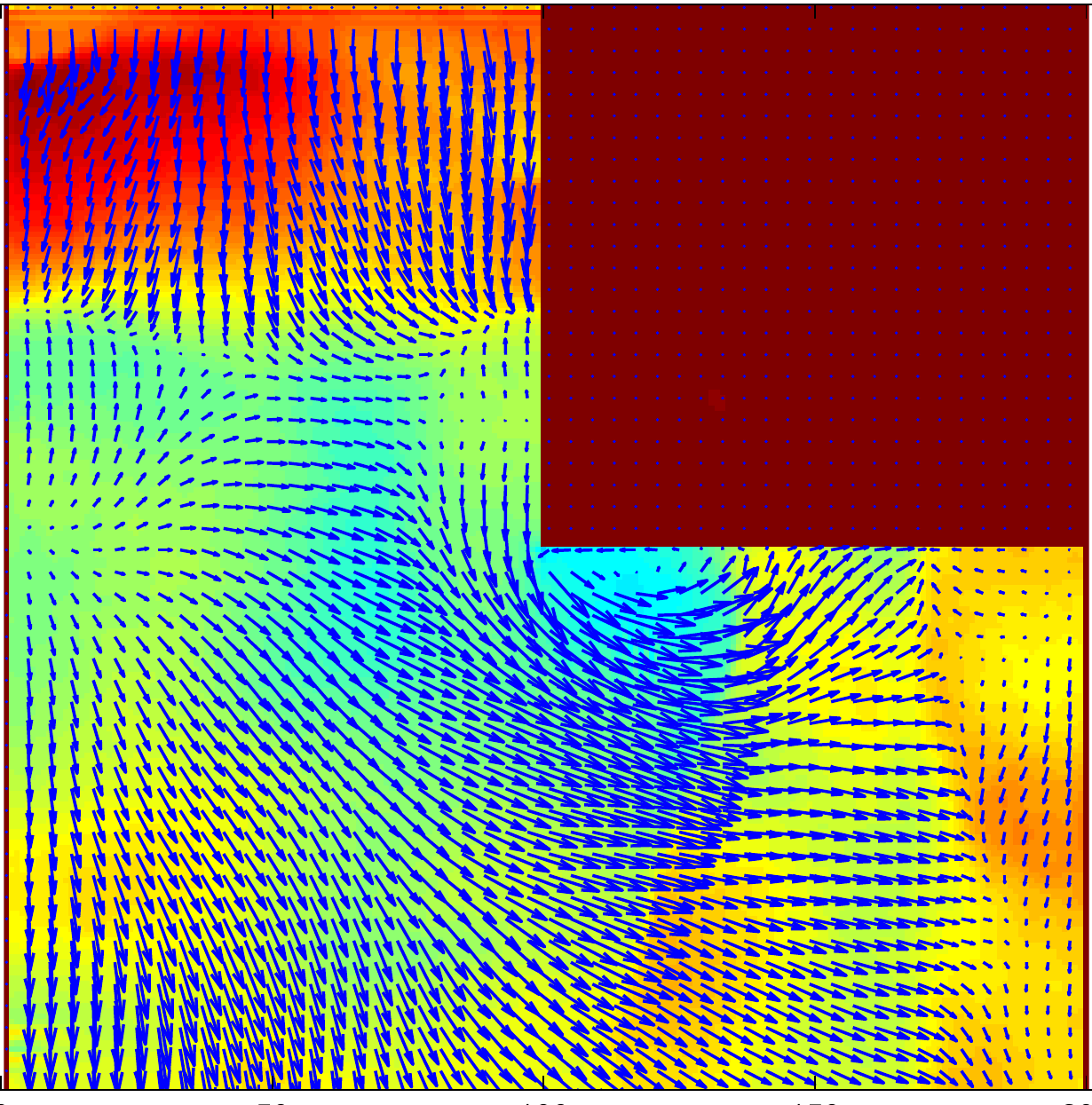}\\
\includegraphics[trim = 49mm 78.5mm 36mm 92mm, clip,angle=90, width=0.31\textwidth]{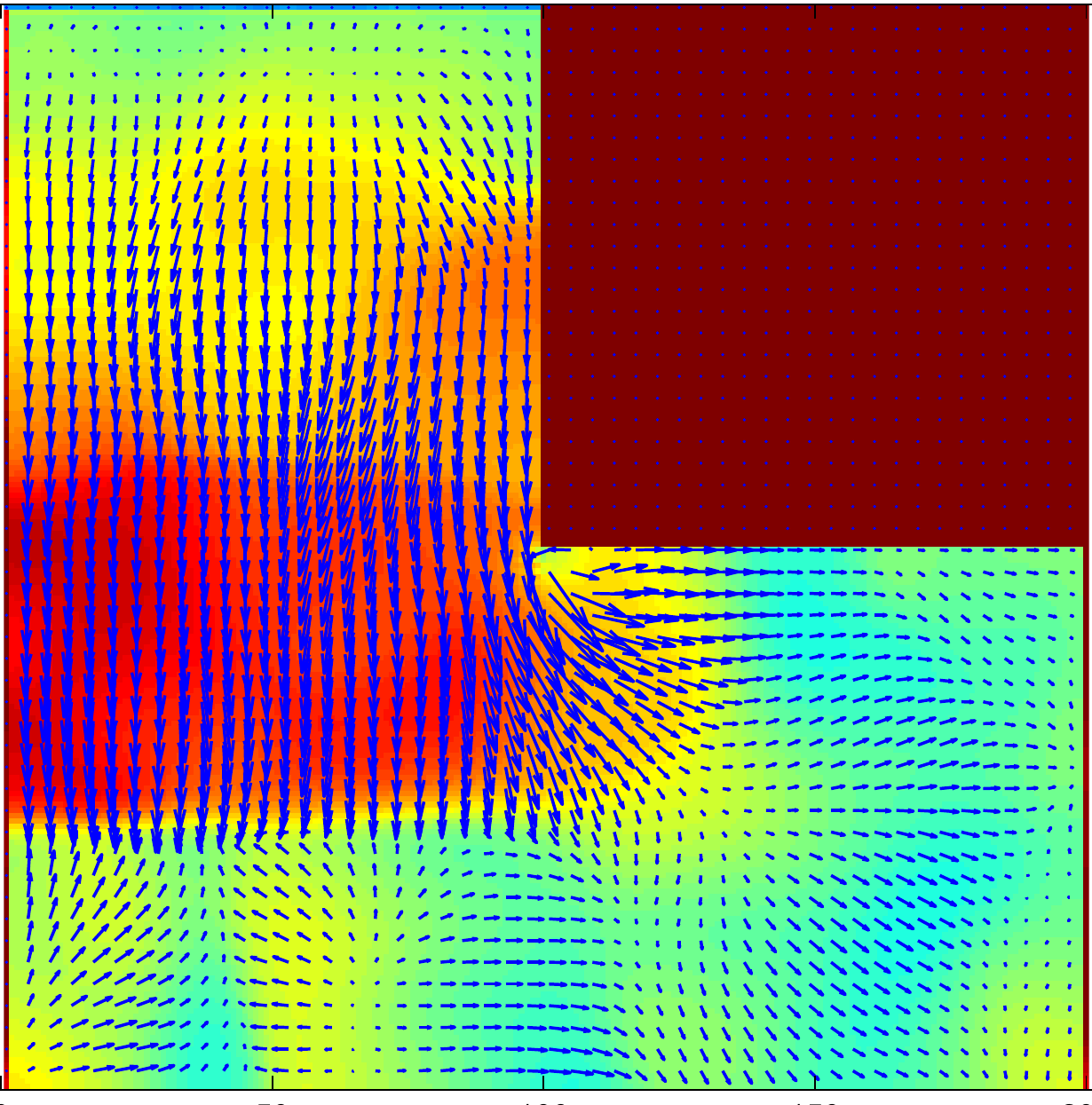}
\includegraphics[trim = 49mm 78.5mm 36mm 92mm, clip,angle=90, width=0.31\textwidth]{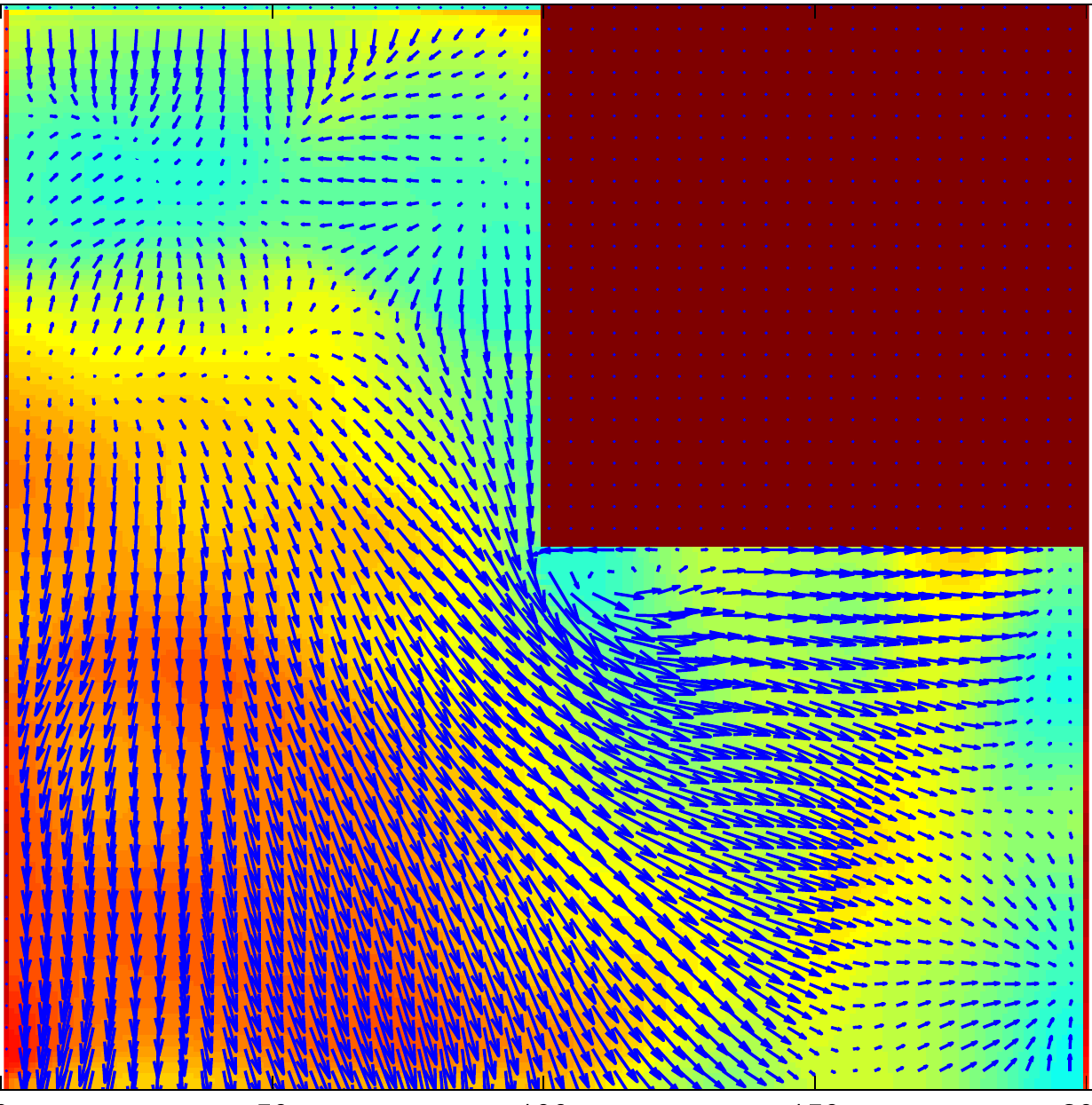}
\includegraphics[trim = 49mm 78.5mm 36mm 92mm, clip,angle=90, width=0.31\textwidth]{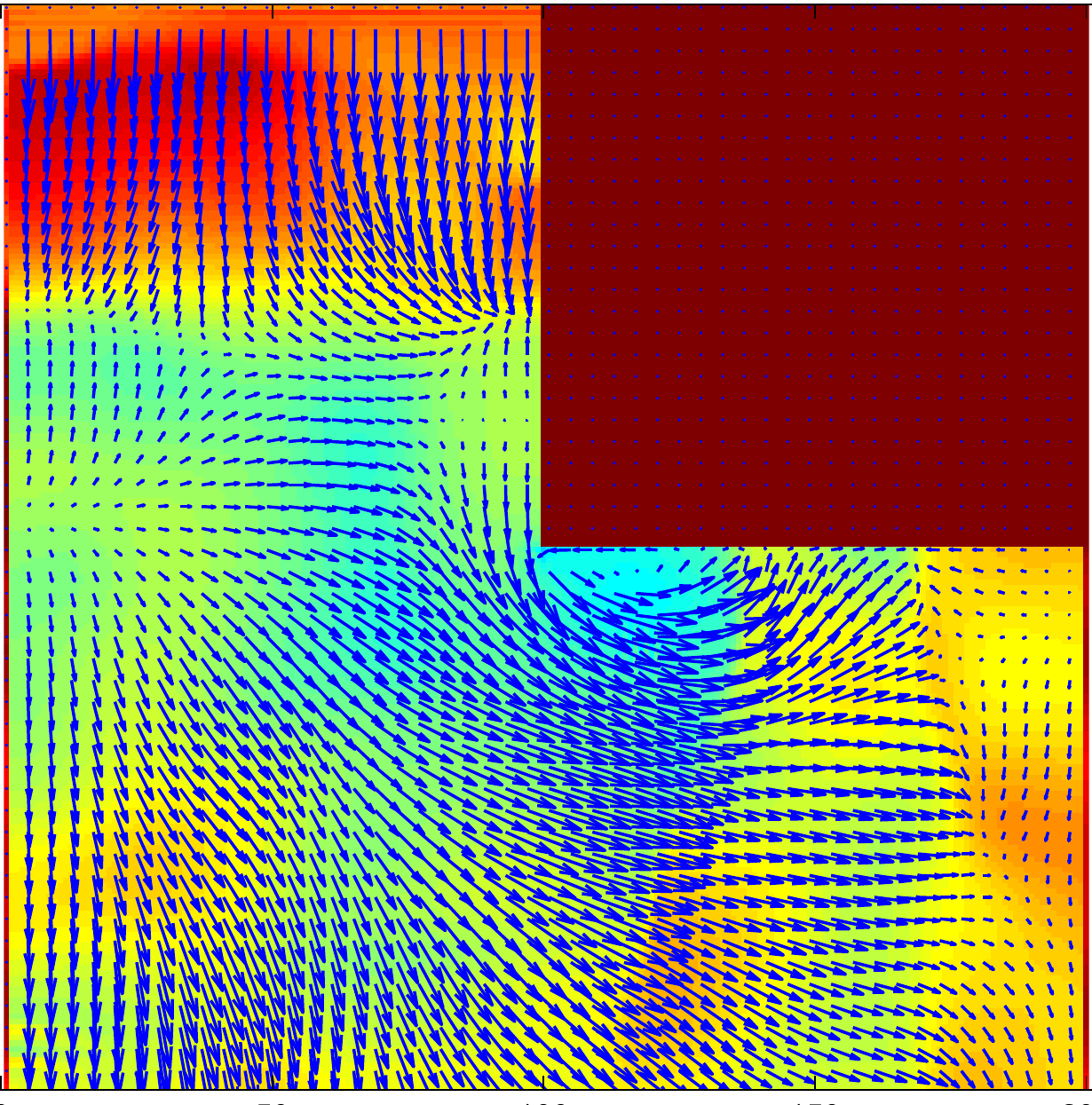}\\
\includegraphics[trim = 35mm 185mm 35mm 100mm, clip,width=0.4\textwidth]{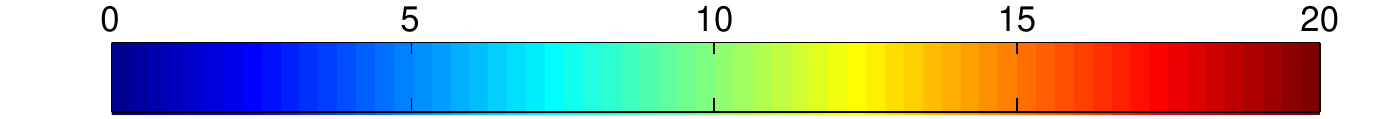}\\
\caption{Comparisons of elevation colormaps in millimeters and of free-surface motion vector fields, for the suddenly expanding flume flow with uniform inlet velocity profile: (a) without stochastic forcing; (b) with stochastic forcing; Top row, estimations using twoObs; Bottom row, true states; From left to right, time $t\,u_0/L=1.19$, $t\,u_0/L=1.57$ and $t\,u_0/L=1.98$.}
\label{fig:channel:theRegimeOne}
\end{figure}

\begin{figure}[h!]
\centering
{\small (a)}\\
\includegraphics[trim = 49mm 78.5mm 36mm 92mm, clip,angle=90, width=0.31\textwidth]{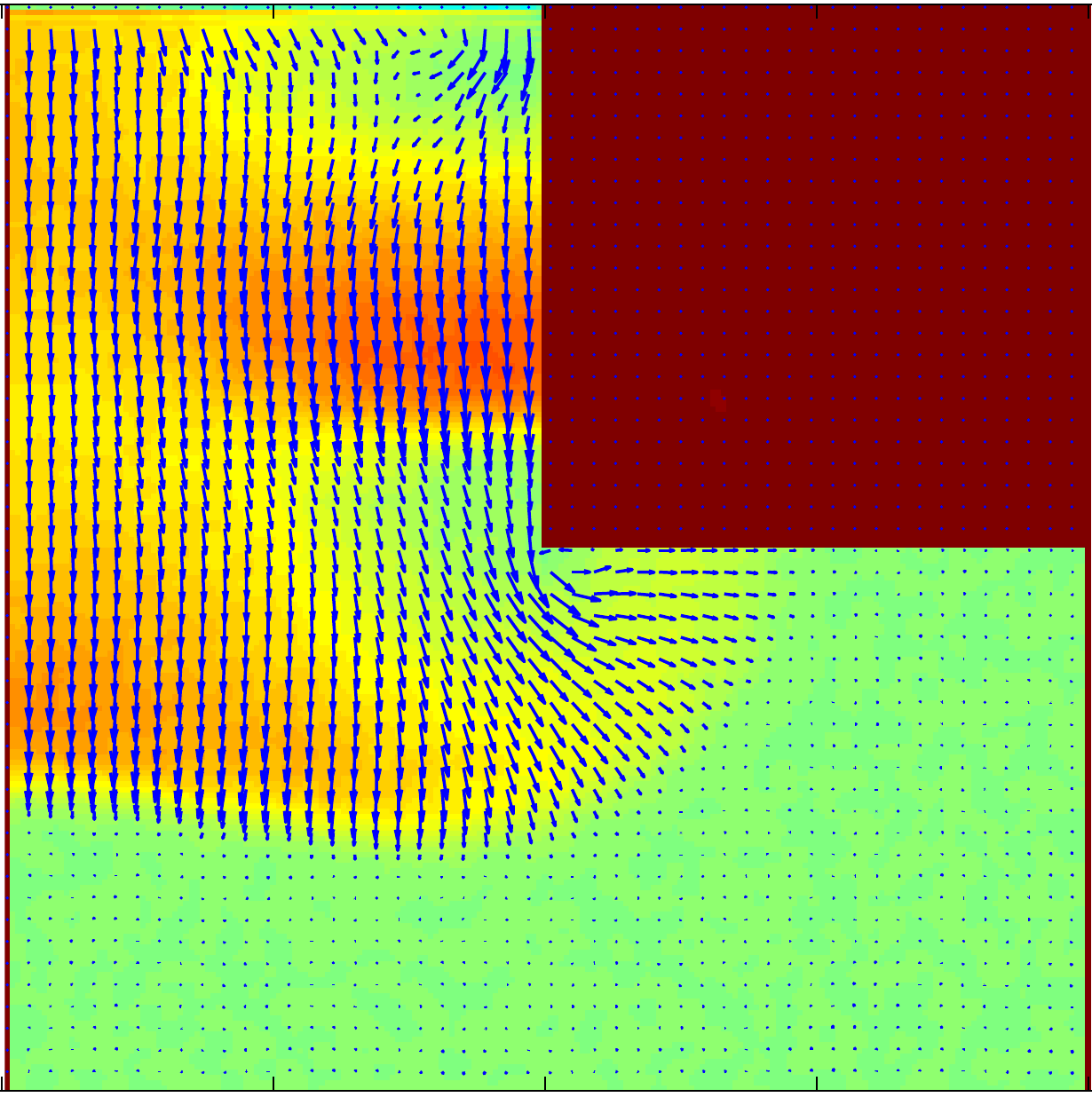}
\includegraphics[trim = 49mm 78.5mm 36mm 92mm, clip,angle=90, width=0.31\textwidth]{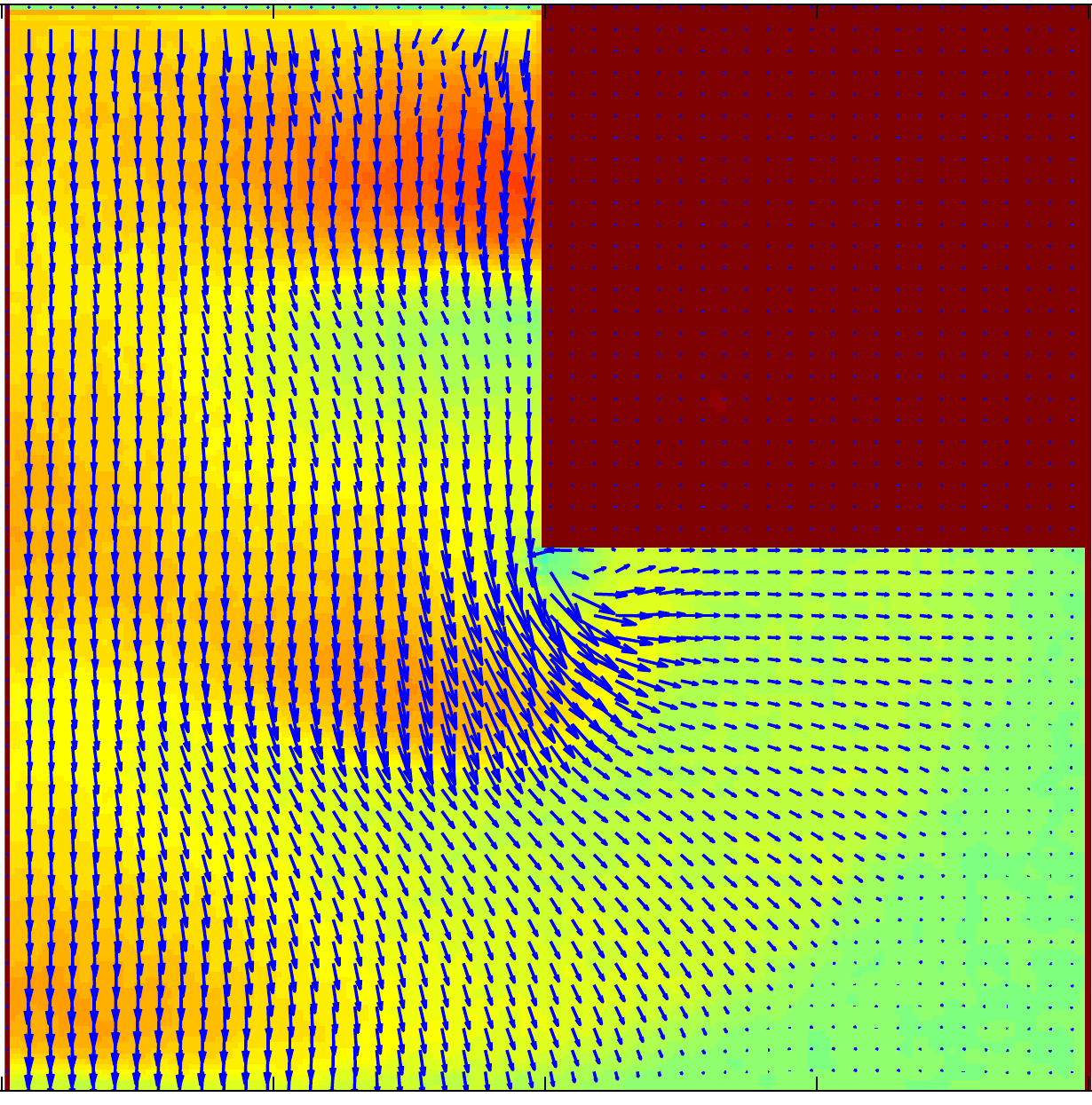}
\includegraphics[trim = 49mm 78.5mm 36mm 92mm, clip,angle=90, width=0.31\textwidth]{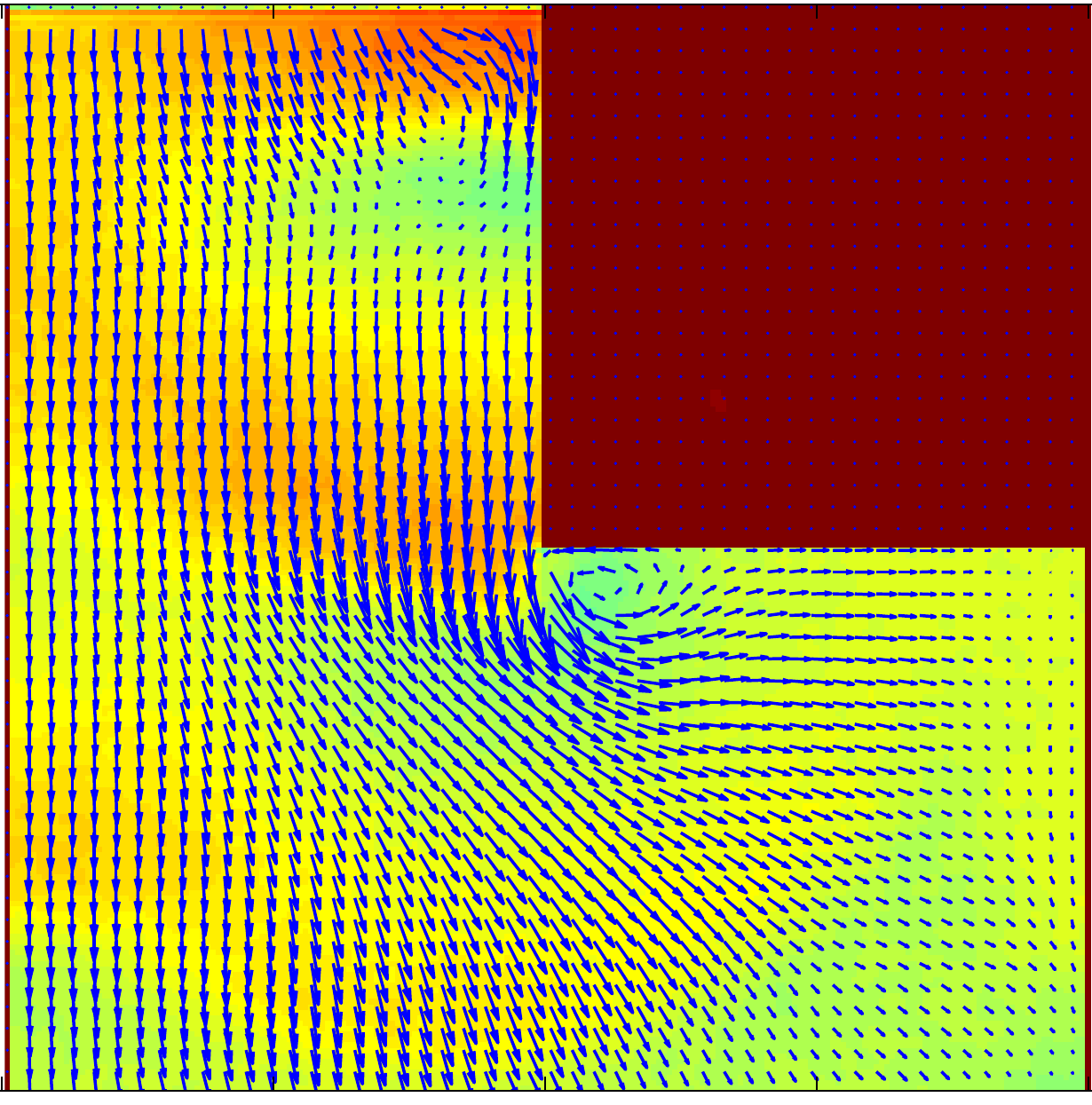}\\
\includegraphics[trim = 49mm 78.5mm 36mm 92mm, clip,angle=90, width=0.31\textwidth]{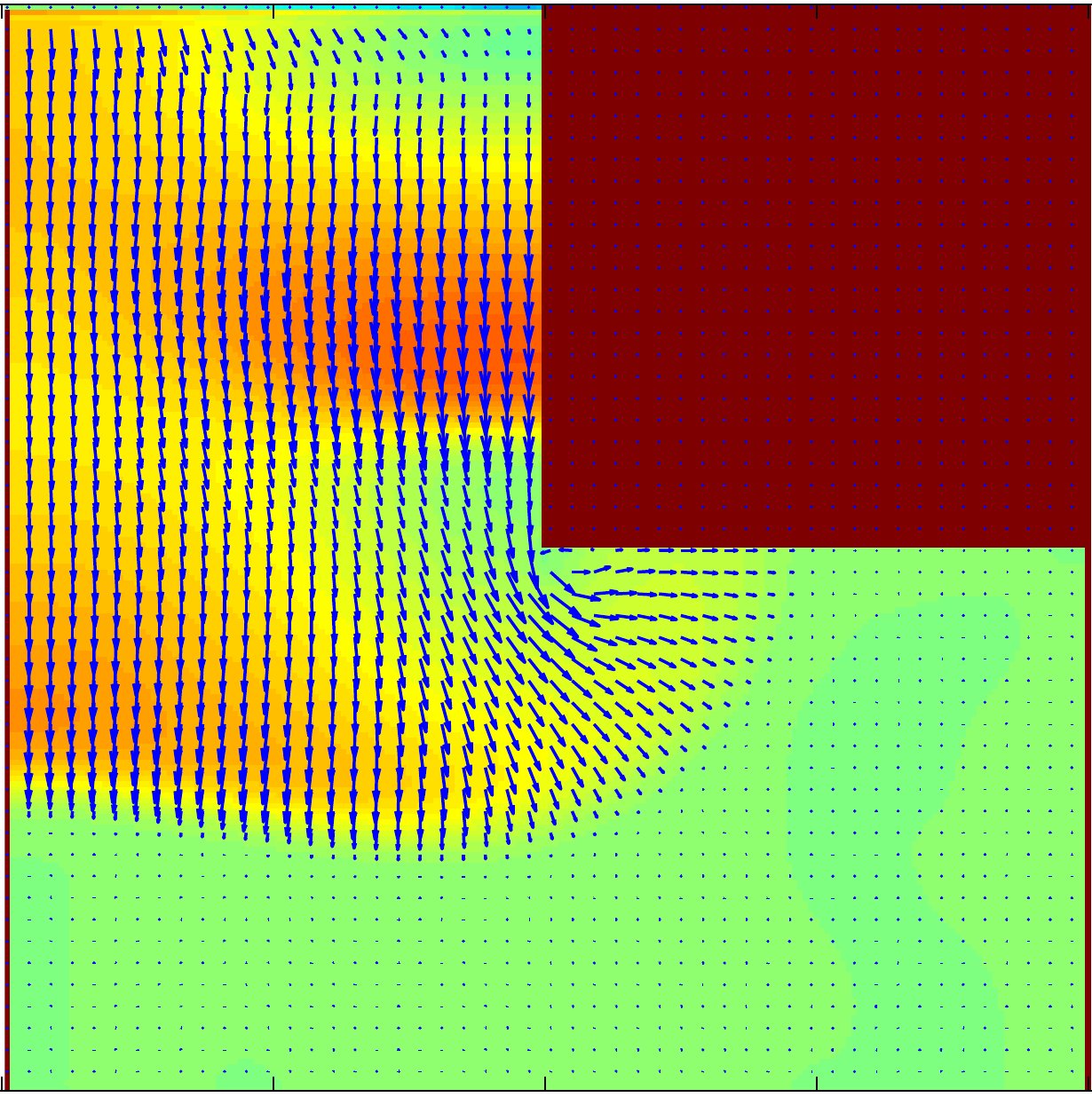}
\includegraphics[trim = 49mm 78.5mm 36mm 92mm, clip,angle=90, width=0.31\textwidth]{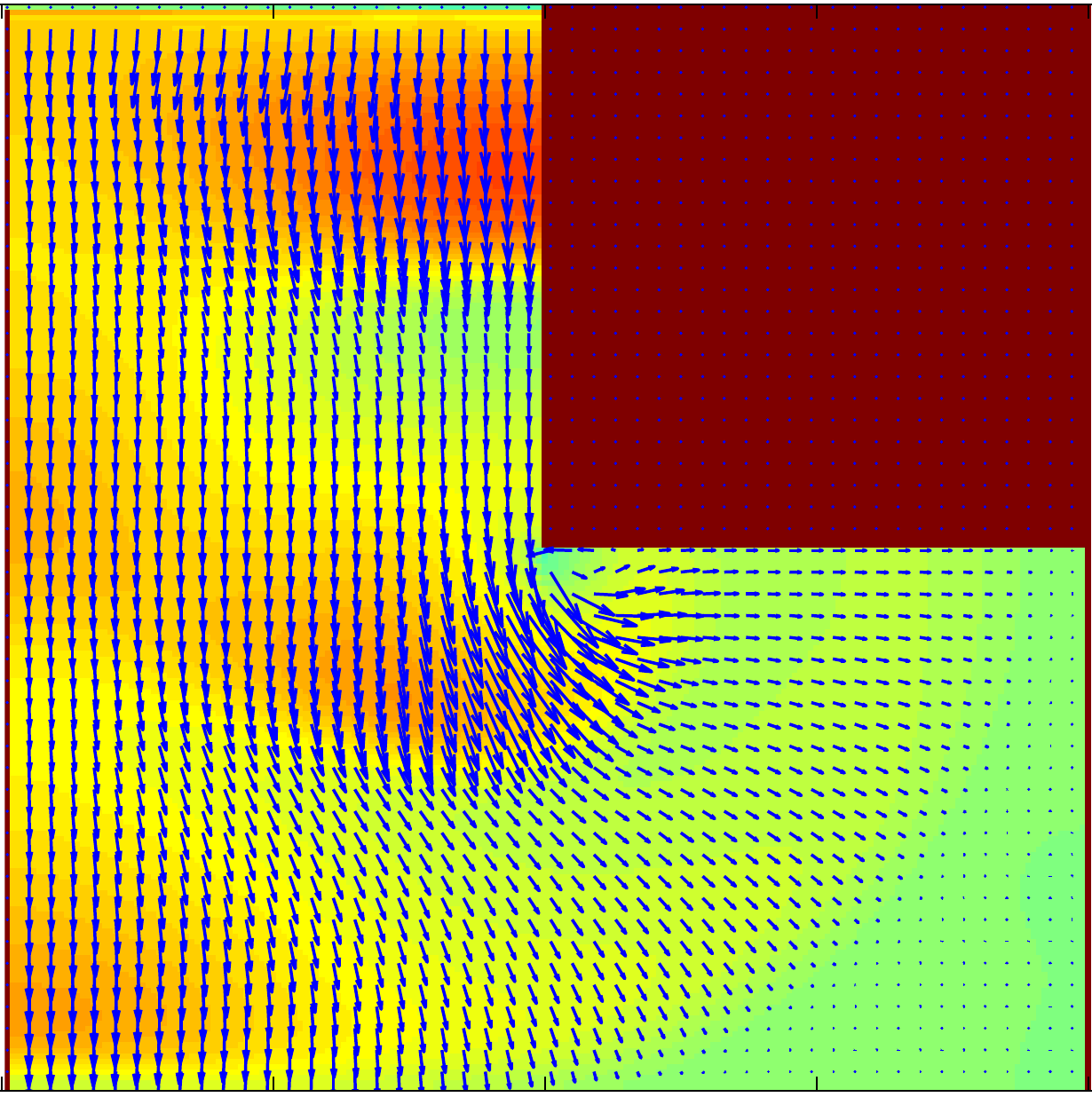}
\includegraphics[trim = 49mm 78.5mm 36mm 92mm, clip,angle=90, width=0.31\textwidth]{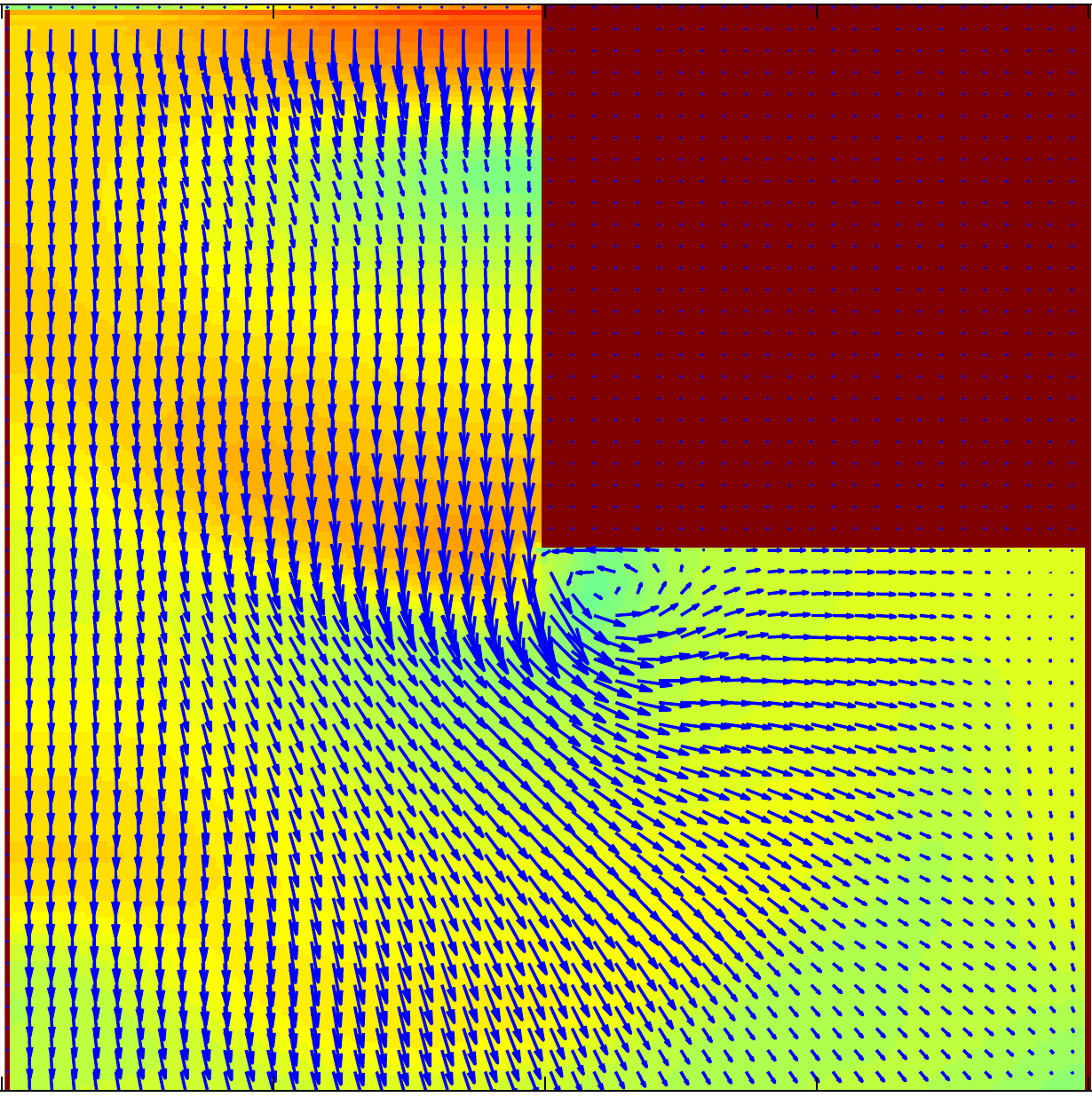}\\
{\small (b)}\\
\includegraphics[trim = 49mm 78.5mm 36mm 92mm, clip,angle=90, width=0.31\textwidth]{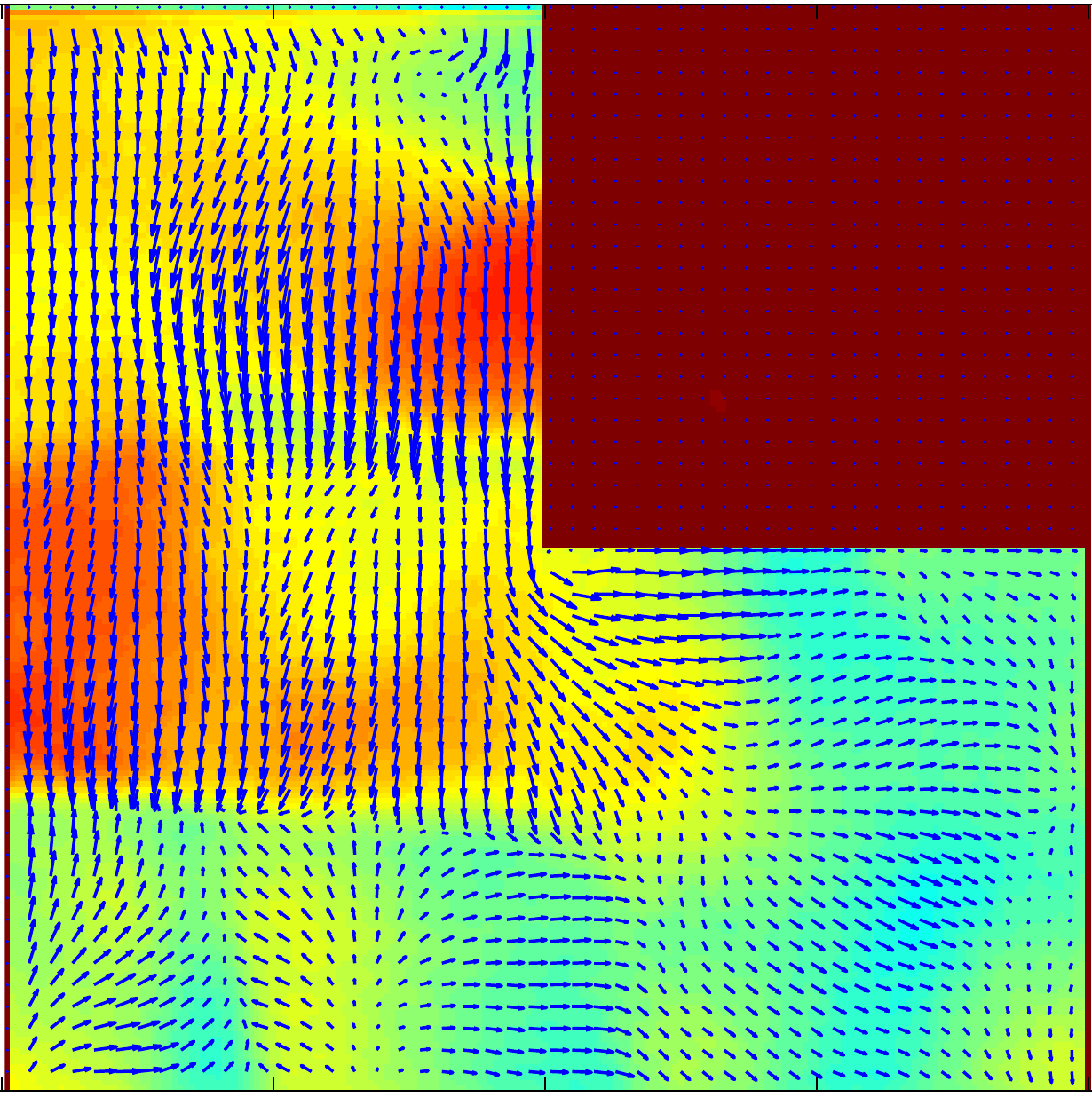}
\includegraphics[trim = 49mm 78.5mm 36mm 92mm, clip,angle=90, width=0.31\textwidth]{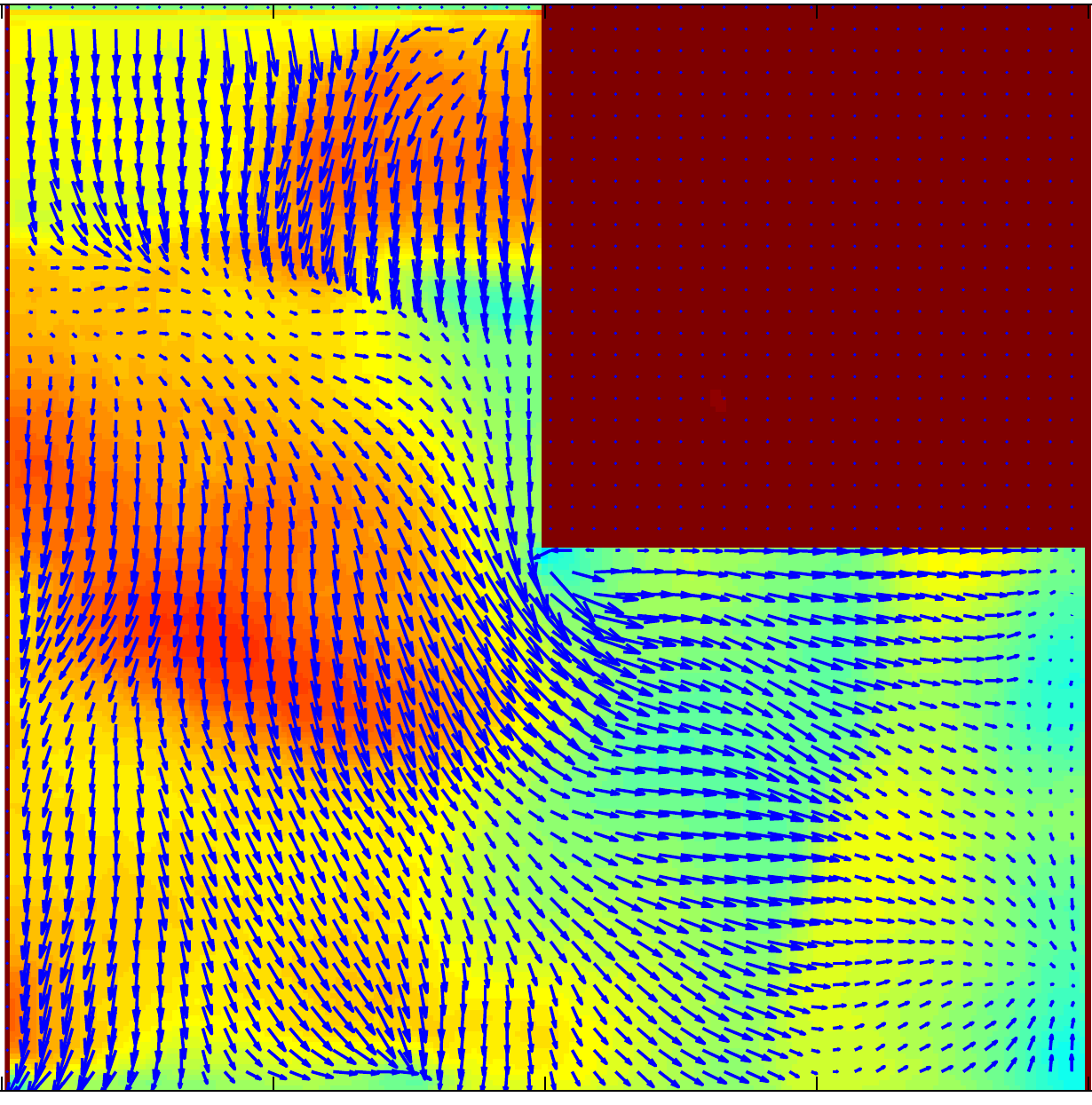}
\includegraphics[trim = 49mm 78.5mm 36mm 92mm, clip,angle=90, width=0.31\textwidth]{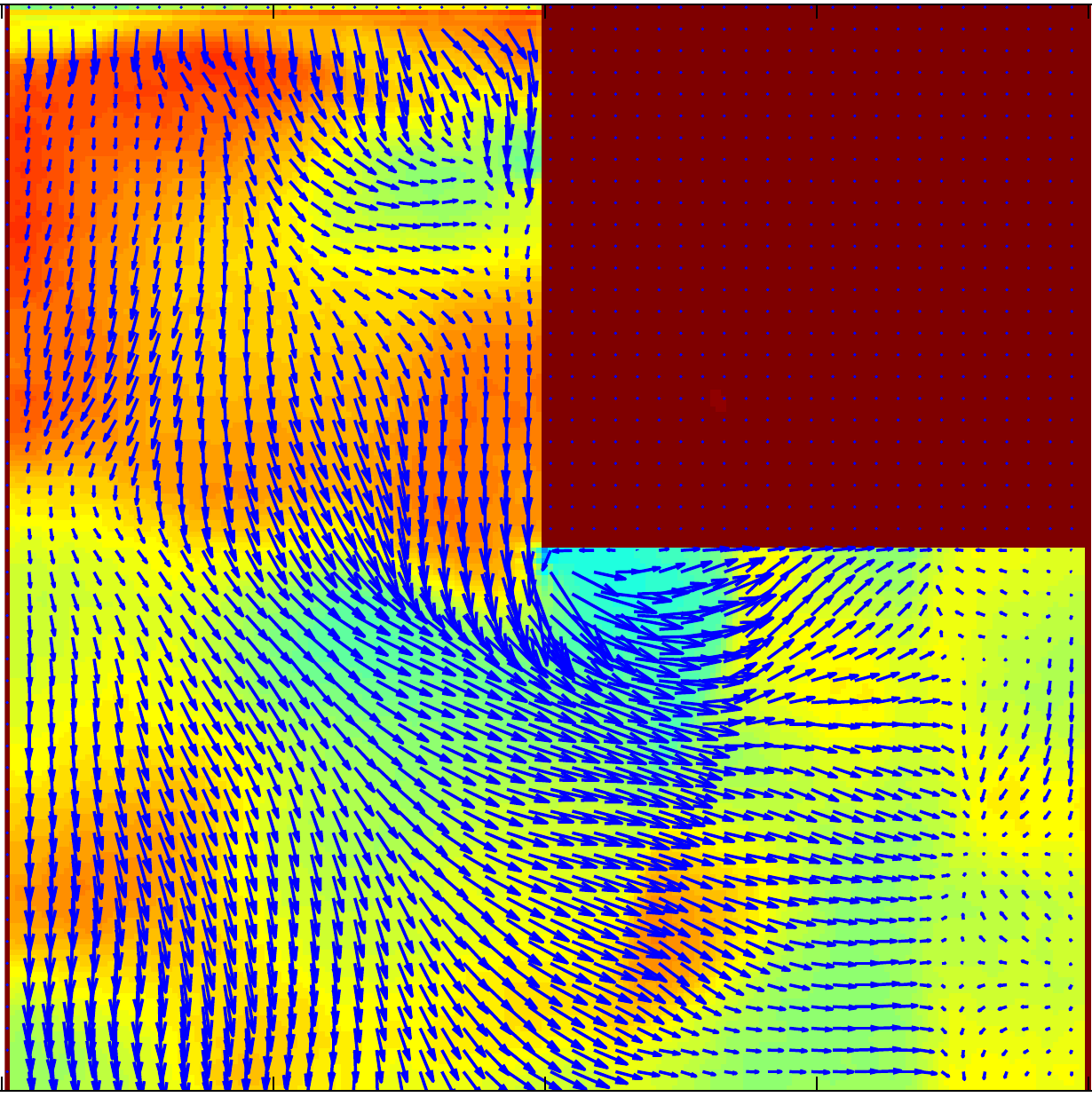}\\
\includegraphics[trim = 49mm 78.5mm 36mm 92mm, clip,angle=90, width=0.31\textwidth]{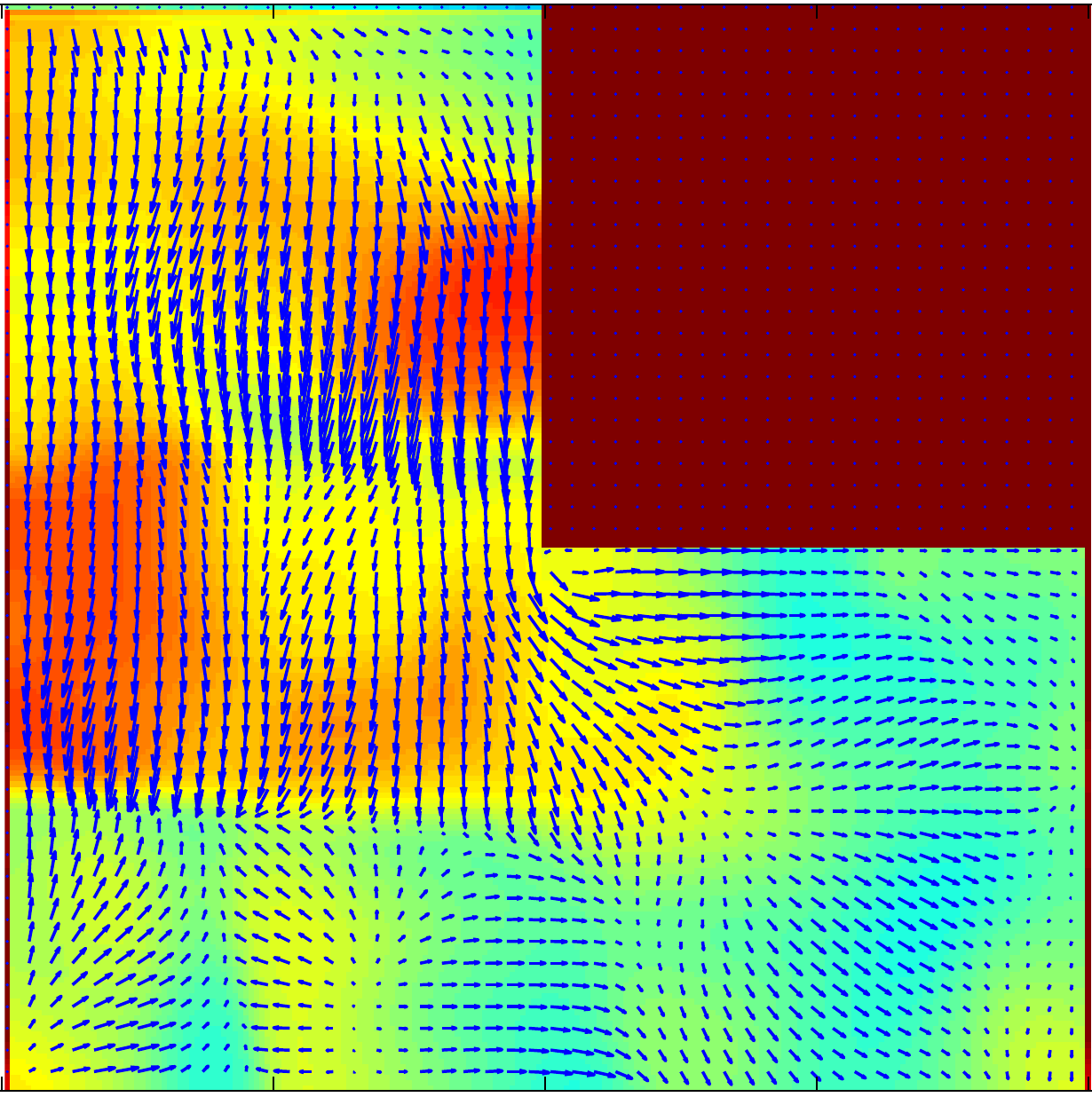}
\includegraphics[trim = 49mm 78.5mm 36mm 92mm, clip,angle=90, width=0.31\textwidth]{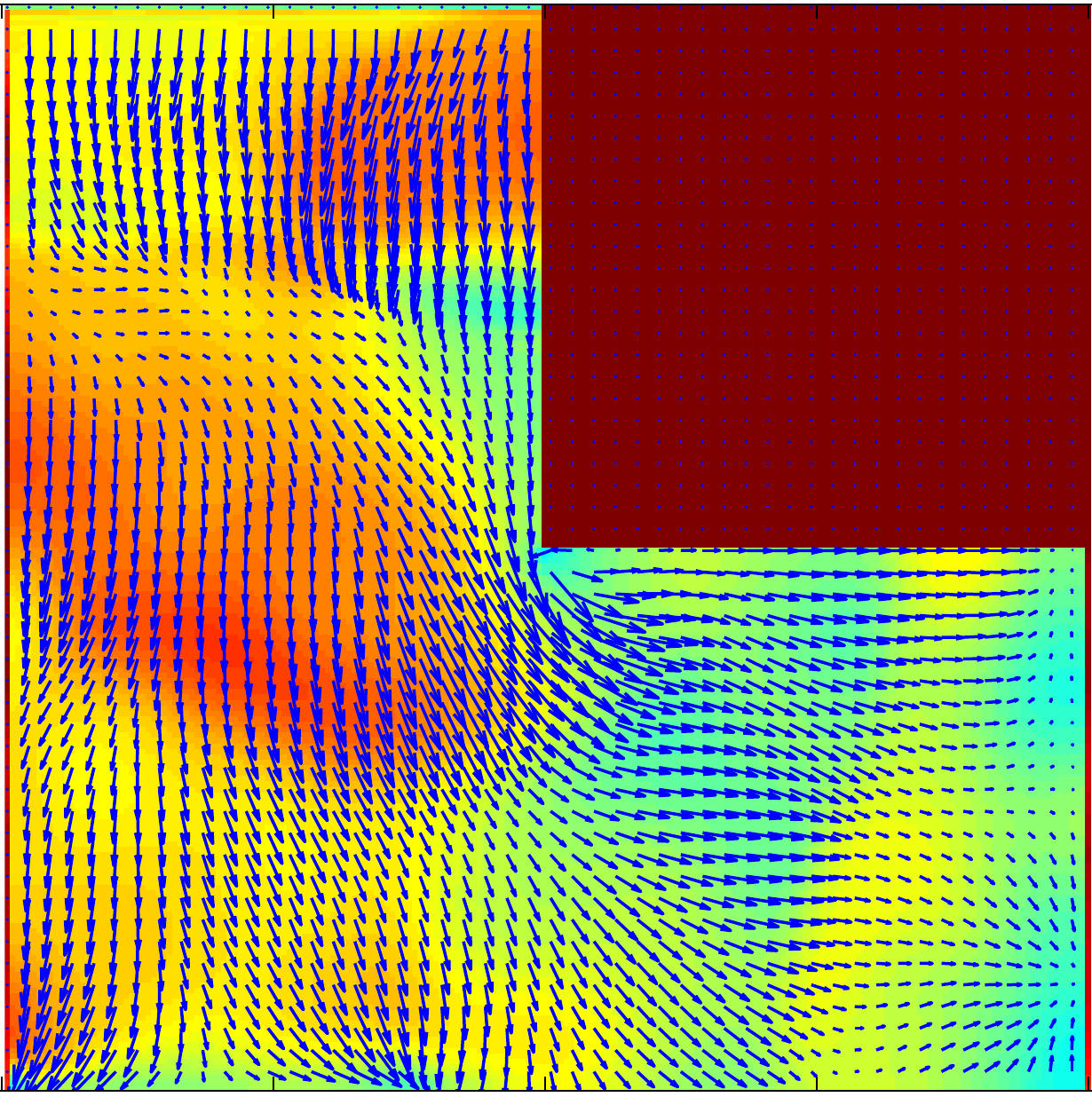}
\includegraphics[trim = 49mm 78.5mm 36mm 92mm, clip,angle=90, width=0.31\textwidth]{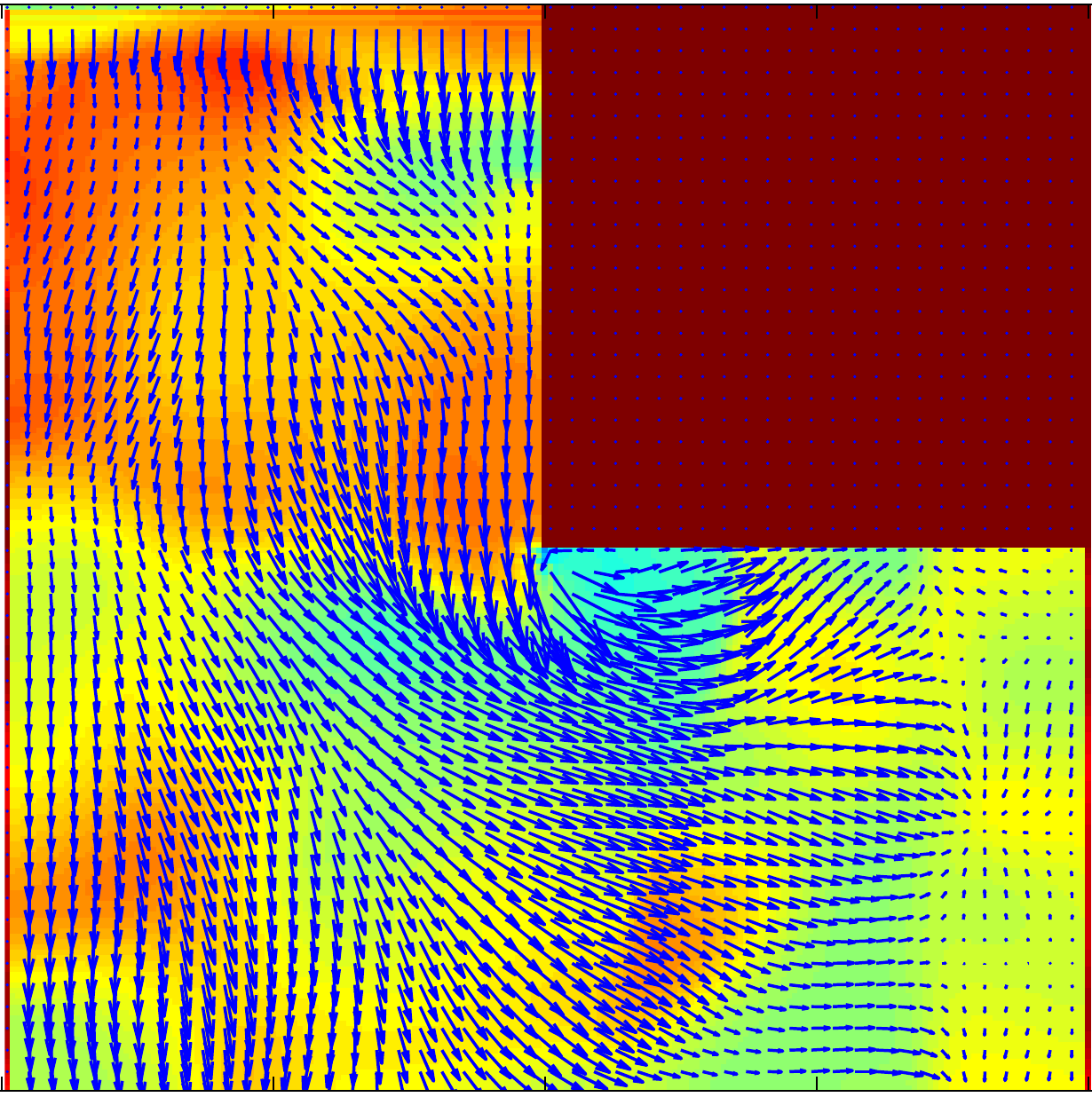}\\
\includegraphics[trim = 35mm 185mm 35mm 100mm, clip,width=0.4\textwidth]{barChannelState.pdf}
\caption{Comparisons of elevation colormaps in millimeters and of free-surface motion vector fields, for the suddenly expanding flume flow with half-bell shape inlet velocity profile: (a) without stochastic forcing; (b) with stochastic forcing; Top row, estimations using twoObs; Bottom row, true states; From left to right, time $t\,u_0/L=1.19$, $t\,u_0/L=1.57$ and $t\,u_0/L=1.98$.}
\label{fig:channel:theRegimeTwo}
\end{figure}

\Table{\label{tab:flume_errors} Ratios between assimilation and free-run errors for the suddenly expanding flume flow at time $t\,u_0/L=1.98$.}
\br
& &\centre{2}{oneObs}&\centre{2}{twoObs}\\
\ns
Inlet velocity& Stochastic &\crule{4}\\
profile &  forcing$^{\rm a}$ & $\mathcal{E}_{\hat h}/\mathcal{E}_{h}$ & $\mathcal{E}_{\hat{\bar{u}},\hat{\bar{v}}}/\mathcal{E}_{\bar{u},\bar{v}}$ & $\mathcal{E}_{\hat h}/\mathcal{E}_{h}$ & $\mathcal{E}_{\hat{\bar{u}},\hat{\bar{v}}}/\mathcal{E}_{\bar{u},\bar{v}}$\\
\mr
Homogeneous & No & 0.05 & 0.08 & 0.04 & 0.06\\
Homogeneous            & Yes & 0.19 & 0.21 & 0.16 & 0.19\\
Hall-bell shape            & No & 0.17 & 0.25 & 0.15 & 0.26\\
Hall-bell shape& Yes & 0.30 & 0.57 & 0.20 & 0.53\\
\br
\end{tabular}
\item[] $^{\rm a}$ included in the true states $\tilde \x_{0 \ldots t}$ and therefore in the observations $\y_{1 \ldots t}$.
\end{indented}
\label{tad:scoreOpenChannel}
\end{table}

\begin{figure}[h!]
\centering
{\small (a)}\\
\includegraphics[trim = 49mm 78.5mm 36mm 92mm, clip,angle=90, width=0.31\textwidth]{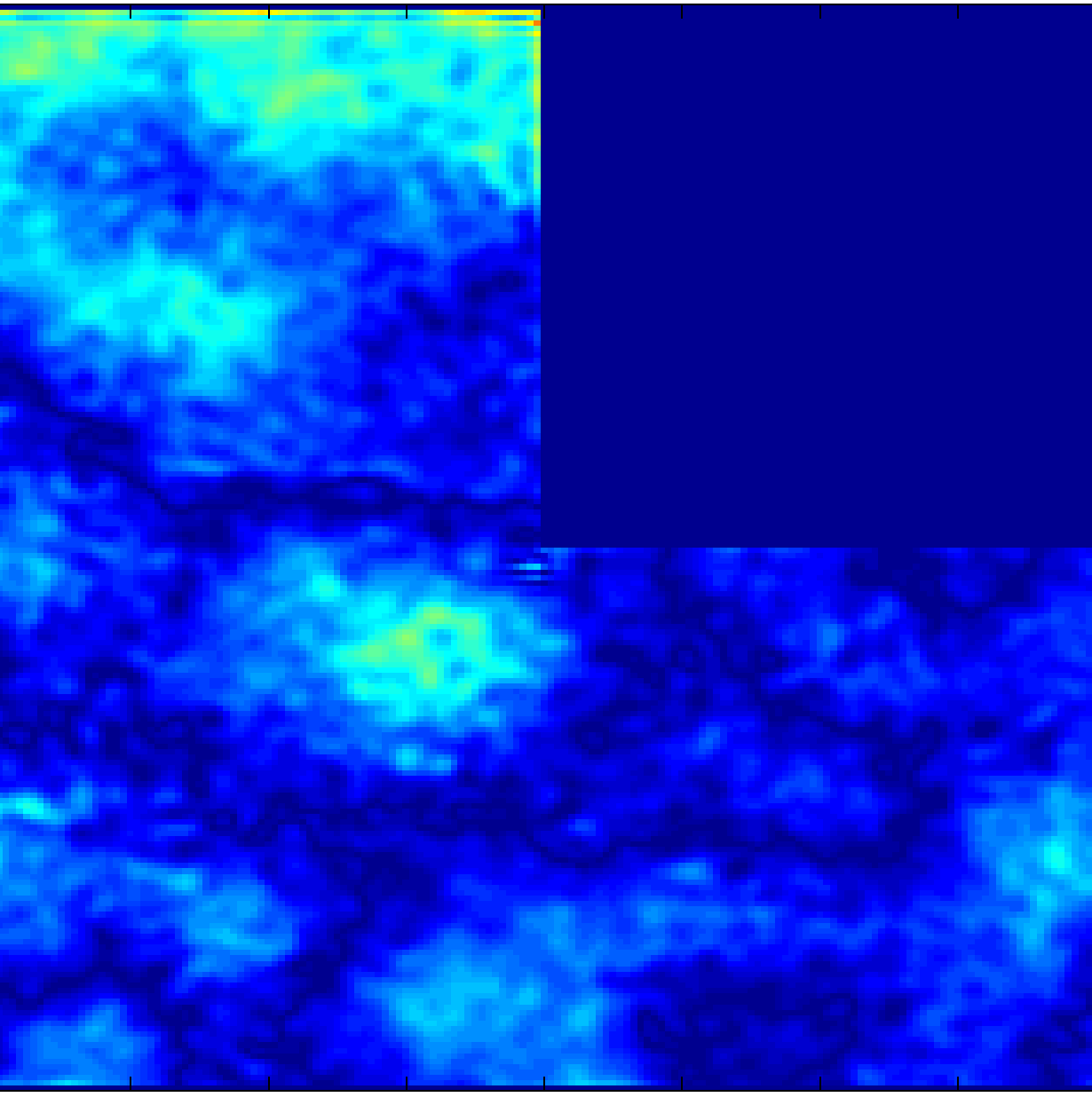}
\includegraphics[trim = 49mm 78.5mm 36mm 92mm, clip,angle=90, width=0.31\textwidth]{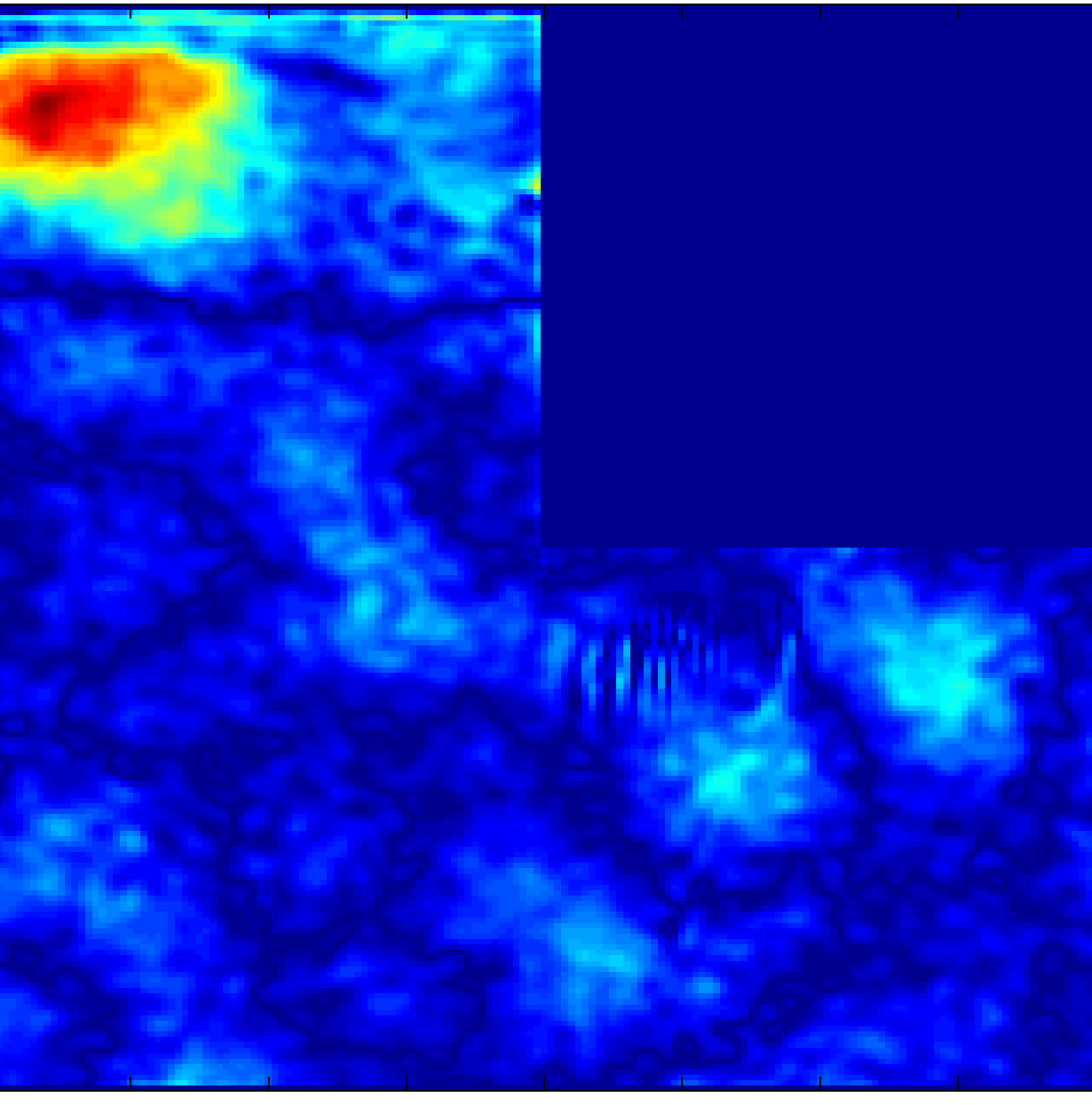}
\includegraphics[trim = 43mm 185mm 35mm 100mm, clip, angle=90,width=0.0285\textwidth]{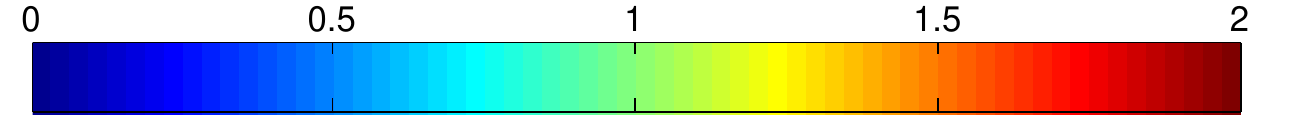}\\
\includegraphics[trim = 49mm 78.5mm 36mm 92mm, clip,angle=90, width=0.31\textwidth]{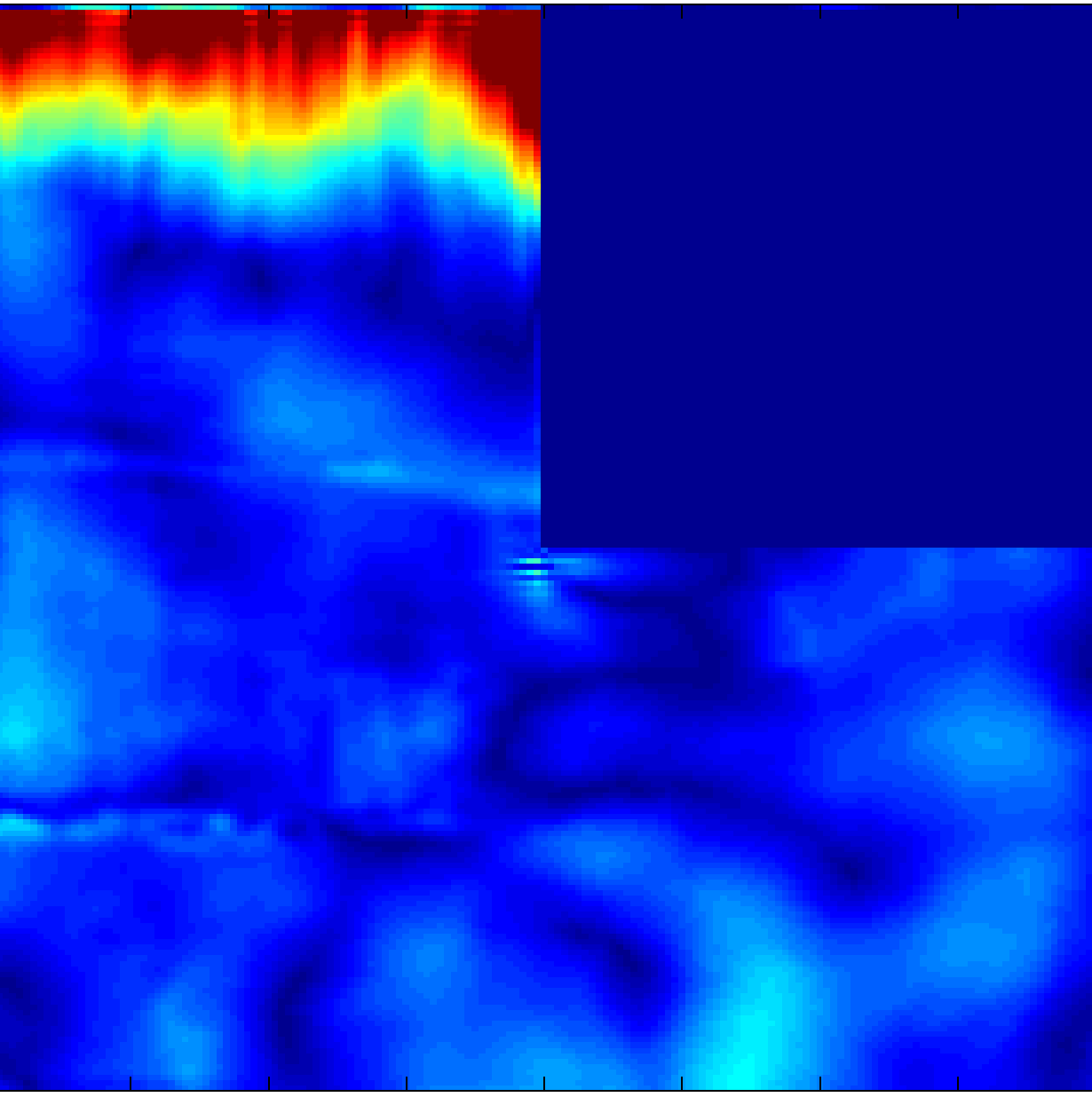}
\includegraphics[trim = 49mm 78.5mm 36mm 92mm, clip,angle=90, width=0.31\textwidth]{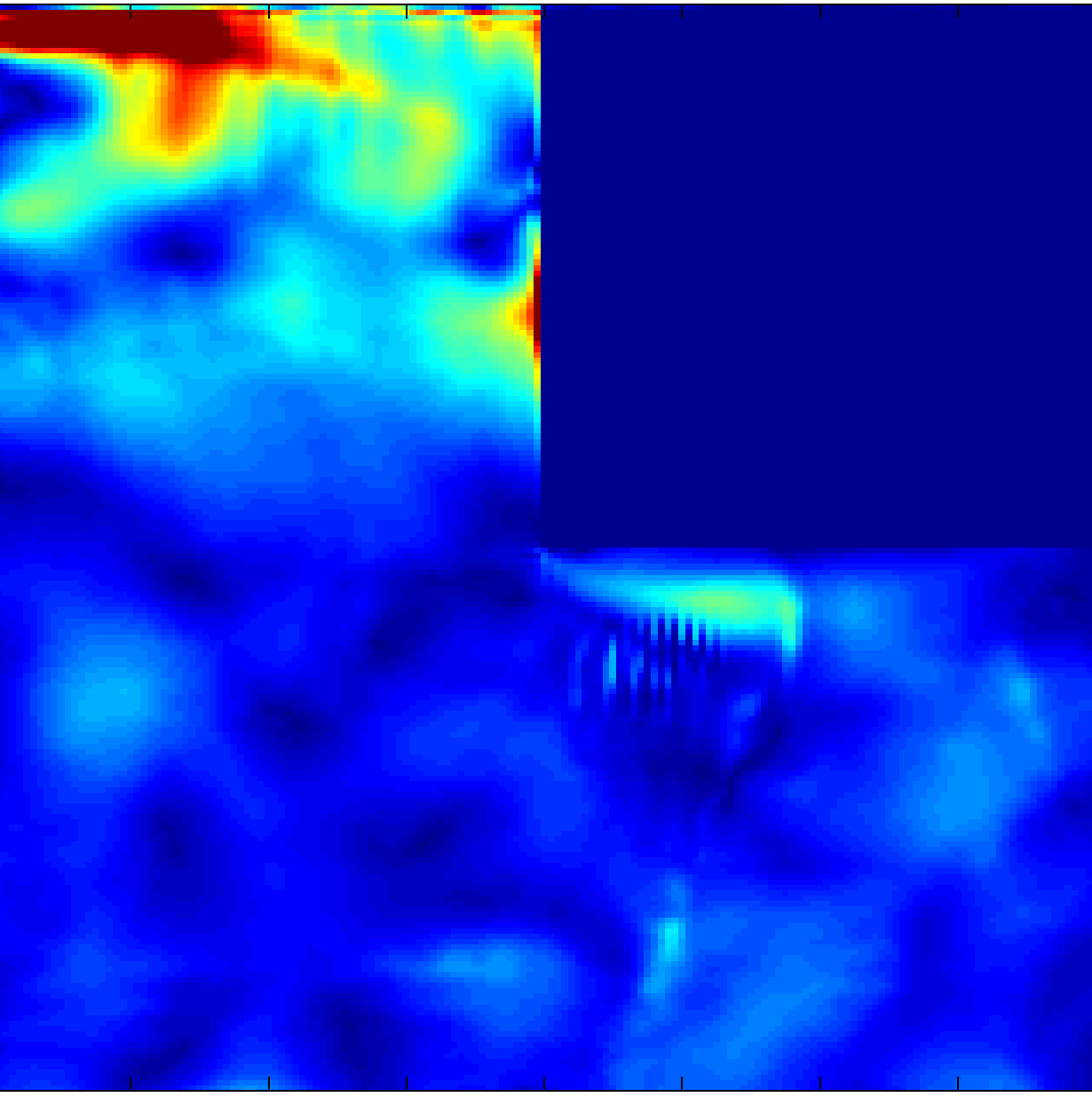}
\includegraphics[trim = 43mm 185mm 35mm 100mm, clip, angle=90,width=0.0285\textwidth]{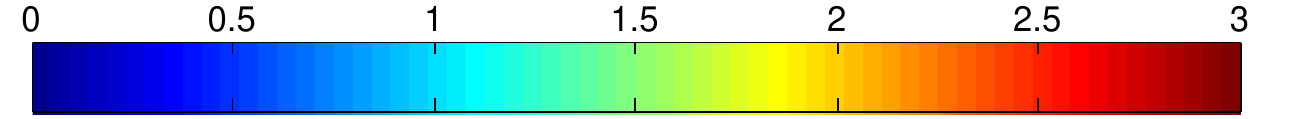}\\
{\small (b)}\\
\includegraphics[trim = 49mm 78.5mm 36mm 92mm, clip,angle=90, width=0.31\textwidth]{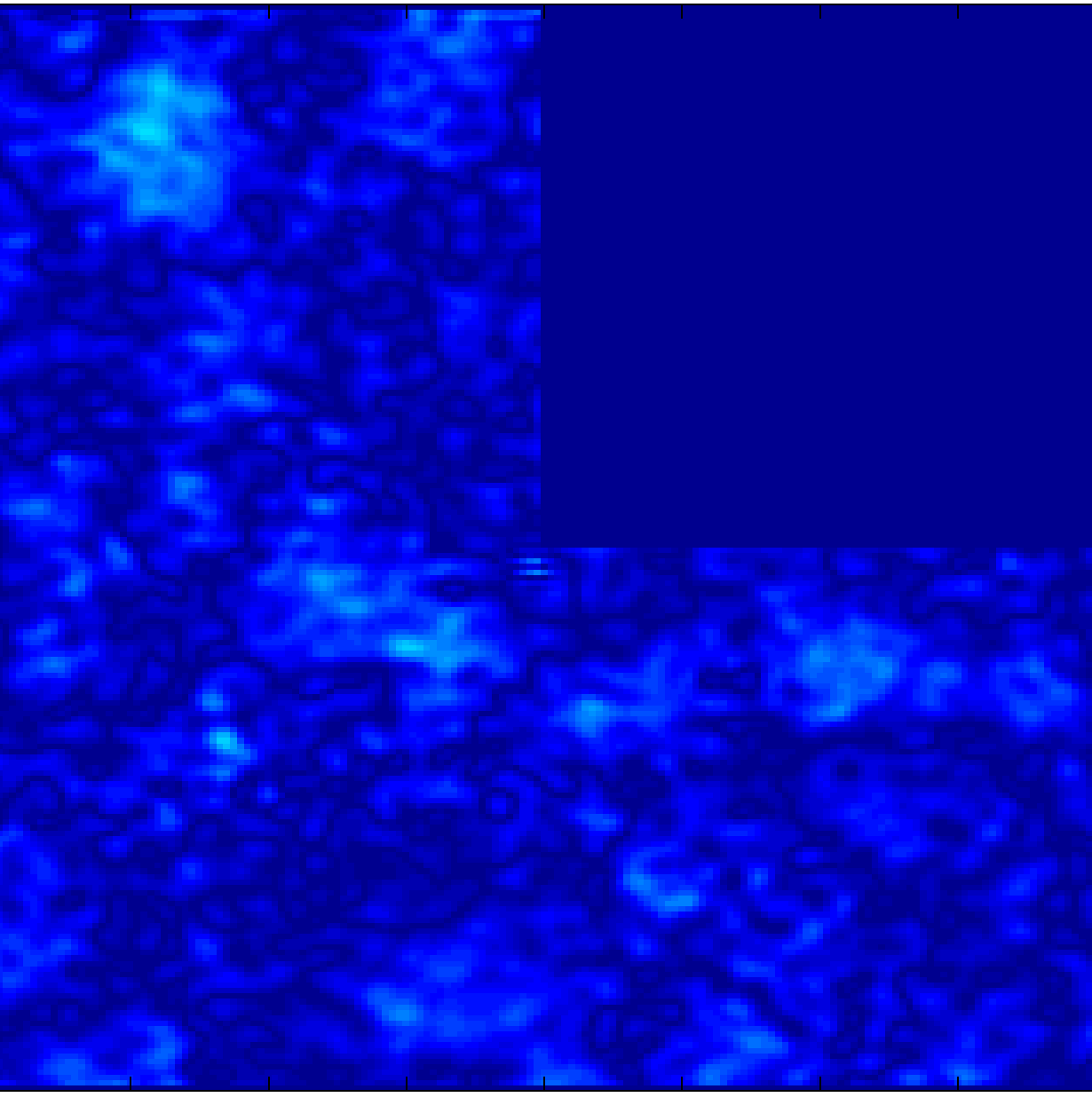}
\includegraphics[trim = 49mm 78.5mm 36mm 92mm, clip,angle=90, width=0.31\textwidth]{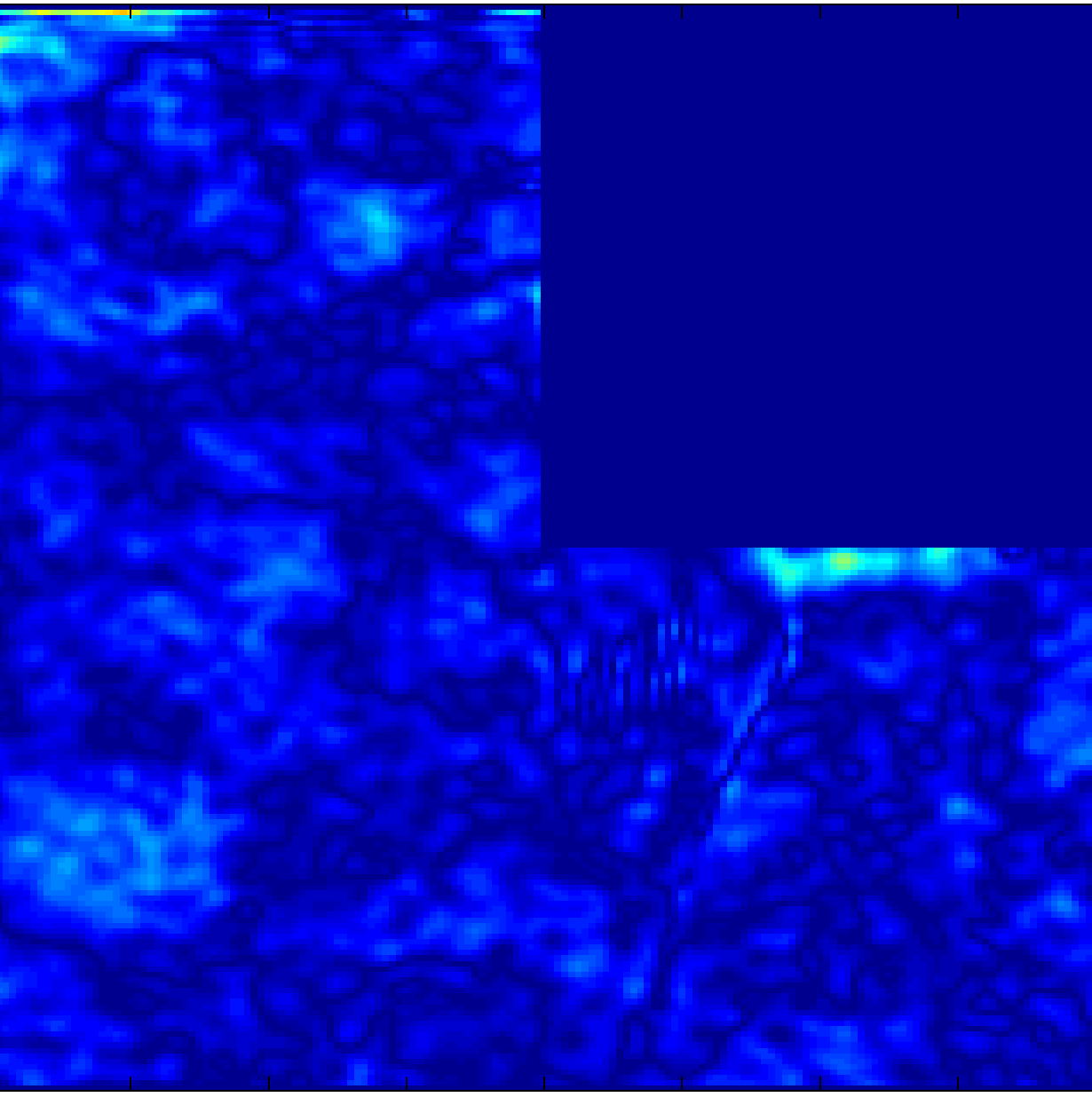}
\includegraphics[trim = 43mm 185mm 35mm 100mm, clip, angle=90,width=0.0285\textwidth]{barChannelErrorElevation.pdf}\\
\includegraphics[trim = 49mm 78.5mm 36mm 92mm, clip,angle=90, width=0.31\textwidth]{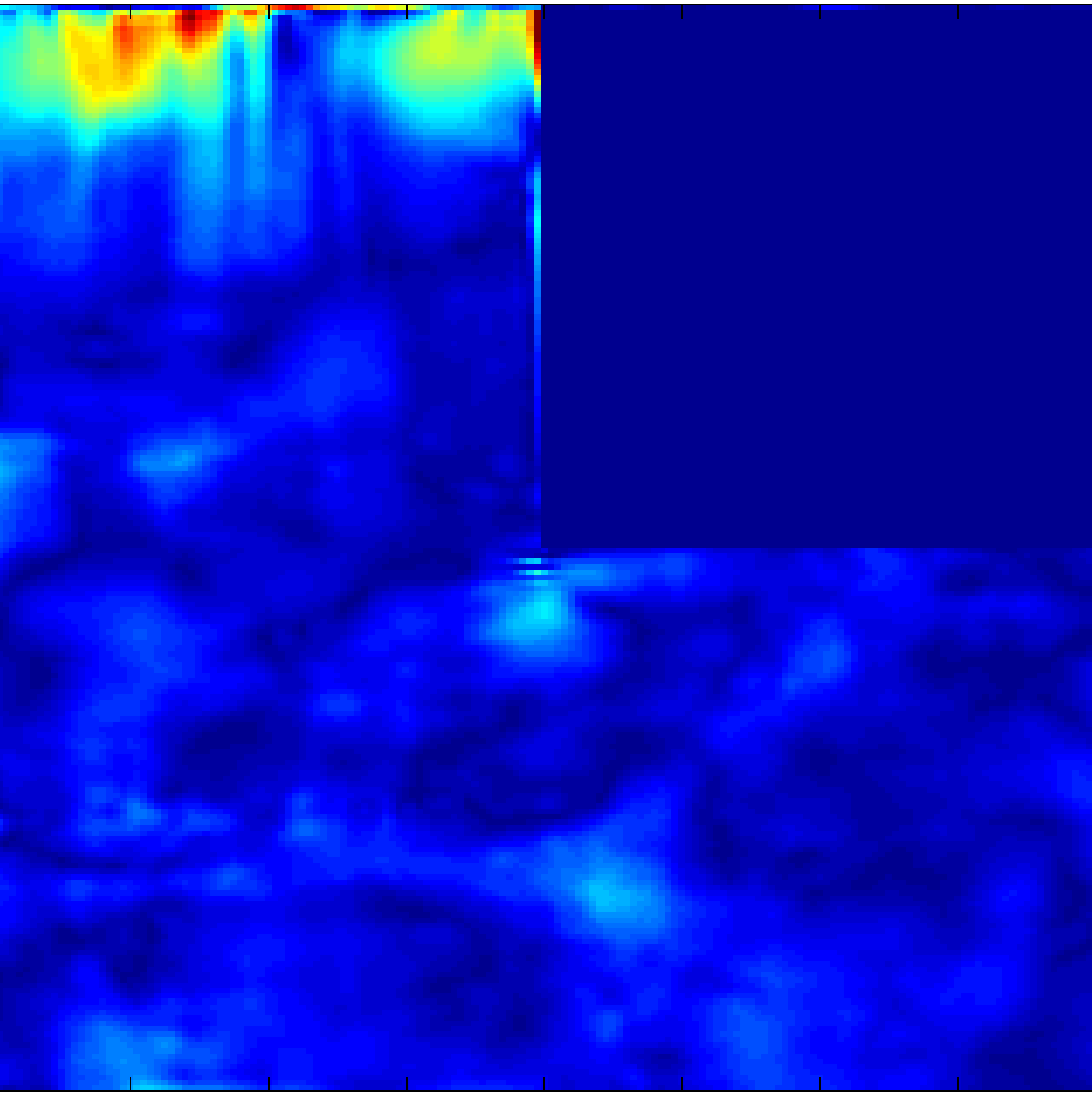}
\includegraphics[trim = 49mm 78.5mm 36mm 92mm, clip,angle=90, width=0.31\textwidth]{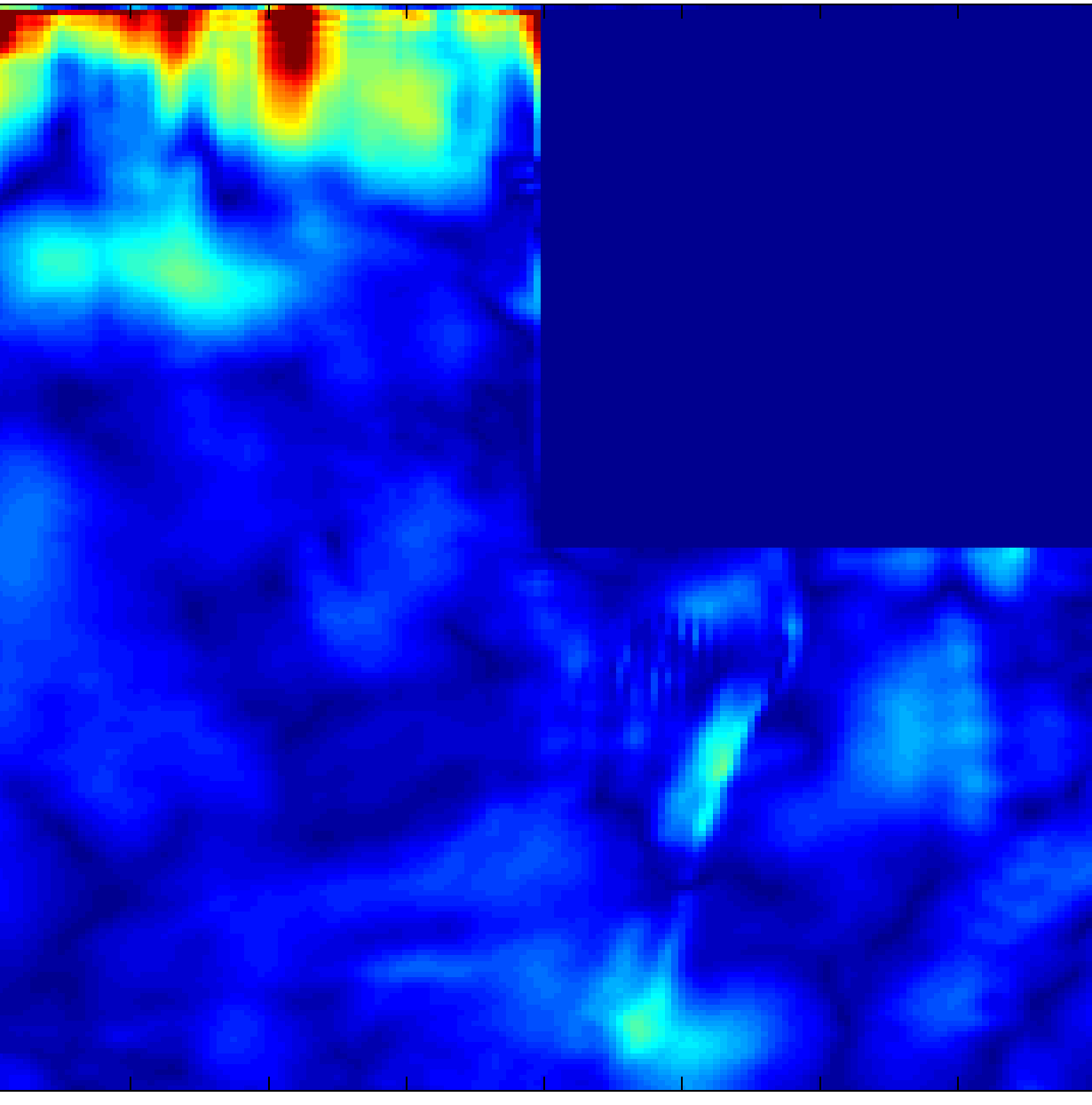}
\includegraphics[trim = 43mm 185mm 35mm 100mm, clip, angle=90,width=0.0285\textwidth]{barChannelErrorDisplacement.pdf}
\caption{Error maps for the suddenly expanding flume flow with stochastic forcing: (a), estimations for oneObs; (b), estimations for twoObs. Left, $t\,u_0/L=1.19$. Right, $t\,u_0/L=1.98$. First rows: elevation errors $\mathcal{E}_{\hat h}~\mbox{in~[\%]}$. Second rows:  velocity errors $\mathcal{E}_{\hat{\bar{u}},\hat{\bar{v}}}~\mbox{in~[\%]}$.}
\label{fig:errorMapSingleVsTwo}
\end{figure}

Results showing the particle filter estimations for an homogeneous inlet velocity profile are given in figure \ref{fig:channel:theRegimeOne} together with the true states. The free surface elevation and motion were reconstructed well for all the time steps considered, but were in less agreement near the inlet area especially when the surface dynamic was more complex due to the stochastic forcing. As reflected in figure \ref{fig:channel:theRegimeTwo} this behaviour was emphasized with more complex half-bell shape inlet velocity profile. The errors of the estimations performed in the configuration with stochastic forcing and uniform inlet velocity profile are displayed in figure \ref{fig:errorMapSingleVsTwo}. In the inlet region the highest errors reached $20\%$ and $43\%$ for the elevation and for the motion, respectively. However, the use of two observations (twoObs) instead of only one (oneObs) improved significantly the agreement for both the elevation and the motion. This demonstrated the capability of the proposed particle filter scheme to deal with complex free surface flow configurations, and as it was expected the use of more observations was necessary to reconstruct better the flow. Table \ref{tad:scoreOpenChannel} summarizes the influence of the inlet velocity profile shape, of the stochastic forcing and of the number of observations considered, on the estimations accuracy. These results also indicate the gain obtained using the observations driving the model instead of running the model alone. The ratios between assimilation and free-run errors for the velocity $\mathcal{E}_{\hat U,\hat V}/\mathcal{E}_{U,V}$ and for the elevation $\mathcal{E}_{\hat h}/\mathcal{E}_{h}$ increased with the complexity of the flow and decreased when using two observations instead of one. This illustrates the importance of coupling the data and the dynamical model to reconstruct the overall flow state (elevation and motion) with a single partial observation (elevation at the inflow location at a given time).

\section{Experimental demonstration}\label{sec:experimental_valid}
We now characterize the ability of the proposed WEnKF scheme, to reconstruct free surface real flows geometry and motion, given only Kinect based measurements of the elevation. 
\subsection{Kinect depth sensor}\label{ssec:kinect}
Capturing the geometry of a 3D object is of great interest for many applications. Different strategies (e.g. stereovision, shape from deformations) handling different technologies (e.g. binocular system, structured light) have lead to a large variety of sensors having different characteristics. We report the reader to the work of \citeasnoun{Herbort11a} for an introduction to depth sensors. Note that recently stereoscopic particle images velocimetry approaches have been proposed to estimate simultaneously morphology and velocity of moving free surfaces \cite{turney_etal_2009,gomit_etal_2013,chatellier_etal_2013}. 
In this work, we choose to use the Kinect sensor to estimate the geometry of a free surface flow for the first time \cite{combes_etal_2011}. More recently \citeasnoun{mankoff_russo_2013} introduced the Kinect sensor to the earth science community. This depth sensor has the advantage to be cheap (about 150 \euro), to provide high-level programming interfaces and not to need any specific calibration. These three characteristics make it a good candidate for practical use. It is partly composed of a $320 \times 240$ RGB sensor at $30~\mbox{Hz}$, a $320 \times 240$ infrared sensor at $30~\mbox{Hz}$ and an infrared pattern projector. Range images are obtained from the so-called light-coding technique: the projection of the infrared pattern on the object under study is captured by the infrared sensor and the analysis of this projection is used to recover the geometry of the object. This approach is close to the optical profilometry technique proposed by \citeasnoun{cobelli09a} for the measurement of water waves and in practice provides range images of $640\times480$ pixels at $30~\mbox{Hz}$. To the best of our knowledge, the Kinect sensor is not well-documented and the main available information are shared by users from the web. Note that in the paper of \citeasnoun{mankoff_russo_2013} hardware and software are described and a code for data processing is provided.

\begin{figure}[h]
\centering
\includegraphics[width=0.19\textwidth]{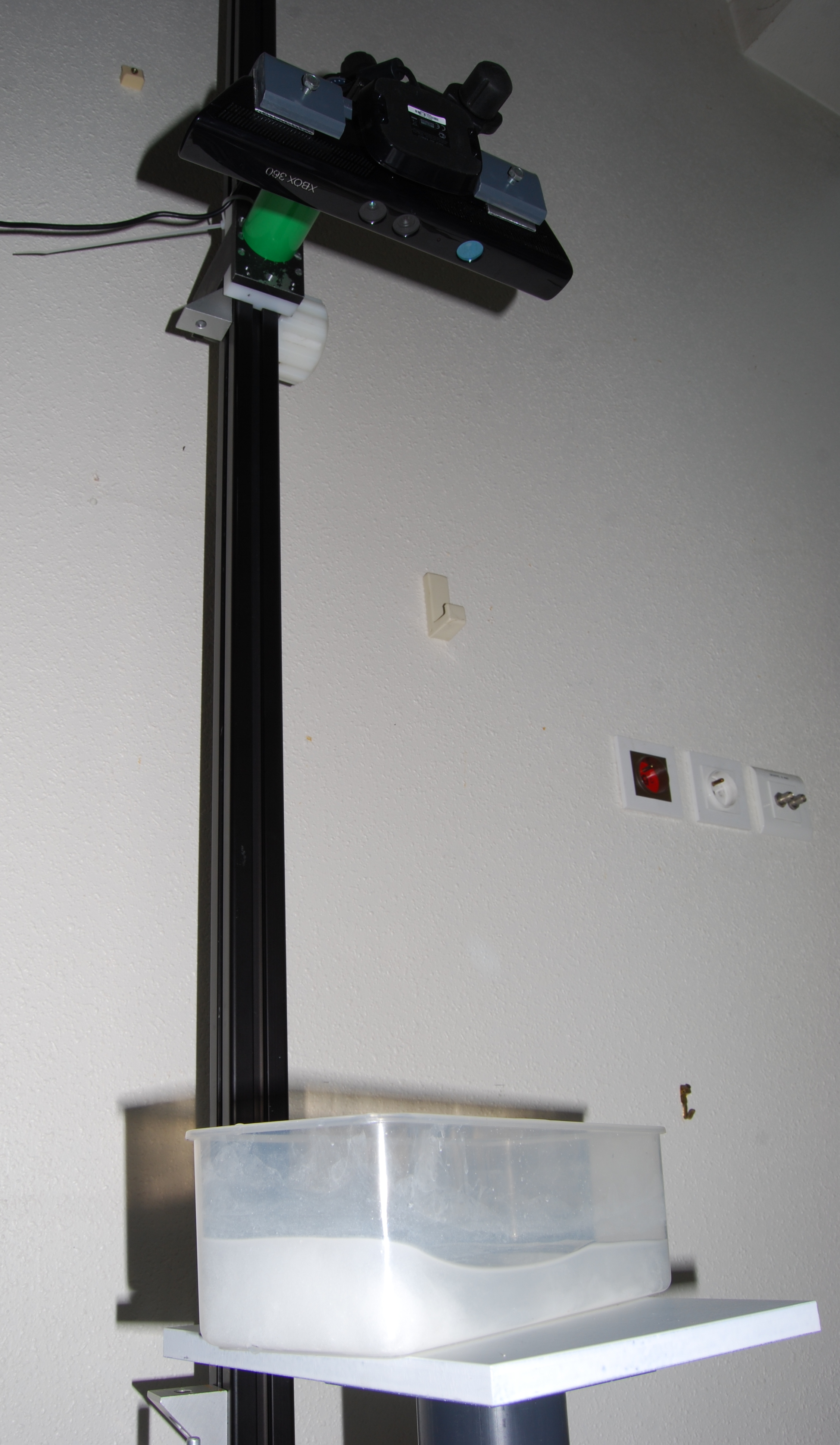}
\hspace*{2cm}
\includegraphics[width=0.55\textwidth]{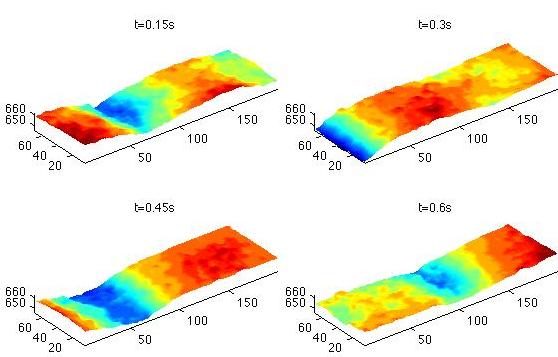}
\caption{3D water surface geometry estimated with a Kinect sensor: left, picture of the experimental setting when observing a moving wave in a water tank; right, 3D reconstruction of the wave moving in the tank from Kinect acquisitions at $t=0.15, 0.3, 0.45$ and $0.6$ seconds. All spatial scales are in millimeters. From \protect\citeasnoun{combes_etal_2011}.}
\label{fig:3Dreconstr}
\end{figure}

 In this work, we do not focus on the technology of the Kinect and of its subsequent limitations and facilities but use it, as best as we can, in a "black-box" way. For that purpose, we investigated its accuracy to estimate range images of solid and liquid smooth surfaces. Liquid's light diffusivity was enhanced by the addition of white dye for the use of the Kinect sensor. For a sake of simplicity, our study was restricted to surfaces located between 680 and 780 mm from the device. However, note that the devise is able to estimate depth data up to 13 meters leading to an observable surface of about $10~\mbox{m} \times 14~\mbox{m}$ with a magnification of about $2~\mbox{cm}$ near the optical axis and $20~\mbox{cm}$ at the periphery of the observable window. A more complete description of our experiments can be found in \citeasnoun{combes_etal_2011}. The main conclusion of our studies were the following: On solid surface, the Kinect sensor displayed a measurement uncertainty of $0.9~\mbox{mm}$ for both flat and sinus-like surfaces; The sensor captured successfully sinus-like varying elevations with spatial periods smaller than 20 mm and amplitudes smaller than 2 mm; Measurement errors coming from observations of solid and liquid surfaces were comparable when the attenuation coefficient of the liquid was larger than 113~${\rm m}^{-1}$. As an illustration, figure \ref{fig:3Dreconstr} shows a 3D temporal reconstruction of a water wave moving in a rectangular flat bottom tank, captured by the Kinect. 

\subsection{Wave in a rectangular flat bottom tank}
The real experiments carried out in this study consisted in observing the free surface of a fluid contained in a rectangular flat bottom tank of size $Lx \times
Ly = 250~\mbox{mm} \times 100~\mbox{mm}$ as illustrated in figure \ref{fig:3Dreconstr}. More specifically we observed the evolution of a unidirectional wave generated by an initial free surface height difference $h_0=1~\mbox{cm}$. In the following the characteristic velocity $u_0$ is considered as an approximation of wave phase velocity $\sqrt{g\,h_0}$. The wave propagation was simulated with the shallow-water numerical model in a computational domain $L_x \times L_y$ discretized on a square grid of $n_x \times n_y= 222 \times 88$ points and with a time step $\Delta t\,u_0/L_x=0.0042$.

The Kinect depth observations were characterized by a high level of noise and exhibited large regions of missing data on the boundaries due to light reflections on the tank’s wall. The observed free surface behaved roughly as an unidirectional wave along the $x$ axis. The initial state $\x_{\rm init}$ was considered with a flat surface and a null velocity although the real state was a moving wave. The observations $\y_{1 \ldots t}$ were assimilated every $10 \Delta t\,u_0/L$, leading to an observations Strouhal number $St_{\rm obs}=L/(10 \Delta t\,u_0) \simeq 24$, that was rather high. We tested ensemble size from 20 to 200 and 
in practice when $N>100$ we do no more observe any significant reduction the estimation error. As a result, we use $N=100$.  Assimilating each observation took about 10 minutes on a standard personal computer.

\begin{figure}[h!]
\centering
\includegraphics[trim = 80.4mm 87mm 91.5mm 82mm,clip,width=0.285\textwidth,angle=0,height=0.4\textwidth]{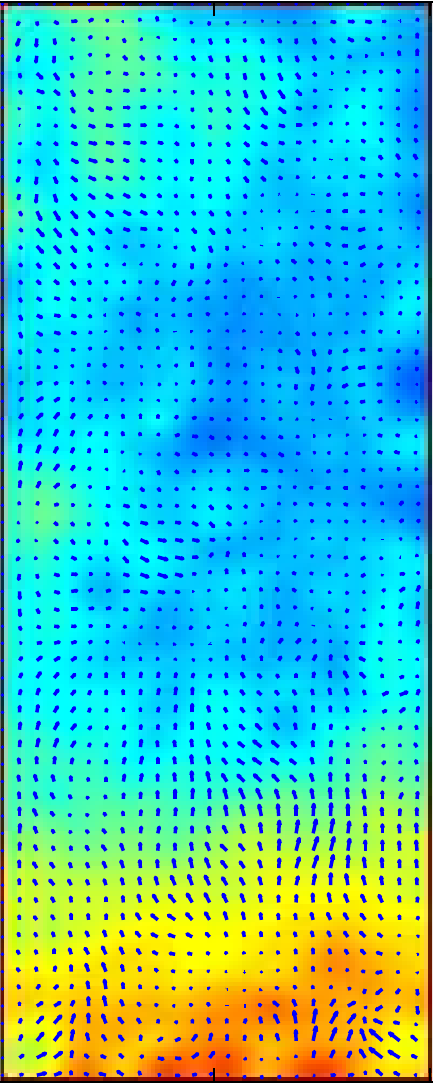}
\includegraphics[trim = 80.4mm 87mm 91.5mm 82mm,clip,width=0.285\textwidth,angle=0,height=0.4\textwidth]{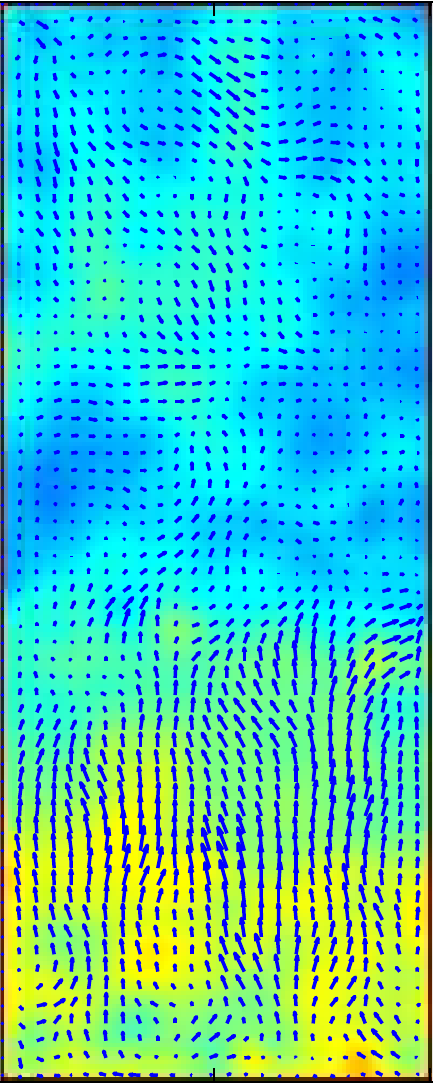}
\includegraphics[trim = 80.4mm 87mm 91.5mm 82mm,clip,width=0.285\textwidth,angle=0,height=0.4\textwidth]{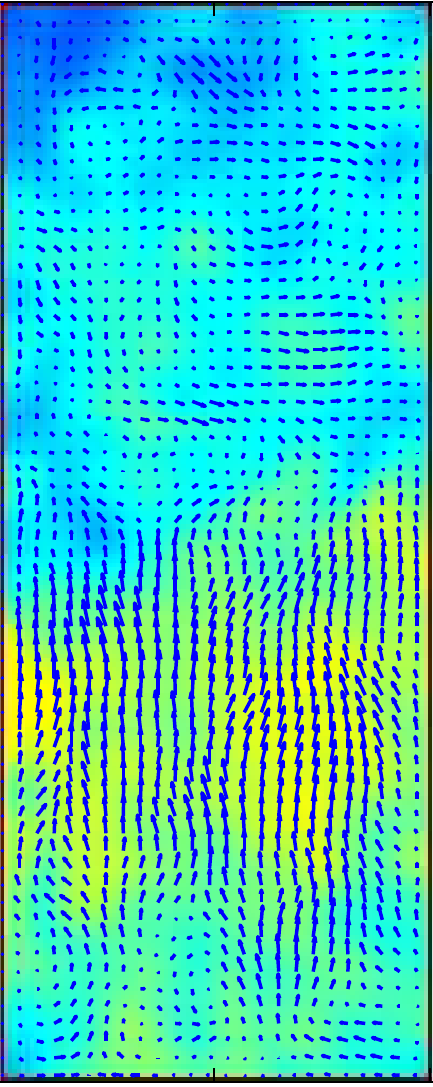}
\includegraphics[trim = 80.4mm 87mm 91.5mm 82mm,clip,width=0.285\textwidth,angle=0,height=0.4\textwidth]{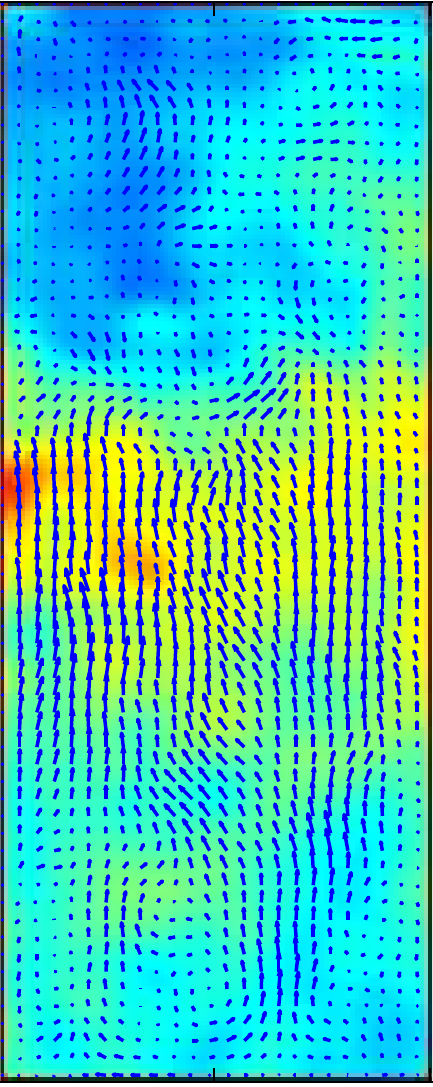}
\includegraphics[trim = 80.4mm 87mm 91.5mm 82mm,clip,width=0.285\textwidth,angle=0,height=0.4\textwidth]{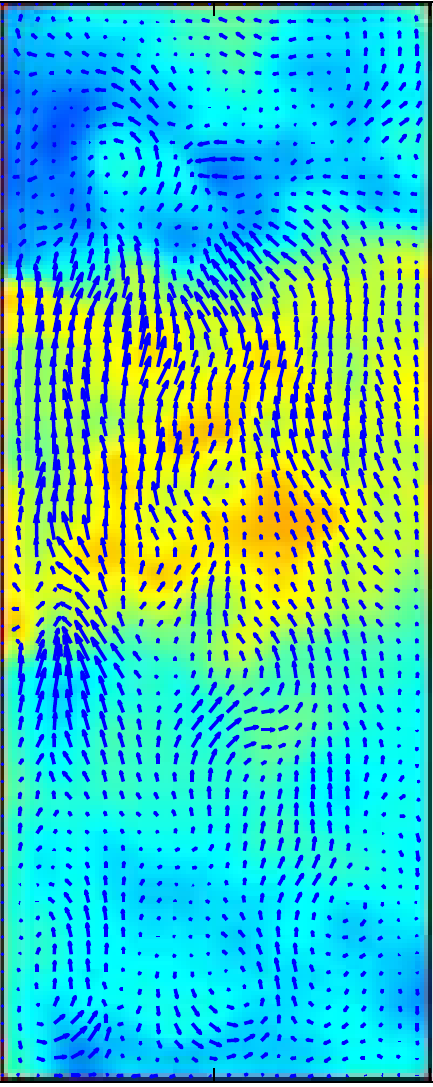}
\includegraphics[trim = 80.4mm 87mm 91.5mm 82mm,clip,width=0.285\textwidth,angle=0,height=0.4\textwidth]{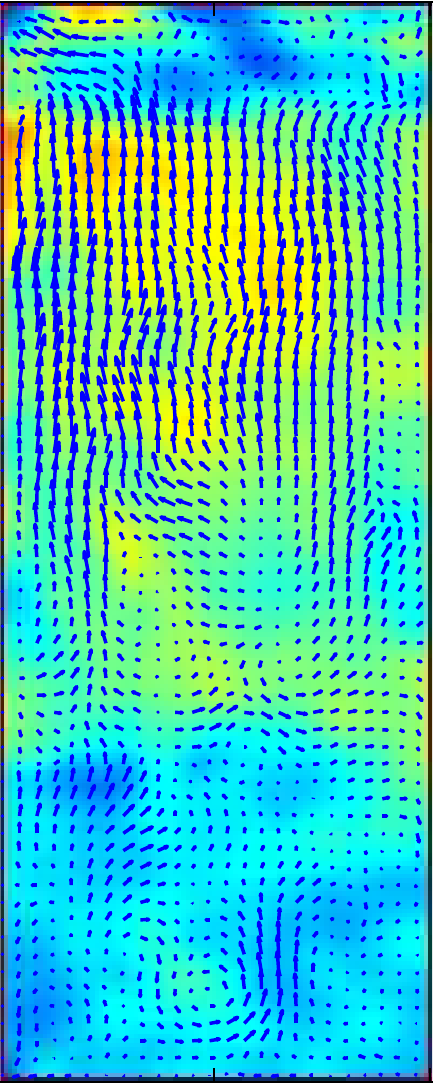}
\includegraphics[trim = 80.4mm 87mm 91.5mm 82mm,clip,width=0.285\textwidth,angle=0,height=0.4\textwidth]{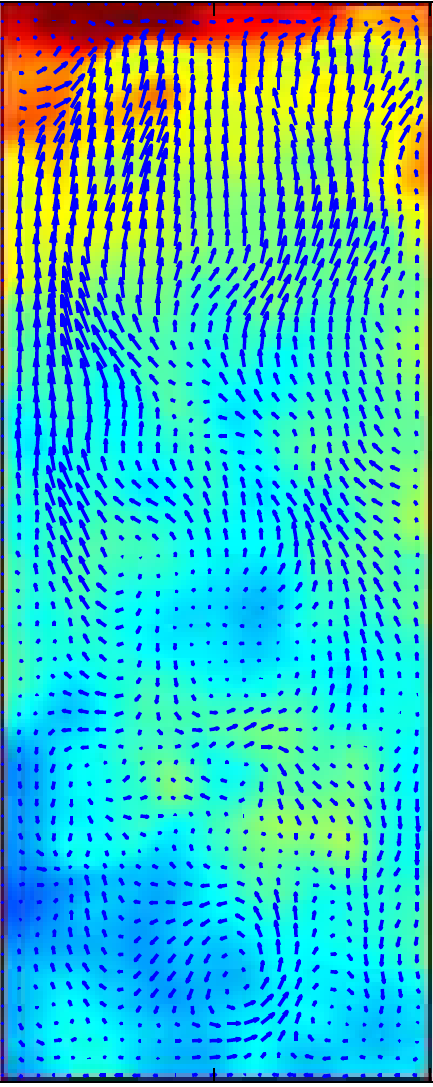}
\includegraphics[trim = 80.4mm 87mm 91.5mm 82mm,clip,width=0.285\textwidth,angle=0,height=0.4\textwidth]{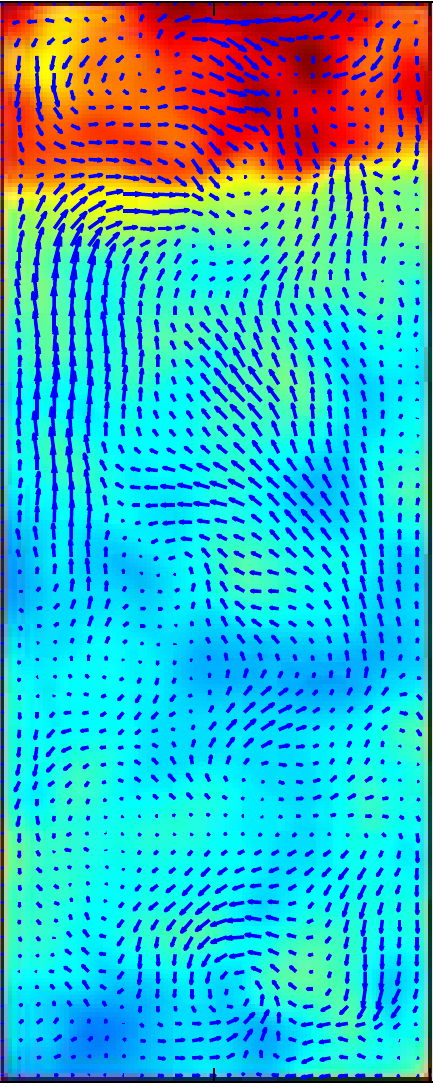}
\includegraphics[trim = 80.4mm 87mm 91.5mm 82mm,clip,width=0.285\textwidth,angle=0,height=0.4\textwidth]{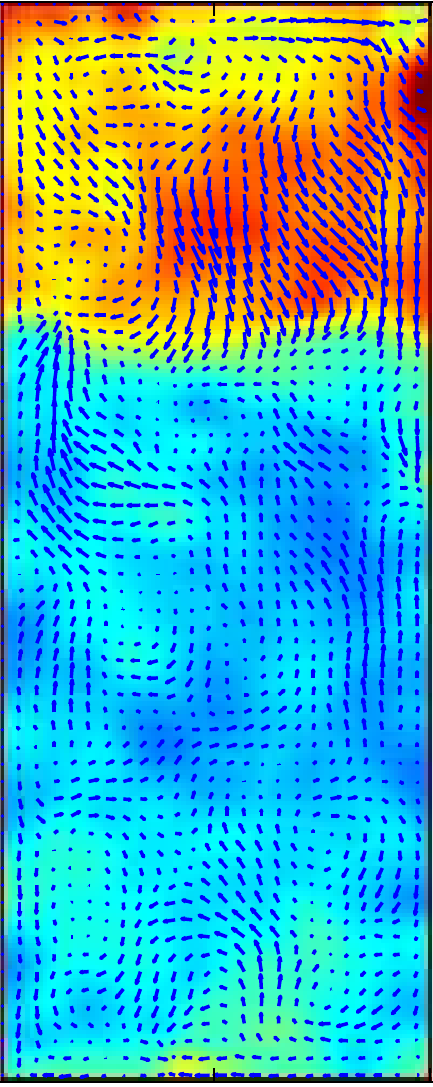}
\includegraphics[trim = 80.4mm 87mm 91.5mm 82mm,clip,width=0.285\textwidth,angle=0,height=0.4\textwidth]{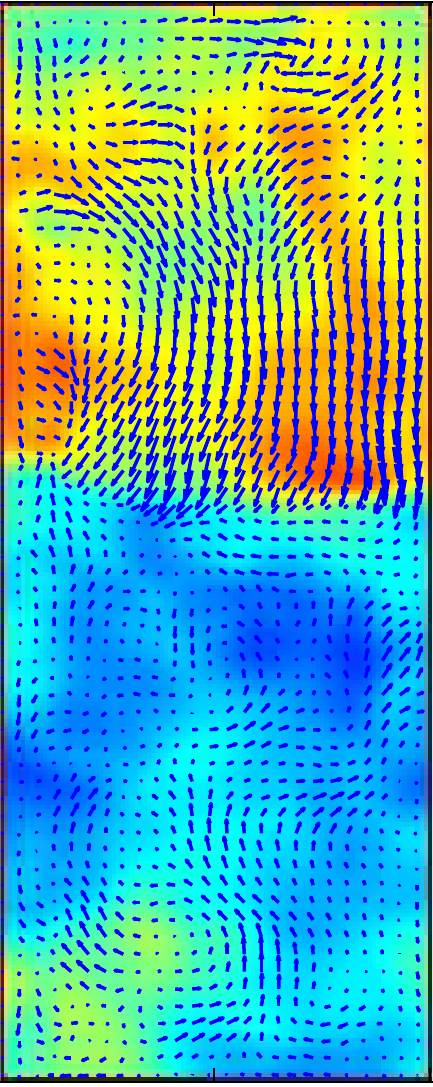}
\includegraphics[trim = 80.4mm 87mm 91.5mm 82mm,clip,width=0.285\textwidth,angle=0,height=0.4\textwidth]{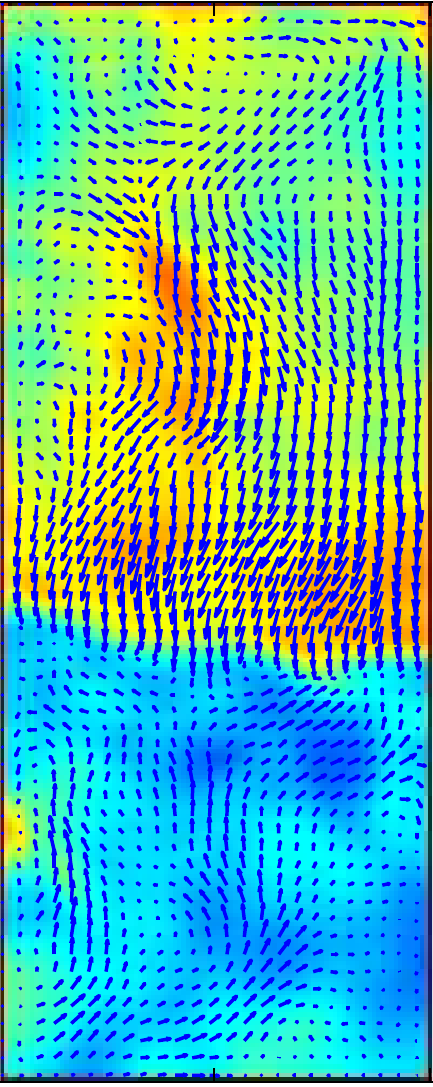}
\includegraphics[trim = 80.4mm 87mm 91.5mm 82mm,clip,width=0.285\textwidth,angle=0,height=0.4\textwidth]{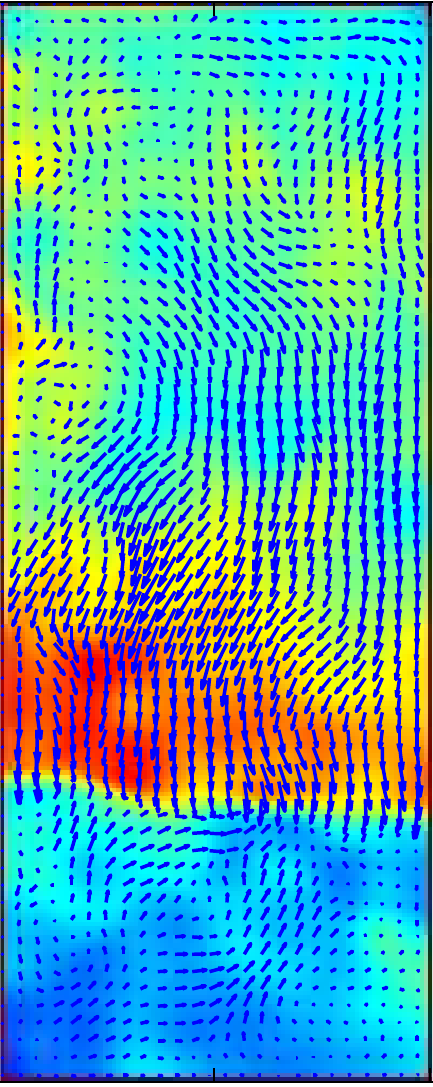}
\vskip -0.5cm
\includegraphics[trim = 174mm 80mm 20mm 80mm,clip,angle=-90,width=0.4\textwidth]{stateAfterCorrection02UANDLevel.pdf}
\caption{Elevation colormaps in millimeters and free-surface motion vector fields for a wave in a rectangular flat bottom tank. From left to right and top to bottom, snapshots of the estimation displayed every 20 time step $\Delta t u_0/L_x$.}
\label{fig:illusRealFlow1}
\end{figure}

Figure \ref{fig:illusRealFlow1} shows estimated free-surface elevation colormap and motion vector fields for a water wave propagating in a rectangular flat bottom tank. This time sequence in a plan view exhibits one back and forth of the wave. 
The reconstructed height sequence is a compromise between noisy Kinect observations and an imperfect dynamical model, exhibiting the spatiotemporal complexity of the free surface motion. The predicted free surface motion fields indicated the highest velocities on the wave crest and recirculation regions on either side upstream and downstream of the wave. Providing a physically consistent initial state for such a real flow was not straightforward: we could use the measured elevation however the associated motion fields was difficult to model. Interestingly, although we initiated the assimilation process with a flat surface and a null velocity, as time proceeds the particle filter rapidly provided accurate estimations. The whole free surface geometry dynamics was remarkably reconstructed from noisy measures of the elevation only.

\section{Conclusion \& perspectives}
\label{sec:concl}
Most of the time there are only temporal sequences of sparse data coming from lagrangian or eulerian measurements of the flow elevation to characterize a free surface flow dynamics. Although sparse and noisy, these observations when properly coupled with a shallow water model contain enough information to reconstruct the whole free surface flow dynamics. 

This paper has presented the first free surface flow reconstruction using Kinect based depth measurements only. The Kinect sensor was able to capture observations of wave-like surfaces with wavelengths and amplitudes sufficiently small to support applications such as flow monitoring or medium to large scale flows characterization. The sequential data assimilation algorithm, based on a particle filter stochastic approach, has been validated on two numerical cases and a real experiment in laboratory. Results have shown remarkable reconstruction of both elevation and velocities of the free surface flows. In addition, the influence of noise and outliers in the input depth data has been assessed, indicating that the method has exhibited good robustness properties up to large deteriorations of the data quality. Another feature of the proposed WEnKF scheme has been the quasi non-sensitivity to the initial state perturbation levels. When starting from flat surface and null velocity the technique has rapidly and accurately recovered the dynamics of complex free surface flows. The enhancement of the particle filter with two observations instead of classically using only one has improved significantly the agreement for both the elevation and the motion. Finally, in contrast with data assimilation methods considering ``perfect'' models, the stochastic data-assimilation scheme carried out in this study has yielded remarkable estimations, given the complexity of the flow configurations and the limitations of the dynamical model.

Note that all the presented results characterized acquisitions within an area near the image center and from a short range. Future works will consist in investigating the data assimilation technique over larger ranges (we recall that the Kinect is able to estimate depth data up to 13 meters leading to an observable surface of about $10~\mbox{m}\times 14~\mbox{m}$ with a magnification of about 2 cm near the optical axis and $20~\mbox{cm}$ at the periphery of the observable window) and under more practical conditions such as discharge estimation. Future works will also consist in estimating parameters of the model such as the bed roughness in addition to the model noise term in order to reduce the departure of our model to the actual dynamic of the fluid and ultimately to estimate more accurately the states of the system. 


\References

\harvarditem{B\'{e}langer \harvardand\ Vincent}{2005}{belanger05a}
B\'{e}langer E \harvardand\ Vincent A  2005 Data assimilation {(4D-VAR)} to
  forecast flood in shallow-waters with sediment erosion {\em J. Hydrol.} {\bf
  300}(1--4)~114--125

\harvarditem{Bradford \harvardand\ Sanders}{2002}{bradford_etal_2002}
Bradford S~F \harvardand\ Sanders B~F  2002 Finite-volume model for
  shallow-water flooding of arbitrary topography {\em J. Hydraul. Eng.} {\bf
  128}~289--298

\harvarditem{Buhmann}{2003}{buhmann03a}
Buhmann M~D  2003 {\em Radial {B}asis {F}unctions} Cambridge University Press

\harvarditem{Castaings {\it et~al\/}}{2006}{castaings06a}
Castaings W, Dartus D, Honnorat M, Le~Dimet F~X, Loukili Y \harvardand\ Monnier
  J  2006 {Automatic differentiation: a tool for variational data assimilation
  and adjoint sensitivity analysis for flood modeling} {\em Lecture Notes in
  Computational Science and Engineering} {\bf 50}~249--262

\harvarditem{Chatellier {\it et~al\/}}{2013}{chatellier_etal_2013}
Chatellier L, Jarny S, Gibouin F \harvardand\ David L  2013 A parametric
  {PIV/DIC} method for the measurement of free surface flows {\em Exp. Fluids}
  {\bf 54}(3)~1488--1502

\harvarditem{Cobelli {\it et~al\/}}{2009}{cobelli09a}
Cobelli P, Maurel A, Pagneux V \harvardand\ Petitjeans P  2009 Global
  measurement of water waves by {F}ourier transform profilometry {\em Exp.
  Fluids} {\bf 46}~1037--1047

\harvarditem{Colburn {\it et~al\/}}{2011}{colburn_etal_2011}
Colburn C, Cessna J \harvardand\ Bewley T  2011 State estimation in
  wall-bounded flow systems. {P}art 3. {T}he ensemble {K}alman filter {\em J.
  Fluid Mech.} {\bf 682}~289--303

\harvarditem{Comb\`es {\it et~al\/}}{2011}{combes_etal_2011}
Comb\`es B, Guibert A, M\'emin E \harvardand\ Heitz D  2011 Free-surface flows
  from {K}inect: feasability and limits {\em in} `The Forum on Recent
  Developments in Volume Reconstruction Techniques Applied to 3D Fluid and
  Solid Mechanics (FVR2011)' Chasseneuil, France

\harvarditem{Cuzol {\it et~al\/}}{2007}{cuzol_etal_2007}
Cuzol A, Hellier P \harvardand\ M\'emin E  2007 A low dimensional fluid motion
  estimator {\em Int. J. Computer Vision} {\bf 75}(3)~329--349

\harvarditem{Cuzol \harvardand\ M\'emin}{2009}{cuzol_memin_2009}
Cuzol A \harvardand\ M\'emin E  2009 A stochastic filtering technique for fluid
  flows velocity fields tracking {\em IEEE Trans. on Pattern Analysis and
  Machine Intelligence} {\bf 31}(7)~1278--1293

\harvarditem{D'Adamo {\it et~al\/}}{2007}{dadamo_etal_2007}
D'Adamo J, Papadakis N, M\'emin E \harvardand\ Artana G  2007 Variational
  assimilation of {POD} low-order dynamical systems {\em J. Turb.} {\bf
  8}(9)~1--22

\harvarditem{Ding {\it et~al\/}}{2004}{ding04a}
Ding Y, Jia Y \harvardand\ Wang S~S~Y  2004 {Identification of Manning's
  Roughness Coefficients in Shallow Water Flows} {\em Journal of Hydraulic
  Engineering} {\bf 130}(6)~501--510

\harvarditem{Doucet {\it et~al\/}}{2000}{doucet00a}
Doucet A, Godsill S \harvardand\ Andrieu C  2000 On sequential monte carlo
  sampling methods for bayesian filtering {\em {STATISTICS} {AND} {COMPUTING}}
  {\bf 10}(3)~197--208

\harvarditem{Evensen}{1994}{Evensen94a}
Evensen G  1994 Sequential data assimilation with a nonlinear quasi-geostrophic
  model using monte carlo methods to forecast error statistics {\em J.
  Geophysical Research} {\bf 99}(C5)~10,143--10,162

\harvarditem{Evensen}{2003}{evensen03a}
Evensen G  2003 The {E}nsemble {K}alman {F}ilter: theoretical formulation and
  practical implementation {\em Ocean Dynamics} {\bf 53}(4)~343--367

\harvarditem{Foures {\it et~al\/}}{2014}{foures_etal_2014}
Foures D, Dovetta N, Sipp D \harvardand\ Schmid P  2014 A data-assimilation
  method for {R}eynolds-averaged {N}avier–{S}tokes-driven mean flow
  reconstruction {\em J. Fluid Mech.} {\bf 759}~404--431

\harvarditem{Golub \harvardand\ Van~Loan}{1996}{golub96a}
Golub G \harvardand\ Van~Loan C  1996 {\em Matrix {C}omputations} Johns Hopkins
  Studies in the Mathematical Sciences Johns Hopkins University Press

\harvarditem{Gomit {\it et~al\/}}{2013}{gomit_etal_2013}
Gomit G, Chatellier L, Calluaud D \harvardand\ David L  2013 Free surface
  measurement by stereo-refraction {\em Exp. Fluids} {\bf 54}(6)~1540--1550

\harvarditem{Gordon {\it et~al\/}}{1993}{gordon93a}
Gordon N, Salmond D \harvardand\ Smith A  1993 Novel approach to
  {nonlinear/non-Gaussian} {B}ayesian state estimation {\em Radar and Signal
  Processing, {IEE} Proceedings F} {\bf 140}(2)~107--113

\harvarditem{Gronskis {\it et~al\/}}{2013}{gronskis_etal_2013}
Gronskis A, Heitz D \harvardand\ M\'emin E  2013 Inflow and initial conditions
  for direct numerical simulation based on adjoint data assimilation {\em J.
  Comp. Phys.} {\bf 242}(6)~480--497

\harvarditem{Hartnack {\it et~al\/}}{2005}{hartnack05a}
Hartnack J, Madsen H \harvardand\ Sorensen J  2005 Data assimilation in a
  combined {1D-2D} flood model {\em in} `Innovation, advances and
  implementation of flood forecasting technology'

\harvarditem{Heitz {\it et~al\/}}{2010}{heitz_etal_2010}
Heitz D, M\'emin E \harvardand\ Schn\"orr C  2010 Variational fluid flow
  measurements from image sequences: synopsis and perspectives {\em Exp.
  Fluids} {\bf 48}(3)~369--393

\harvarditem{Herbort \harvardand\ W\"{o}hler}{2011}{Herbort11a}
Herbort S \harvardand\ W\"{o}hler C  2011 An introduction to image-based {3D}
  surface reconstruction and a survey of photometric stereo methods {\em 3D
  Research} {\bf 2}~1--17

\harvarditem{Honnorat {\it et~al\/}}{2009}{honnorat09a}
Honnorat M, Monnier J \harvardand\ Dimet F~X~L  2009 Lagrangian data
  assimilation for river hydraulics simulations {\em Computing and
  Visualization in Science} {\bf 12}~235--246

\harvarditem{Honnorat {\it et~al\/}}{2010}{honnorat10a}
Honnorat M, Monnier J, Riviere N, Huot E \harvardand\ Le~Dimet F~X  2010
  {Identification of equivalent topography in an open channel flow using
  Lagrangian data assimilation} {\em Computing and Visualization in Science}
  pp~1--8

\harvarditem{Hostache {\it et~al\/}}{2010}{hostache10a}
Hostache R, Lai X, Monnier J \harvardand\ Puech C  2010 Assimilation of
  spatially distributed water levels into a shallow-water flood model. {P}art
  {II}: {U}se of a remote sensing image of {M}osel {R}iver {\em J. Hydrol.}
  {\bf 390}(3-4)~257--268

\harvarditem{Hoteit {\it et~al\/}}{2008}{Hoteit08a}
Hoteit I, Pham D~T, Korres G \harvardand\ Triantafyllou G  2008 `Particle
  {K}alman filtering for data assimilation in {M}eteorology and
  {O}ceanography.'

\harvarditem{Houtekamer \harvardand\ Mitchell}{2001}{houtekamer01a}
Houtekamer P \harvardand\ Mitchell H  2001 A {S}equential {E}nsemble {K}alman
  {F}ilter for {A}tmospheric {D}ata {A}ssimilation {\em Monthly Weather Review}
  {\bf 129}(1)~123--137

\harvarditem{Jazwinski}{1970}{Jazwinski70a}
Jazwinski A  1970 {\em Stochastic Processes and Filtering Theory} {{Academic}
  Press}

\harvarditem{Lai \harvardand\ Monnier}{2009}{lai09a}
Lai X \harvardand\ Monnier J  2009 Assimilation of spatially distributed water
  levels into a shallow-water flood model. {P}art {I}: {M}athematical method
  and test case {\em J. Hydrol.} {\bf 377}(1-2)~1--11

\harvarditem{Le~Gland {\it et~al\/}}{2009}{legland09a}
Le~Gland F, Monbet V \harvardand\ Tran V~D  2009 {Large {S}ample {A}symptotics
  for the {E}nsemble {K}alman {F}ilter} Research Report RR-7014 INRIA

\harvarditem{Madsen {\it et~al\/}}{2003}{madsen03a}
Madsen H, Hansen F \harvardand\ Damgard J  2003 Data assimilation in the {MIKE
  11} flood forecasting system using kalman filtering {\em in} `{General
  Assembly of the International Union of Geodesy and Geophysics}' pp~75--81

\harvarditem{Mankoff \harvardand\ Russo}{2013}{mankoff_russo_2013}
Mankoff K~D \harvardand\ Russo T~A  2013 The {K}inect: a low-cost,
  high-resolution, short-range {3D} camera {\em Earth Surf. Process. Landforms}
  {\bf 38}(99)~926--936

\harvarditem{Monnier \harvardand\ Gejadze}{2007}{monnier07a}
Monnier J \harvardand\ Gejadze I  2007 {On a 2D zoom for 1D shallow-water
  model: coupling and data assimilation} {\em Comp. Meth. Appl. Mech. Eng.
  (CMAME).} {\bf 196}(45--48)~4628--4643

\harvarditem{Mons {\it et~al\/}}{2014}{mons_etal_2014}
Mons V, Chassaing J~C, Gomez T \harvardand\ Sagaut P  2014 Is isotropic
  turbulence decay governed by asymptotic behavior of large scales? {A}n
  eddy-damped quasi-normal {M}arkovian-based data assimilation study {\em Phys.
  Fluids} {\bf 26}(11)~115105

\harvarditem{Papadakis \harvardand\ M\'emin}{2008}{papadakis_memin_2008}
Papadakis N \harvardand\ M\'emin E  2008 Variational assimilation of fluid
  motion from image sequences {\em SIAM J. Imaging Sci.} {\bf 1}(4)~343--363

\harvarditem{Papadakis {\it et~al\/}}{2010}{Papadakis10a}
Papadakis N, M\'emin E, Cuzol A \harvardand\ Gengembre N  2010 Data
  assimilation with the {W}eighted {E}nsemble {K}alman {F}ilter {\em Tellus A}
  {\bf 62}(5)~673--697

\harvarditem{Powell {\it et~al\/}}{2008}{powell08a}
Powell B, Arango H, Moore A, Lorenzo E~D, Milliff R \harvardand\ Foley D  2008
  4dvar data assimilation in the intra-americas sea with the regional ocean
  modeling system (roms) {\em Ocean Modelling} {\bf 25}(34)~173--188

\harvarditem{Roux \harvardand\ Dartus}{2006}{roux06a}
Roux H \harvardand\ Dartus D  2006 Use of parameter optimization to estimate a
  flood wave: {P}otential applications to remote sensing of rivers {\em J.
  Hydrol.} {\bf 328}~258--266

\harvarditem{Salman {\it et~al\/}}{2006}{salman06a}
Salman H, Kuznetsov L, Jones C~K~R~T \harvardand\ Ide K  2006 A method for
  assimilating lagrangian data into a shallow-water-equation ocean model {\em
  Monthly weather review} {\bf 134}~1081--1101

\harvarditem{Strub, Percelay, Tossavainen \harvardand\ Bayen}{2009}{strub09a}
Strub I, Percelay J, Tossavainen O \harvardand\ Bayen A  2009 {C}omparison of
  {T}wo {D}ata {A}ssimilation {A}lgorithms for {S}hallow {W}ater {F}lows {\em
  Networks and Heterogeneous Media} {\bf 4}(2)~409--430

\harvarditem{Strub, Percelay, Stacey \harvardand\ Bayen}{2009}{strub09b}
Strub I~S, Percelay J, Stacey M~T \harvardand\ Bayen A~M  2009 Inverse
  estimation of open boundary conditions in tidal channels {\em Ocean
  Modelling} {\bf 29}(1)~85--93

\harvarditem{Suzuki}{2012}{suzuki_2012}
Suzuki T  2012 Reduced-order {K}alman-filtered hybrid simulation combining
  particle tracking velocimetry and direct numerical simulation {\em J. Fluid
  Mech.} {\bf 709}~249--288

\harvarditem{Tinka {\it et~al\/}}{2009}{tinka09a}
Tinka A, Strub I, Wu Q \harvardand\ Bayen A  2009 Quadratic {P}rogramming based
  data assimilation with passive drifting sensors for shallow water flows {\em
  in} `International conference on decision and control' pp~7614 --7620

\harvarditem{Titaud {\it et~al\/}}{2010}{titaud10a}
Titaud O, Vidard A, Souopgui I \harvardand\ Le~Dimet F~X  2010 {Assimilation of
  Image Sequences in Numerical Models} {\em Tellus Series A : Dynamic
  meteorology and oceanography} {\bf 62}(1)~30--47

\harvarditem{Turney {\it et~al\/}}{2009}{turney_etal_2009}
Turney D~E, Anderer A \harvardand\ Banerjee S  2009 A method for
  three-dimensional interfacial particle image velocimetry ({3D-IPIV}) of an
  air–water interface {\em Meas. Sci. Technol.} {\bf 20}(4)~045403

\harvarditem{Verlaan \harvardand\ Heemink}{1996}{verlaan96a}
Verlaan M \harvardand\ Heemink A  1996 {Data Assimilation Schemes for
  Non-Linear Shallow Water Flow Models} {\em in} `Advances in Fluid Mechanics'
  Vol. 122 pp~277--286

\harvarditem{Vreugdenhil}{1995}{vreugdenhil95a}
Vreugdenhil C  1995 {\em Numerical {M}ethods for {S}hallow-{W}ater {F}low}
  Springer

\harvarditem{Wolf {\it et~al\/}}{2000}{wolf00a}
Wolf T, Senegas J, Bertino L \harvardand\ Wackernagel H  2000 Application of
  data assimilation to three-dimensional hydrodynamics: The case of the {Odra
  Lagoonn} {\em in} `Geostatistics for Environmental Applications' Kluwer
  Academic pp~157--168

\harvarditem{Yang {\it et~al\/}}{2015}{yang_etal_2014}
Yang Y, Robinson C, Heitz D \harvardand\ M\'emin E  2015 Enhanced
  ensemble-based {4DV}ar scheme for image assimilation {\em Comp. Fluids} {\bf
  115}~201--210

\endrefs

\end{document}